\documentclass[12pt,preprint]{aastex63}
\usepackage{graphicx}
\usepackage[]{xcolor}

\received{...}
\revised{...}
\accepted{...}
\submitjournal{ApJ}

\begin{document}

\title{The Highly Self-Absorbed Blazar, PKS\,1351$-$018}

\correspondingauthor{Brian Punsly}
\email{brian.punsly@cox.net}

\author{Brian Punsly}
\affiliation{1415 Granvia Altamira, Palos Verdes Estates CA, USA 90274}
\affiliation{ICRANet, Piazza della Repubblica 10 Pescara 65100, Italy}
\affiliation{ICRA, Physics Department, University La Sapienza, Roma, Italy}

\author[0000-0003-3079-1889]{S{\'a}ndor Frey}
\affiliation{Konkoly Observatory, ELKH Research Centre for Astronomy and Earth Sciences, Konkoly Thege Mikl{\'o}s {\'u}t 15-17, 1121 Budapest, Hungary}
\affiliation{Institute of Physics, ELTE E{\"o}tv{\"o}s Lor{\'a}nd University, P{\'a}zm{\'a}ny P{\'e}ter s{\'e}t{\'a}ny 1/A, 1117 Budapest, Hungary}

\author{Cormac Reynolds}
\affiliation{CSIRO Astronomy and Space Science, PO Box 1130, Bentley WA 6102, Australia}

\author{Paola Marziani}
\affiliation{INAF, Osservatorio Astronomico di Padova, Italia}

\author{Alexander Pushkarev}
\affiliation{Crimean Astrophysical Observatory, Nauchny 298409, Crimea, Russia}
\affiliation{Astro Space Center of Lebedev Physical Institute, Profsoyuznaya 84/32, Moscow 117997, Russia}

\author{Sina Chen}
\affiliation{Physics Department, Technion Haifa, 32000, Israel}

\author{Shang Li}
\affiliation{School of Physics and Materials Science, Anhui University, Hefei 230601, China}

\author{Preeti Kharb}
\affiliation{National Centre for Radio Astrophysics, Tata Institute of Fundamental Research, Post Bag 3, Ganeshkhind, Pune 411007, India}

\begin{abstract}
PKS\,1351$-$018 at a redshift of $z=3.71$ is one of the most luminous, steady synchrotron sources with a luminosity $> 10^{47}$\,erg~s$^{-1}$. The synchrotron luminosity
does not seem to vary by more than $\sim 25\%$ over 35 years. In order to appreciate this remarkable behavior, if it were at $z=0.5$, it would have a flux density at 15 GHz
in a range of $110 - 137$\,Jy over 11 yrs. In spite of this steady behavior, two strong $\gamma$-ray flares $\lesssim 10^{49}$\,erg~s$^{-1}$ were detected in 2011 and 2016.
There is blazar-like behavior coexisting with the steady behavior. This study is aimed at elucidating the dual nature of this source. We find that the radio source is
extremely compact with a bright core and a steep spectrum secondary component, 12\,mas away, that appears to be constant in position and flux density in six epochs from
1995 to 2018. We estimate that a jet with a time averaged power of $(5.2 \pm 3.2) \times 10^{45}$\,erg~s$^{-1}$
terminates in this lobe that is advancing $\gtrsim 0.9 c$ at a deprojected distance of $1-3$\,kpc from the central engine. This is the rare case of a young ($\sim
6000$\,yr), very powerful radio source that is viewed a few degrees from the jet axis. We find evidence of a high velocity (4000\,km~s$^{-1}$), high ionization wind
emanating form a luminous quasar. The young radio jet appears to experience modest bending as it navigates through the intense quasar environment.
\end{abstract}

\section{Introduction}
\label{intro}

The quasar PKS\,1351$-$018 was identified as one of the most luminous of the known ``ultra-luminous radio cores'' with a synchrotron luminosity of $> 10^{47}$\,erg~s$^{-1}$ \citep{pun95}. This is a high redshift quasar at $z=3.71$ \citep{osm94}. The flux density decreases sharply at frequencies below 6.5 GHz in the quasar rest frame \citep{spo85}. Thus, at one time, it was considered a candidate Gigahertz Peaked Spectrum (GPS) radio source with an observed spectral peak near 1.4\,GHz \citep{spo85,ode91}. In spite of the spectral turnover below 6.5 GHz, it was later rejected as a GPS quasar due to the broad spectral peak \citep{dev97}. PKS\,1351$-$018 also has a steady spectrum based on the twenty-one 5.0 GHz (23.5 GHz in the quasar rest frame) flux density measurements from the Australia Telescope Compact Array (ATCA) calibrator web-page\footnote{\url{http://www.narrabri.atnf.csiro.au/calibrators/}} over a
15-yr period. The mean flux density is 930\,mJy with a standard deviation of 43\,mJy. The measured data variation of $\pm4.7\%$ is similar to the $5\%$ uncertainty that is estimated for individual ATCA flux density measurements \citep{mur10}. Similarly, but statistically less significant, there are eight 22.4 GHz (106 GHz in the quasar rest frame) ATCA calibrator observations over 11 yr with a mean of 543\,mJy and a standard deviation of 35\,mJy or $6.5\%$. In spite of this steady behavior, PKS\,1351$-$018 was detected in $\gamma$-rays by the Large Area Telescope (LAT) on board the \textit{Fermi} satellite \citep{ack17,li18,sah20}. It is this dichotomy, a steady behavior near the spectral peak at cm wavelengths and the extreme blazar-like strong gamma-ray flares and enormous synchrotron luminosity, that has motivated the following detailed study of this extremely powerful jet source in the early Universe.

The paper begins with an in-depth study of the radio light curves to look for evidence of blazar-like phenomenon in Section~\ref{lightcurve}. In
Section~\ref{imaging}, we consider radio interferometer imaging in order to look for blazar-like structure changes and to define the source size. In Section~\ref{sed}, we
construct the synchrotron spectral energy distribution (SED). We analyze the optical spectrum and use it to define the energetics of the accretion flow in
Section~\ref{energetics}.
We follow this up with a depiction of the $\gamma$-ray flares (Section~\ref{gammaray}). In Sections~\ref{radiofit} and \ref{northlobe} we develop simple models of the
stationary secondary component at $\sim 12.5$\,mas from the nucleus. We are then able to bound the jet power. Throughout this paper, we adopt the following
cosmological parameters: $H_{0}=69.6$\,km~s$^{-1}$~Mpc$^{-1}$, $\Omega_{\Lambda}=0.714$, and $\Omega_{m}=0.286$ and use Ned Wright's Javascript Cosmology Calculator website
\citep{wri06}. In our adopted cosmology we use a conversion of 7.31\,pc to 1\,mas.

\section{5 GHz Light Curve Analysis}
\label{lightcurve}

\begin{figure*}
\begin{center}
\includegraphics[width= 0.95\textwidth,angle =0]{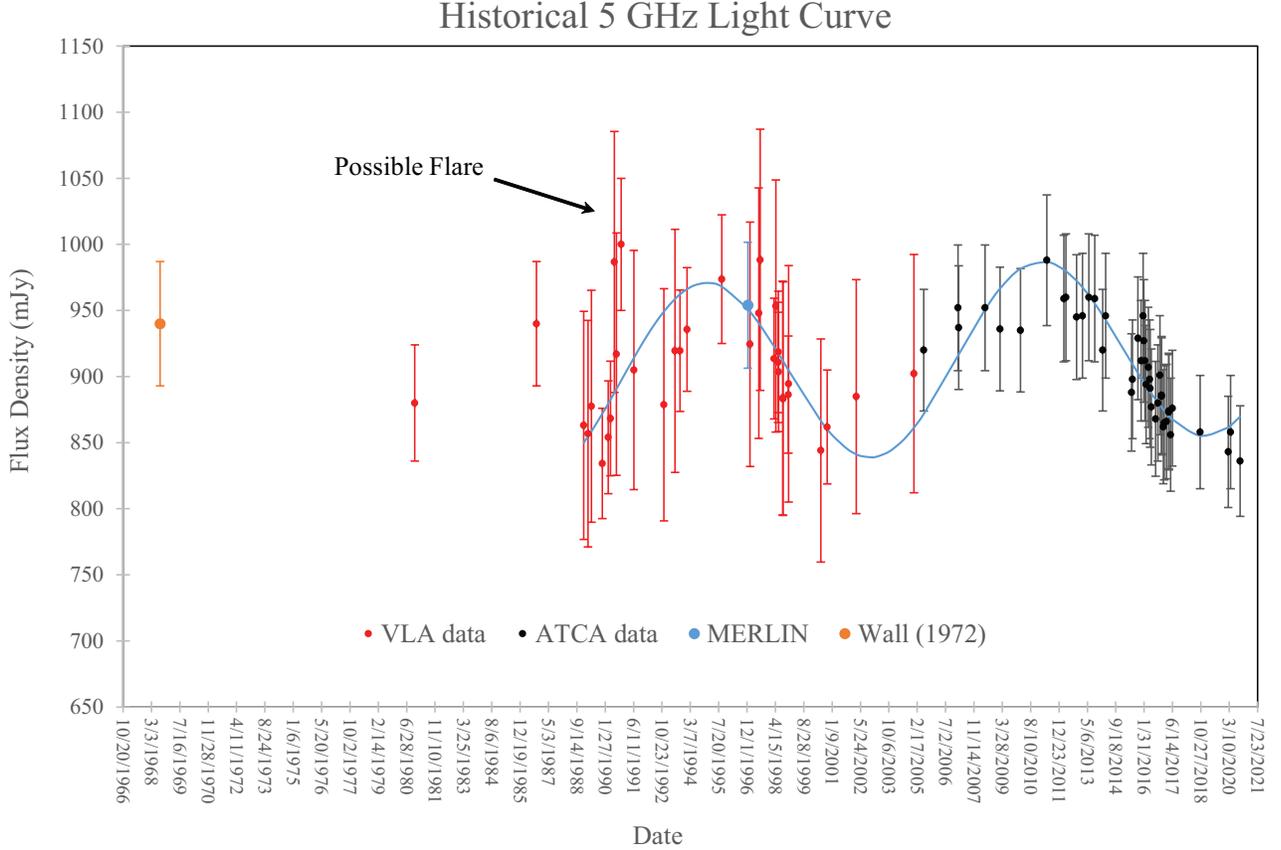}
\caption{The highest density time sampling at any radio frequency is obtained at 5 GHz. The light curve is essentially in two halves, one is based primarily on VLA data and
the latter half is based on ATCA data. This eliminates any perceived short term variability that is a consequence of absolute flux density calibration scales from different
telescopes. We have verified in multiple epochs that the VLA flux density is consistently lower than the quasi-simultaneous ATCA flux density. There is a slow gradual
variation that is visually highlighted by a sine curve with no physical meaning or statistical significance ascribed to it -- only for a visual aid. There is also a sign of a flare in 1990. Flares are difficult to verify, since the measurement uncertainty is large enough to be consistent with enormous
intrinsic flux density changes at this very high redshift. The MERLIN observation is described in Section 3.}
\label{fig:radiolightcurve}
\end{center}
\end{figure*}

Light curves can indicate evidence of blazar-like behavior. Rapid variations can be used to estimate the brightness temperature, $T_{\mathrm{b}}$.
Figure~\ref{fig:radiolightcurve} shows the most densely sampled radio light curve that we could create. There are more archival data at 5\,GHz (corresponding to 23.5\,GHz in
the quasar rest frame) than at other frequencies. If there is a blazar coexisting with strong, persistent emission, it can be significantly variable at higher frequency such as 1.25\,cm wavelength in the quasar rest frame \citep{tor01}. Figure~\ref{fig:radiolightcurve} illustrates the difficulties associated with the analysis. PKS\,1351$-$018 has modest
variation on the order of $5-10\%$ of the background of a quiescent flux density of $\sim 900$\,mJy. Yet, the uncertainty in the individual observations is $5-10\%$. The data for Figure~\ref{fig:radiolightcurve} can be found in Appendix A.

In order to get access to a large volume of unpublished Very Large Array (VLA) data (in the time frame before ATCA data became
available), we relied on the National Radio Astronomy Observatory (NRAO) VLA Archive Survey (NVAS) Images Pilot page\footnote{\url{http://archive.nrao.edu/nvas/read.shtml}}.
We downloaded calibrated visibility FITS files and performed a self-calibration of the phase. This is a bright source with no confusing nearby sources, which lends itself to successful self-calibration. In general, we do not know the history of
these data-sets, for instance, the observer's intent with respect to the science goals and the flux density accuracy goal or if accurate automated ``bad data flagging'' was implemented properly. For outlier data points (candidates for rapid change), we considered things such as image rms noise, signatures of phase noise or side-lobes in the images of this unresolved source and elevation above the horizon. For crucial, suspect data-sets, we had to reduce the data by hand and flag bad antennas and reprocess the images. In order to minimize these issues we chose to avoid early VLA images that were not already present in the literature. In the end, the most conservative flux density calibration uncertainty that we can choose is $10\%$ in general (Lorant Sjouwerman\footnote{Contact person for NVAS} private
communication). Although looking at the scatter of the points, this might be too large for most observations. However, if we were able to ascertain (i.e. from published
papers or direct contact with the principal investigator) that the data reduction proceeded without these potential issues, an uncertainty in the flux density measurements is $5\%$ based on the VLA manual\footnote{\url{https://science.nrao.edu/facilities/vla/docs/manuals/oss/performance/fdscale}} \citep[see also][]{per13}. We did not find all
these circumstances, since it is sometimes difficult to recreate the past. This seemed like a reasonable compromise, NVAS FITS files greatly reduce the time required to
analyze a large volume of data at the expense of a larger uncertainty.

We also wanted to increase the density of the time sampling of the ATCA observations after 2007. There were numerous observations in projects C2898 (2014) and C2914 (2016).
The reductions of these data were generously provided for the purposes of this paper by Jamie Stevens (ATCA Senior System Scientist). The dense data sampling highlights the
gradual changes in the flux density. This was true even in the summer of 2016 when there was a large $\gamma$-ray flare (see Section~\ref{gammaray}).

With that qualifier on the large VLA uncertainty aside, note that there are not the recurring large abrupt changes in amplitude in Figure~\ref{fig:radiolightcurve} that one
sometimes observes with a blazar \citep{tor01,hov09}. But, there is some modest slow variation in amplitude. We have superimposed a sine wave on the densely sampled data from 1988 to 2020 in order to draw
one's eye to the gentle waves of variability that seem to exist within the data set even though the error bars are of similar amplitude. We do not, in any way, suggest that
there is periodic behavior. This is only a device to emphasize the subtle long term variations. We can crudely estimate changes (peak to peak) of $\sim 140$\,mJy over time
frames of $\sim 8$\,yr. This crude estimate is adequate for our purposes. The time variable brightness temperature is estimated in \citet{hov09} as
\begin{equation}
T_{\mathrm{b}} =1.548\times 10^{-32}\frac{\Delta S_{\nu} d_{L}^{2}}{\nu^{2}t^{2}(1+z)}\;,
\label{eq:Tb}
\end{equation}
where $\Delta S_{\nu}$ (measured in Jy) is the change in flux density observed at frequency $\nu$ (measured in GHz) in a time frame $t$ (measured in days in the observer's frame) and $d_{L}$ is the luminosity distance measured in meters. The slow wave modulation yields
$T_{\mathrm{b}} = 2.26\times 10^{12}\,\mathrm{K}$. When $T_{\mathrm{b}}> 10^{12}\,\mathrm{K}$, the inverse Compton catastrophe occurs. Most of the electron energy is radiated
in the inverse Compton regime. The radio synchrotron spectrum from the jet is diminished in intensity to unobservable levels
\citep{kel69}. In order to explain the observed radio synchrotron
jet in such sources, Doppler boosting is customarily invoked to
resolve the paradox. The minimum Doppler factor, $\delta$, required to avoid the inverse Compton catastrophe is \citep{hov09}
\begin{equation}
\delta >\left[{\frac{T_{\mathrm{b}}}{10^{12}\,\mathrm{K}}}\right]^{0.33}\;.
\label{eq:doppler}
\end{equation}
We note that slightly smaller values for the denominator ($T_{\mathrm{b}} \sim 1-5\times 10^{11}\,\mathrm{K}$) based on the equipartition assumption are often invoked
\citep{rea94}. However, based on component sizes and flux density measured with Very Long Baseline Interferometry (VLBI) we estimate $T_{\mathrm{b}} \approx
10^{12}\,\mathrm{K}$, in the next section on radio images. Furthermore, this paper does not assume equipartition in the jet, as we discuss in Section~\ref{northlobe}. From
Equations~(\ref{eq:Tb}) and (\ref{eq:doppler}), we get a bound of $\delta > 1.31$, not the kind of large Doppler factor expected for the strong $\gamma$-ray flare
\citep{sah20}.

\begin{figure*}
\begin{center}
\includegraphics[width= 0.95\textwidth,angle =0]{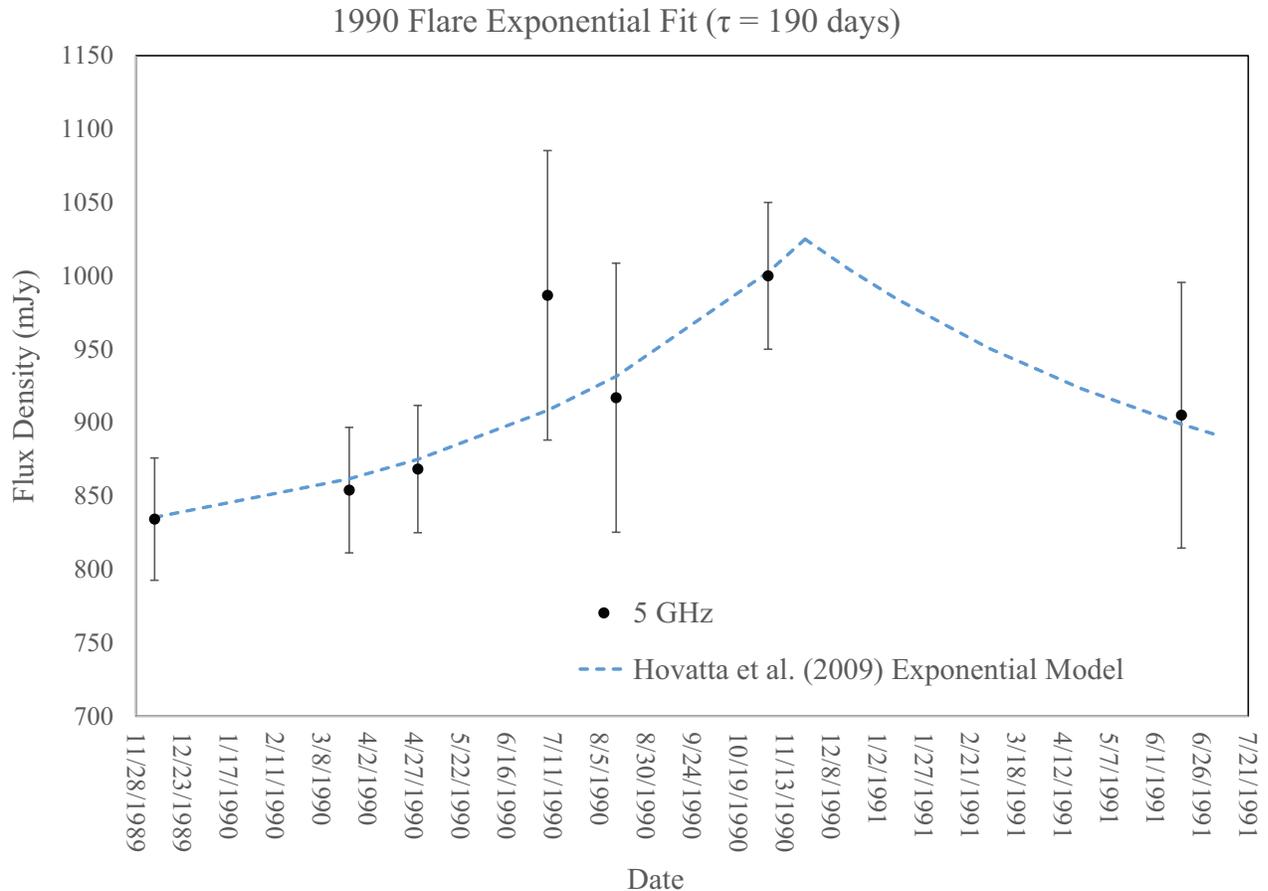}

\caption{Evidence of a 5 GHz flare is presented in this closeup of the corresponding region in Figure~\ref{fig:radiolightcurve}. We fit the flux densities with a
blazar flare model from the literature that is described in the text.} \label{fig:radiolightcurve-1990}
\end{center}
\end{figure*}

The best evidence for a flare in Figure~\ref{fig:radiolightcurve} is the indicated region during 1990. This requires a special consideration of the data in order to verify that this sparsely sampled event is not a manifestation of flawed observations or data reductions. We consider a magnified view of this region for a detailed analysis in Figure~\ref{fig:radiolightcurve-1990}.

Before defining the flare, we noted that the 1990 August 14 data had 6 antennas for which the gain amplitude did not track the other 20 antennas, for whatever reason. The
data were processed by hand by Lorant Sjouwerman instead of the automated NVAS routine. These antennas were flagged and the data reduction repeated based on the other 20
antennas followed by a phase self-calibration. Figure~\ref{fig:radiolightcurve-1990} indicates our best estimates of the flux density and uncertainty. We fit this with a
blazar flare model developed in \citet{val99} and \citet{hov09}. The authors successfully describe flares by an exponential rise followed by an exponential decay with a time
constant 1.3 times the time constant of the rising phase. We fit the data by finding the aforementioned model that minimizes the excess variance, $\Sigma_{\mathrm{rms}}$, of
the fit to the 7 data points that define the flare in Figure~\ref{fig:radiolightcurve-1990} \citep{nan97}:
\begin{equation}
\Sigma_{\mathrm{rms}}^{2}=\frac{1}{N}\sum_{i=1}^{N}
\frac{(S_{i}-f_{i})^2- \sigma_{i}^{2}}{f_{i}^{2}}\; ,
\label{equ:variance}
\end{equation}
where $i$ labels one of the $N$ measured flux densities, $f_{i}$ is the expected value of this flux density from the \citet{hov09} model, $S_{i}$ is the measured flux density
and $\sigma_{i}$ is the uncertainty in this measurement. The best fit is one that peaks on 1990 November 24 at 1025\,mJy, corresponding to a flare peak of 225\,mJy above the baseline. The time constant of the rise is $\tau = 190$\,d. \citet{hov09} use $\tau =t$ in Equation~\ref{eq:Tb}. They also identify the flare peak (225\,mJy in this case) with $\Delta S_{\nu}$ in Equation~(\ref{eq:Tb}). We formally interpret Equation~(\ref{eq:Tb}) as the change in flux density in a time $t$. For $t=\tau$, the maximum $\Delta S_{\nu}$ during the rise is $\Delta S_{\nu} = (1-e^{-1})\,225
\, \mathrm{mJy} = 142 \, \mathrm{mJy}$. This yields $\delta > 8.1$ from Equation~(\ref{eq:doppler}).
\par This minimum value of Doppler factor, $\delta_{\mathrm{min}}$, can be used to restrict the line of sight (LOS) to the emitting region of the jet. First, we express the
Doppler
factor in terms of kinematic quantities,
\begin{equation}
\delta = \frac{\gamma^{-1}}{1-\beta \cos{\theta}},\; \gamma^{-2} = 1- \beta^{2}\;,
\label{eq:kinematics}
\end{equation}
where $\beta$ is the normalized three-velocity of bulk motion, the associated Lorentz factor is $\gamma$, and $\theta$ is the angle of the motion to the LOS to the observer
\citep{lin85}. For each value of $\delta_{\mathrm{min}}$, one can vary $\beta$ in Equation~(\ref{eq:kinematics}) to find the maximum value of $\theta$,
$\theta_{\mathrm{max}}\{\delta_{\mathrm{min}}[(T_{\mathrm{b}})]\}$, that is compatible with
$\delta_{\mathrm{min}}$ \citep{gho07}:
\begin{equation}
\theta_{\mathrm{max}}\{\delta_{\mathrm{min}}[(T_{\mathrm{b}})]\}=\mathrm{Max}_{\mid_\beta}\left(\arccos\left\{\left[1-\left(\frac{\sqrt{1-\beta^{2}}}{\delta_{\mathrm{min}}[(T_{\mathrm{b}})]}\right)\right]\beta^{-1}\right\}\right)\;.
\label{eq:thetamax}
\end{equation}
From the flare model and Equation~(\ref{eq:thetamax}) we conclude that
\begin{equation}
\theta_{\mathrm{max}} = 7.1\degr.
\label{equ:theta-max}
\end {equation}

Comparing Figures~\ref{fig:radiolightcurve} and \ref{fig:radiolightcurve-1990}, we draw the following conclusion. There is a blazar-like emission region. However, it is
superimposed on the background of a much more luminous mildly variable component. This could be emission from a very luminous sub-kpc scale jet for which $\delta$ is far less
than in the blazar-like region either due to a strong decelerating force and/or a change in the jet propagation direction relative to the LOS. The analysis of the radio
images in the next section can potentially elucidate these circumstances. The apparent lack of many clear instances of blazar-like flares in Figure~\ref{fig:radiolightcurve}(only the one in 1990 in Figure~\ref{fig:radiolightcurve-1990}) is amplified by the time dilation due to the large redshift. In the quasar rest frame, 1989--2020 is only 6.6 years.

\section{Radio Image Analysis}
\label{imaging}

The first radio image that we considered was the 1.4 GHz VLA observation in A-array which found an unresolved core \citep{nef90}. In the absence of diffuse large scale
emission we looked for compact structure on kpc scales. First, we imaged the deepest 5 GHz VLA A-array observation, project AB0560 (1990 March 23, one of the data points in
Figure~\ref{fig:radiolightcurve-1990}). Again, it was an unresolved nucleus. We then looked at the most sensitive 5 GHz observation with the Jansky Very Large Array (JVLA)  in
the A-array, project 16B-130 on 2016 December 3. Matt Smith kindly reduced the data he
observed and created an image also showing nothing but an unresolved core. So we went to higher resolution looking for structure, by searching the Multi-Element Radio
Linked Interferometer (MERLIN) archives for 5 GHz observations. Anita Richards generously provided the image FITS file of the one observation, on 1996 December 1, revealing
an unresolved core. The restoring beam size was $69\,\mathrm{mas} \times 52\,\mathrm{mas}$ at a position angle $\mathrm{PA} = 39\degr$. Thus, we have an upper bound
on the source size that is quite small. Thus motivated, we explore the VLBI observations of this source to look for the structure on scales less than 50\,mas.

\begin{figure*}
\begin{center}
\textbf{Global VLBI 2.3 GHz}\par\medskip
\includegraphics[width= 0.85\textwidth,angle =0]{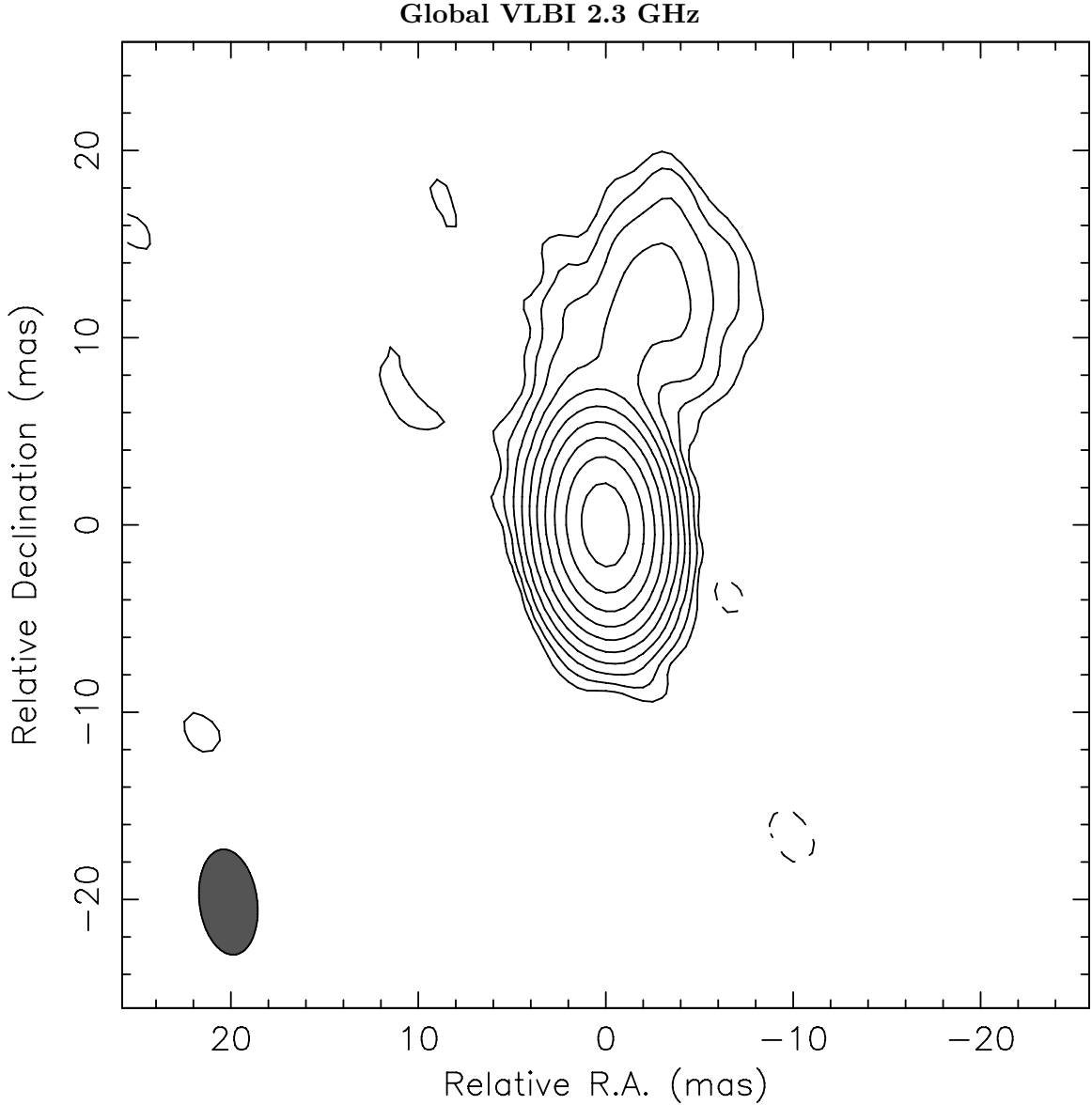}
\caption{The Northern Lobe is very important to our estimate of the jet power. The best image of the feature is reconstructed using the S-band (2.3\,GHz) global VLBI data
taken on 2002 July 24. The peak intensity is 718\,mJy~beam$^{-1}$. The lowest level contour is $\pm 0.92$\,mJy~beam$^{-1}$ and the positive contour levels increase by a
factor of 2. The restoring beam is $3.08\,\mathrm{mas} \times 5.66\,\mathrm{mas}$ at $\mathrm{PA} = 5\degr$. The lobe appears to be at the end of a short, 11.5\,mas jet. The
lobe is resolved (fit with a Gaussian of FWHM of 6.6\,mas from Table~\ref{tab:modelfit-Sband}). It is comparable to the overall jet length projected on the sky plane. It is
also unclear if there is a hotspot. It appears plume-like as opposed to an edge brightened morphology.}
\label{fig:Sband}
\end{center}
\end{figure*}

\subsection{S-band VLBI}

The Astrogeo VLBI FITS image database\footnote{\url{http://astrogeo.org}} contains calibrated VLBI 2.3 GHz data from numerous epochs from 1994 to 2020. Most of the
observations use the 10 station Very Long Baseline Array (VLBA). However, from 1998--2003, the Research and Development VLBA project
employed additionally up to 10 other
antennas in order to create a global VLBI network \citep{pet09,pus12}. The addition of the southern hemisphere stations and the multiple short scans in 24-h long observing
sessions provide the best $(u,v)$ coverage and resolution of any of the observations in the Astrogeo VLBI FITS image database. The observation in Figure~\ref{fig:Sband} from
2002 July 24 is of particular interest. The 2.3 GHz image was published in \citet{pus12}. The 2002 image is the highest sensitivity, high resolution image available at
S-band. The visibility data were fit with Gaussian brightness distribution components with an automated process. We re-analyzed the data with a manual data reduction for the
purposes of this project. There is a prominent diffuse component to the north (which we will call the
``North Lobe''). The details of the Gaussian fit are described in the second entry of Table~\ref{tab:modelfit-Sband}. The method of defining measurement uncertainties can be found in Appendix B.

\begin{table}
\begin{center}
\caption{Gaussian Fits to S-Band VLBI Observations}
    \label{tab:modelfit-Sband}
   \footnotesize{ \begin{tabular}{cccccccc}
        \hline
    Date  & Component  & Flux Density  &  r & Position & FWHM & Axial& $T_{\mathrm{b}}$ \\
        &          &      (mJy)    &        (mas)     & Angle ($\degr$)      &  (mas)         &  Ratio (PA) &  $\,\mathrm{K}$        \\
   \hline
   1997 Jan 11 & Core  & $880 \pm 88$ & 0 & .... & 1.1 & 0 ($-53\degr$)& $8.67 \times 10^{11}$  \\
               &North Lobe  & $30 \pm 13$  & $11.6\pm 1.7$ & $-6$ & 6.1 & 1& $8.76 \times 10^{8}$ \\
   \hline

   2002 Jul 24 & Core  & $766 \pm 77$ & 0 & ... & 1.0 & 1& $8.39 \times 10^{11}$ \\
               &North Lobe  & $34.5 \pm 13.8$ & $11.9\pm 1.1$ & $-9.7$ & 6.6 & 1& $8.50 \times 10^{8}$ \\
   \hline
    2014 Aug 9 & Core  & $787 \pm 79$ & 0 & ... & 1.1 & 1& $7.33 \times 10^{11}$ \\
               &North Lobe  & $32.6 \pm 13.0$  & $12.4\pm 1.7$ & $-11.3$ & 4.8 & 1& $1.54 \times 10^{9}$ \\
   \hline
   \end{tabular}}
\end{center}

\end{table}

The brightness temperature in the last column is computed per the methods of \citet{kel88}:
\begin{equation}
T_{\mathrm{b}}[\mathrm{K}]= 1.22 \times 10^{12}(1+z)\frac{S_{\nu}}{\theta_{1}\theta_{2}\nu^{2}}\;,
\label{eq:VLBI-Tb}
\end{equation}
where $\nu$ is the observed frequency measured in GHz, $\theta_{1}$ ($\theta_{2}$) is the major (minor) axis of the elliptical Gaussian fitted full width at
half-maximum (FWHM) measured in mas, and $S_{\nu}$ is the flux density of the component in Jy.

The earliest entry in Table~\ref{tab:modelfit-Sband} is a VLBA observation from 1997 January 11. It is composed of four 3-min scans spread out over two days to
maximize $(u,v)$ coverage. The fits are not carried out to as many decimal points as our other fitted models \citep{fey00}. The North Lobe flux density is 0.03\,Jy with no
listed uncertainty. Thus, there is an additional uncertainty due to roundoff errors. We also found another image in the Astrogeo VLBI FITS image database with far less dense
coverage in the
$(u,v)$ plane than the 2002 July 24 image that showed the northern lobe prominently (2014 August 9). The $(u,v)$ coverage was fortuitous and the beam shape is not too
elongated as in many epochs. The results of the Gaussian fitting process are listed as the third entry in Table~\ref{tab:modelfit-Sband}. The lobe flux density is probably
the same within uncertainty ($\sim 15-20\% $), but the smaller component size indicates that some diffuse emission was not captured by the sparse $(u,v)$ coverage.
There is strong evidence in Table~\ref{tab:modelfit-Sband} that the North Lobe was stable within VLBI uncertainties for at least 17 years.

Since these are very high resolution observations, we are interested to know how much flux density outside of the unresolved core was missed by the VLBI observations. In order to assess this and being cognizant of possible temporal variability of the core, we looked for quasi-simultaneous ATCA or VLA observations and VLBI observations. The VLBI absolute flux density calibration is less robust based on comparing the scatter of the C-band VLBI flux density to the light curve in Figure~\ref{fig:radiolightcurve}. Thus, it is very desirable to have two quasi-simultaneous VLBI observations with at least one ATCA or VLA observation at the same epoch. We could not find this circumstance in any of the VLBI bands except for S-band. Table~\ref{tab:vlbi-atca} shows the only robust comparison that we could make based on archival data.

\begin{table}
\begin{center}
 \caption{2.3 GHz VLBI Flux Density Compared to ATCA}
 \label{tab:vlbi-atca}
\footnotesize{
\begin{tabular}{ccccc}
 \hline
 Date  & Telescope  & Flux Density (mJy) &  Reference   \\
 \hline
 2018 Jan 18 & VLBI  & $970 \pm 97$\tablenotemark{\tiny{a,b}} & Astrogeo VLBI FITS image database \\
    2018 Jan 22 & ATCA  & $893 \pm 45$\tablenotemark{\tiny{c}}  & ATCA Calibrator Database  \\
   2018 Jan 25 & VLBI  & $923 \pm 92$\tablenotemark{\tiny{a,b}} & Astrogeo VLBI FITS image database\\
   2017 Dec 19\tablenotemark{\tiny{d}} & ATCA  & $881 \pm 44$\tablenotemark{\tiny{c}}  & ATCA Calibrator Database \\ \hline

    \end{tabular}}
\end{center}
\tablenotetext{a}{The sum of the flux densities of the Gaussian fitted nucleus and North Lobe.}
\tablenotetext{b}{10\% uncertainty \citep{hom02,pus12}.}
\tablenotetext{c}{5\% uncertainty \citep{mur10}.}
\tablenotetext{d}{Additional data to corroborate the stability of the ATCA calibration in this time frame.}
\end{table}

Table~\ref{tab:vlbi-atca} seems to indicate that all of the flux density is contained within the VLBI components of the nucleus and North Lobe. If there were a few mJy
resolved out of the North Lobe by the VLBI observations, we cannot confirm or reject this based on the uncertainties of the data in Table~\ref{tab:vlbi-atca}. We conclude
that there is likely no measurable emission between the $\sim 50-60$~mas limit from MERLIN and the $\sim 15$~mas VLBI structure. We assume this to be the case in the
remainder of the paper, up to the uncertainty in the flux density.

\subsection{C-band VLBI}

\begin{figure}[htp!]
\begin{center}
\textbf{VSOP (with VLBA) 4.8 GHz \hspace{4.9cm} VLBA 4.3 GHz}\par\medskip
\includegraphics[width= 0.45\textwidth, bb=0 74 443 512, clip=, angle=0]{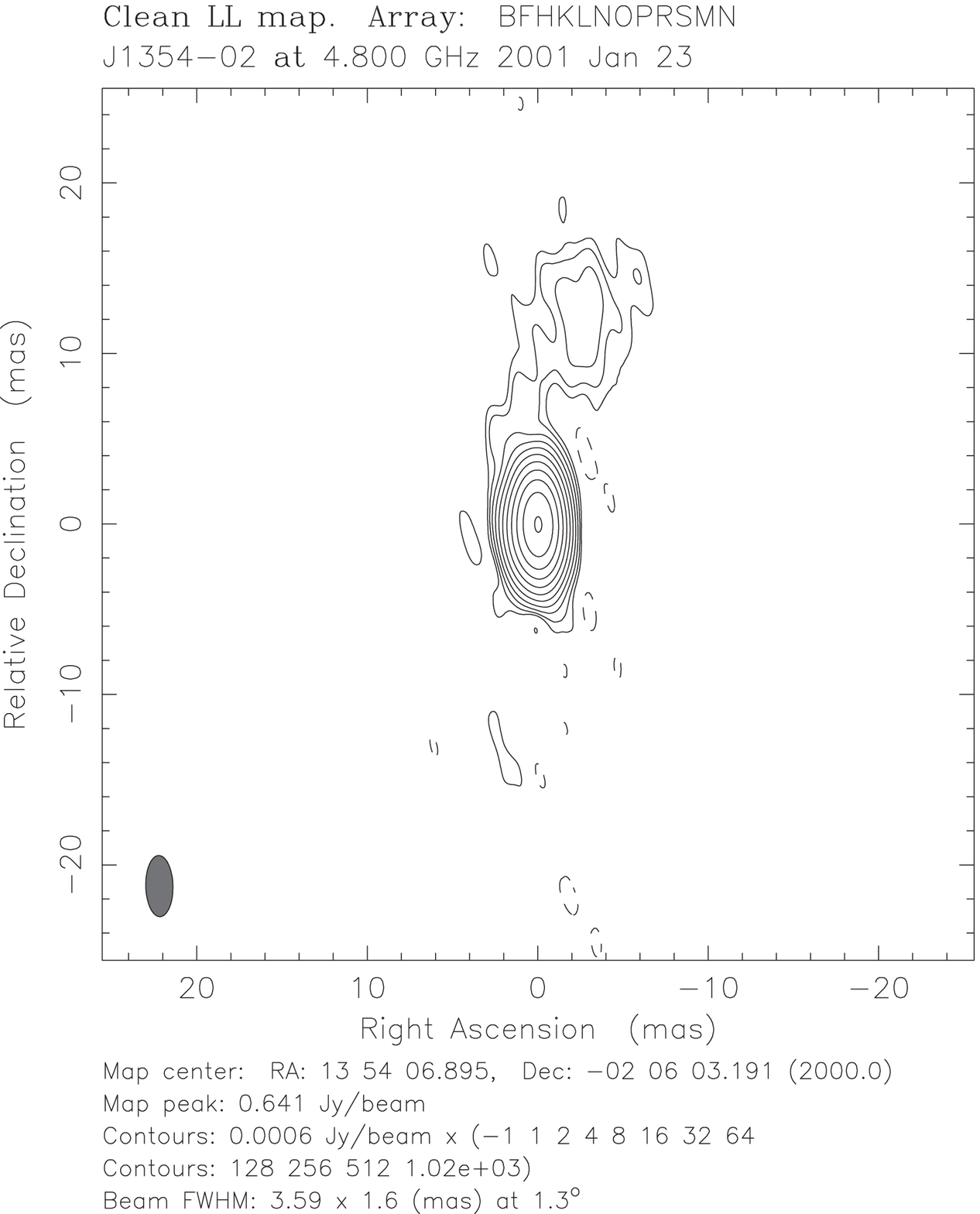}
\includegraphics[width= 0.45\textwidth, bb=0 74 443 512, clip=, angle=0]{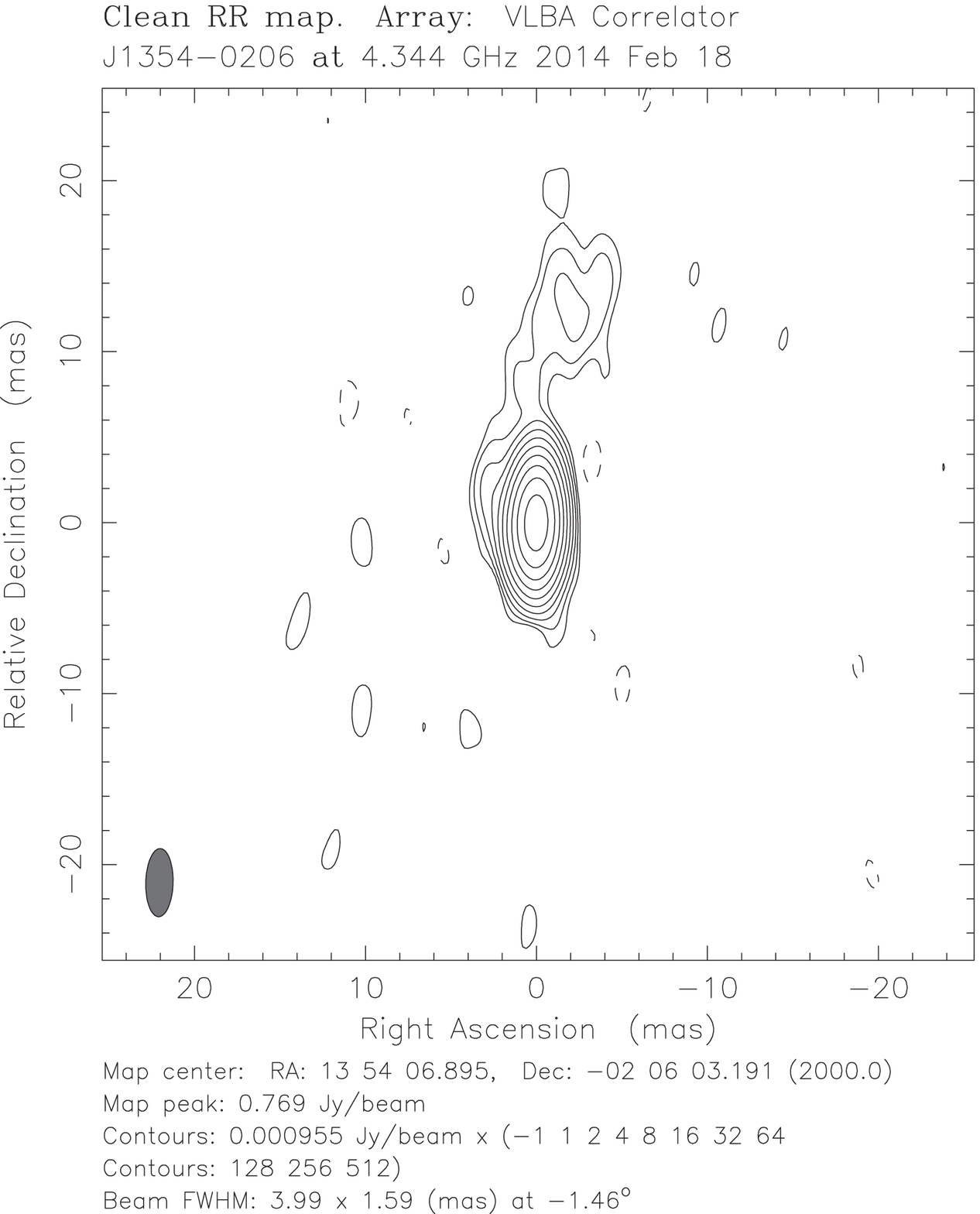}
\caption{This figure compares the
4.8 GHz VSOP (including the VLBA) observation from 2001 January 23 (left) with the 4.3 GHz VLBA observation from 2014 February 18 (right). The latter
observation has far less time on the source, however the two images separated by 13 years look remarkably similar. There are no measurable changes to the North Lobe within
measurement uncertainties as evidenced by Table~\ref{tab:vlbi-Cband-fit}. In the left image, the peak intensity is 641\,mJy~beam$^{-1}$, the lowest level contour is
$\pm0.6$\,mJy~beam$^{-1}$. The restoring beam is $1.6\,\mathrm{mas} \times 3.59\,\mathrm{mas}$ at $\mathrm{PA}=1.3\degr$. In the right image, the peak intensity is
769\,mJy~beam$^{-1}$, the lowest level contour is $\pm0.955$\,mJy~beam$^{-1}$. The restoring beam is $1.59\,\mathrm{mas} \times 3.99\,\mathrm{mas}$ at
$\mathrm{PA}=-1.5\degr$. The positive contour levels increase by a factor of 2.}
\label{fig:Cband}
\end{center}
\end{figure}

\begin{figure*}[htp!]
\begin{center}
\textbf{VSOP (with VLBA) 4.8 GHz \hspace{4.9cm} Global VLBI 5.0 GHz}\par\medskip
\includegraphics[width= 0.5\textwidth,angle =0]{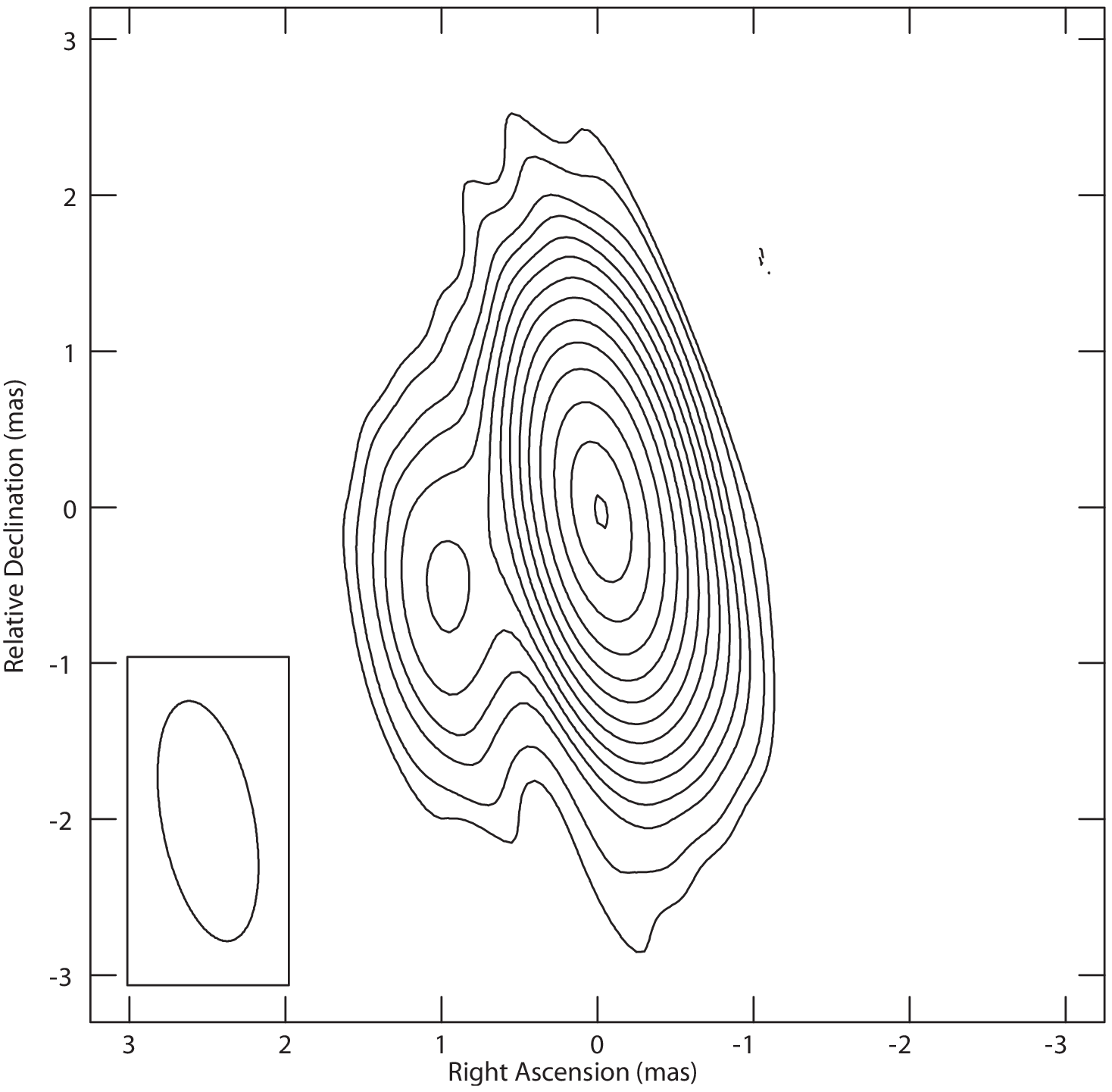}
\includegraphics[width= 0.45\textwidth,angle =0]{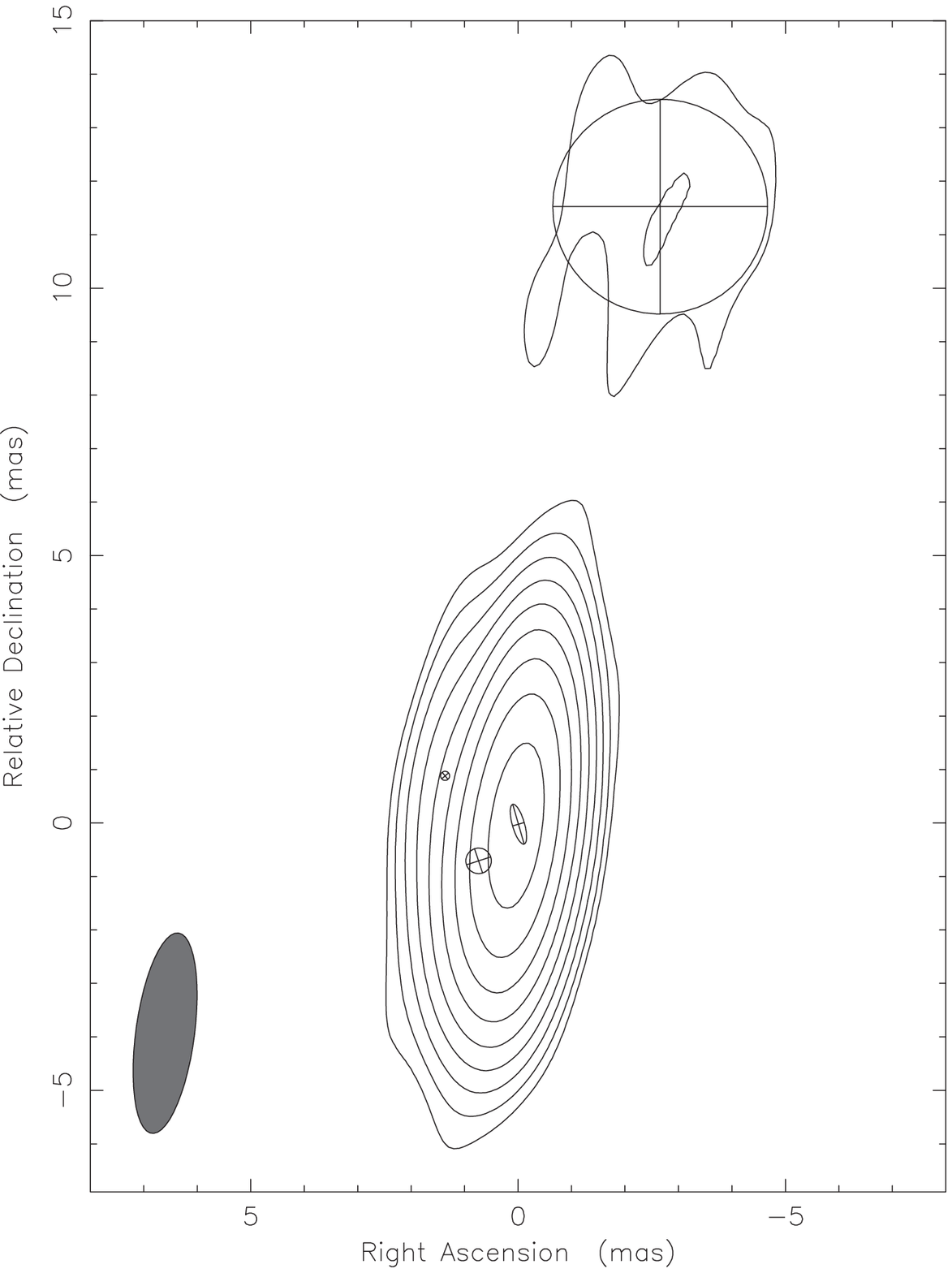}
\caption{\footnotesize{On the left is the 4.8 GHz VSOP image from 2001 January 23, with special weighting to enhance the space--ground baseline data, as reproduced from
\citep{fre02}. Contours are at $-1$, 1, 1.5, 2.5, 3.5, 5, 7, 10, 14, 20, 28, 40, 56, 80, and 99\% of the peak brightness of 547\,mJy~beam$^{-1}$. The restoring beam is
$0.59\,\mathrm{mas} \times 1.56\,\mathrm{mas}$ at $\mathrm{PA}=11\degr$. The image on the right hand side is a 5 GHz global VLBI image from 2000 June 5.
The peak intensity is 663\,mJy~beam$^{-1}$, the lowest level contour is $\pm1.66$\,mJy~beam$^{-1}$. The restoring beam is $1.1\,\mathrm{mas} \times 3.78\,\mathrm{mas}$ at
$\mathrm{PA}=-7.6\degr$. The positive contour levels increase by a factor of 2.
The ellipses with crosses are the Gaussian components from Table~\ref{tab:vlbi-Cband-fit}. Note that both observations detect a Southeast Component.}}
\label{fig:Cband-vsop}
\end{center}
\end{figure*}

There are more VLBI data at C-band than any other frequency band. We were able to find 6 observations with useful data for our purposes. The relevant details are listed in
Table~\ref{tab:vlbi-Cband-fit}. The third column, the restoring beam size, is very important since PKS\,1351$-$018 displays a resolution dependent morphology \citep{fre02}.
The beam position angle is not listed (to save space for more relevant details) as it is always nearly north-south within $15\degr$. All of the data were fit by us except for
the \citet{osu11} data which is from the literature. We refit the \citet{fre97} $(u,v)$ data with circular Gaussian models and multiple nuclear components, so it matches our
data reduction technique for the four more recent data that were fit. The models are fit based on four detected components. As in Figure~\ref{fig:Sband}, the two most
prominent features are the bright nucleus and the diffuse North Lobe. There is also a knot in the northern jet (that is prominent in Figure~\ref{fig:Cband}) as well as a
Southeast Component that is very close to the nucleus (see Figure~\ref{fig:Cband-vsop}). The features that are recovered in the fitting process in the $(u,v)$ plane depend on
the resolution and sensitivity of the observations.

The main purpose of Table~\ref{tab:vlbi-Cband-fit} is to track the location of the North Lobe over time (columns 6 and 7). In terms of $(u,v)$ coverage, the
four early observations are far superior to the last two from the Astrogeo VLBI FITS image database. Surprisingly the 2014 February 18 Astrogeo observation had much more
sensitivity to the North Lobe emission than the other two observations from the Astrogeo VLBI FITS image database (see Figure~\ref{fig:Cband}). The image is from
segment BP177I of the 8th VLBA Calibrator Survey (VCS8) campaign \citep{pet21}. The source was used as an amplitude calibrator and it was observed in one scan on 2014
February 18 that achieved superior dynamic range compared to the other epochs\footnote{Leonid Petrov private communication}.

Compare the 2001 January 23 VLBI Space Observatory Program (VSOP) observation \citep{fre02} with 10 stations of VLBA plus the \textit{HALCA} satellite (the
second entry in Table~\ref{tab:vlbi-Cband-fit}) restored with natural weighting with the best Astrogeo observation from 2014 February 18 in Figure~\ref{fig:Cband}. The
North Lobe position and flux density is very stable over 13 years. There is a faint jet connecting the nucleus to the lobe. In the right hand frame, this jet seems to
begin at a small brighter protrusion of the nuclear contours almost directly northeast of the core. This is the component ``Knot in North Jet'' in
Table~\ref{tab:vlbi-Cband-fit}. Based on Table~\ref{tab:vlbi-Cband-fit} and Figure~\ref{fig:Cband}, due to the positional change from 2001 to 2014, this feature might be
different faint knots in 2001 and 2014. The apparent bending of the jet to the east from 2001 to 2014 might be an indication that there is relativistic motion (Doppler
aberration) in this region of the jet.

At higher resolution, at C-band, as well as at higher frequency (the next subsection), the diffuse radio emission of the North Lobe is resolved out.
However, more compact features are revealed. The left hand frame of Figure~\ref{fig:Cband-vsop} is lifted from \citet{fre02} and represents the same data as in the left hand
side of Figure~\ref{fig:Cband}, but a higher resolution image is made using optimally weighted data on space--ground baselines to enhance angular resolution. The North Lobe
and the Knot in North Jet are resolved out, but a ``Southeast Component'' is revealed. The absolute flux density calibration with VSOP is difficult, but we estimate $\approx
40$\,mJy in the Southeast Component. In order to validate its existence, we present a global VLBI -- VLBA plus 6 European VLBI Network (EVN) telescopes -- image from
a 24-h observation of \citet{osu11} from 2000 June 5, only 7 weeks earlier in the quasar rest frame. The image in the right hand frame of Figure~\ref{fig:Cband-vsop} was
generously created for the
purpose of this study by Shane O'Sullivan. The Gaussian fitted component FWHM from Table~\ref{tab:vlbi-Cband-fit} are represented by circles and ellipses with crosses. These
components seem to capture every observed feature including the North Lobe. There is a strong component to the southeast as well as a very weak compact component to the
northeast, the putative knot in the northern jet. The locations do not line up perfectly with VSOP, but that might be an artifact of trying to fit features on scales smaller
than the synthesized beam. This might also explain the flux density difference in the Southeast Component ($79 \pm 15$\,mJy versus $40 \pm 10$\,mJy). Other contributing
factors could be extreme variability and the absolute flux density calibration of VSOP. In this context, it should be noted that we were unable to retrieve all of the
original files used in the VSOP data reduction. Hence, there is some uncertainty in the absolute flux density calibration in the files we could retrieve. There is a possible
difference with the calibration of \citet{fre02} that indicates $\approx 40$\,mJy in the Southeast Component, we fit 31.2 mJy using the data files that were found. However,
it is still consistent within the uncertainties in Table~\ref{tab:vlbi-Cband-fit}. We investigate this component further with higher frequency VLBI in the next subsection.

\begin{table}[htp!]
\begin{center}
    \caption{Gaussian Fits to C-Band VLBI Observations}
    \label{tab:vlbi-Cband-fit}
   \tiny{ \begin{tabular}{ccccccccccc}
        \hline
        (1)  &(2) & (3) & (4)  & (5)  &  (6) & (7) & (8) & (9) & (10) & (11)\\
        Date  & Array/ & Restoring &Component  & Flux  & $r$ & Position & FWHM & Axial& $T_{\mathrm{b}}$ & Ref.\\
          & Frequency & Beam &  & Density  &  & Angle &  &  & & \\
        &  & (mas) &         &      (mJy)    &        (mas)     & ($\degr$)       &  (mas)         & Ratio/PA  & (K) &       \\
        \hline
        &  &  &  &   &  &  &  &  & & \\
   1995 Jan 28 & VLBA &  $4.9 \times 2.0$ & Core  & $919\pm 92$ & $0$ & ... & 0.57 & 1 & $6.50 \times 10^{11}$ & \tablenotemark{\tiny{a,b}} \\
               & 5 GHz & &North Lobe  & $14.0\pm 5.6$  & $12.4 \pm 0.8$ & $-11.8$ & 2.51 & 1 & $5.11 \times 10^{8}$ & \tablenotemark{\tiny{b}}\\
                &  &  & Knot in North Jet  & not  & detected & ... & ... & ... & ... & \tablenotemark{\tiny{b}} \\
                &  &  & Southeast Component  & $22.4 \pm 7.1 $  & $1.6\pm 0.3$ & $131.0$ & 0.05\tablenotemark{\tiny{c}} & 1 & \tablenotemark{\tiny{c}} &
                \tablenotemark{\tiny{b.d}}\\
               \hline
    2000 Jun 5 & Global VLBI & $3.51 \times 1.10$ & Core  & $681 \pm 68$ & 0 & ... & 0.77 & 0.3/$-164\degr$  & $8.91 \times 10^{11}$  & \tablenotemark{\tiny{g}} \\
               & 5 GHz & & North Lobe  & $16.8\pm 6.7$  & $11.9 \pm 0.7$ & $-12.9$ & 4.02 & 1& $2.42 \times 10^{8}$  &  \\
              &  &  & Knot in North Jet  &$14.3\pm 6.2$  & $1.6\pm 0.3$ & $56.6$ & \tablenotemark{\tiny{c}} & ... & \tablenotemark{\tiny{c}}  &  \\
         &  &  & Southeast Component  & $78.5 \pm 14.5 $  & $1.0\pm 0.5$ & $132.7$ & 0.48 & 1 & $7.88 \times 10^{10}$  & \\
    \hline
    2001 Jan 23\tablenotemark{\tiny{e}} & VSOP & $3.59 \times 1.60$ & Core  & $664\pm 66$ & $0$ & 0 & 0.46 & 1& $7.83 \times 10^{11}$  & \tablenotemark{\tiny{f,b}}\\
               & 4.8 GHz & & North Lobe  & $14.1\pm 5.6$  & $12.5\pm 0.7$ & $-11.0$ & 3.93 & 1& $2.28 \times 10^{9}$  &  \tablenotemark{\tiny{b}}\\
               &  &  & Knot in North Jet  & $4.0\pm 3.0$  & $5.2\pm 0.3$ & $67.0$ & \tablenotemark{\tiny{c}} & ... &\tablenotemark{\tiny{c}}  & \tablenotemark{\tiny{b}} \\
                &  &  & Southeast Component  & $32.1 \pm 8.5 $  & $1.3\pm 0.3$ & $113.0$ & 0.05\tablenotemark{\tiny{c}} & 1 & \tablenotemark{\tiny{c}} &
                \tablenotemark{\tiny{b}} \\
                 \hline
    2001 Jan 23\tablenotemark{\tiny{e}} & VSOP  & $1.56 \times 0.59$ & North Lobe  & not  & detected & ... & ... & ... & ...  & \tablenotemark{\tiny{b}} \\
         & 4.8 GHz &  & Southeast Component  & $40 \pm 10 $  & $1.3\pm 0.3$ & $113.0$ & \tablenotemark{\tiny{c}} & ... & \tablenotemark{\tiny{c}}  &
         \tablenotemark{\tiny{b}}\\
         \hline
    2014 Feb 18 & VLBA & $3.99 \times 1.59$ & Core  & $802\pm 80$ & $0$ & ... & 0.49 & 1 & $1.04 \times 10^{12}$  & \tablenotemark{\tiny{h,b}}\\
               & 4.3 GHz & & North Lobe  & $13.3 \pm 5.3$  & $12.6 \pm 0.8$ & $-12.4$ & 2.04 & 1 & $9.93 \times 10^{9}$  & \tablenotemark{\tiny{b}}  \\
              &  &  & Knot in North Jet  & $8.5 \pm 4.3$  & $5.7 \pm 0.7$ & $18.9$ & 0.68 & 1 & $5.71 \times 10^{10}$  & \tablenotemark{\tiny{b}} \\
         &  &  & Southeast Component\tablenotemark{\tiny{d}}  & $33.1\pm 8.4$  & $1.4\pm 0.3$ & $106.9$ & 0.01 & 1 &  \tablenotemark{\tiny{c,d}}  & \tablenotemark{\tiny{b}}\\
    \hline
    2016 Feb 17 & VLBA & $4.08 \times 1.66$ & Core  & $656\pm 66$ & $0$ & ... & $0.55$ & $1$ & $6.27 \times 10^{11}$  & \tablenotemark{\tiny{h,b}} \\
               & 4.3 GHz & & North Lobe\tablenotemark{\tiny{i}}  & $9.6\pm 3.8$ & $12.8 \pm 0.8$ & $-13.2$ & $1.53$ & 1 & \tablenotemark{\tiny{i}}  & \tablenotemark{\tiny{b}}
               \\
              &  &  & Knot in North Jet\tablenotemark{\tiny{d}}  &$7.1\pm 3.3$  & $6.0 \pm 0.7$ & $15.6$ & $0.34$ & 1 & \tablenotemark{\tiny{c,d}}  & \tablenotemark{\tiny{b}}
              \\
         &  &  & Southeast Component\tablenotemark{\tiny{d}}  & $19.7 \pm 5.5 $  & $1.5 \pm 0.4$ & $114.0$ & $0.04$ & 1 &\tablenotemark{\tiny{c,d}} &\tablenotemark{\tiny{b}}
         \\
    \hline
    2018 Dec 1 & VLBA & $3.81 \times 1.42$ & Core  & $699 \pm 70$ & $0$ & ... & $0.46$ & $1$ & $9.76 \times 10^{11}$ & \tablenotemark{\tiny{h,b}} \\
               & 4.3 GHz & & North Lobe\tablenotemark{\tiny{i}}  & $11.0 \pm 4.4$ & $12.0 \pm 0.5$ & $-13.5$ & $1.66$ & 1 & \tablenotemark{\tiny{i}}  &
               \tablenotemark{\tiny{b}} \\
              &  &  & Knot in North Jet\tablenotemark{\tiny{d}}  &$13.9 \pm 4.9$  & $5.4 \pm 0.7$ & $19.1$ & $0.01$ & 1 & \tablenotemark{\tiny{c,d}}  &
              \tablenotemark{\tiny{b}} \\
         &  &  & Southeast Component\tablenotemark{\tiny{d}}  & $46.0 \pm 9.0 $  & $1.4\pm 0.4$ & $118.2$ & $0.06$ & 1 & \tablenotemark{\tiny{c,d}}  &\tablenotemark{\tiny{b}}
         \\
    \hline
    \end{tabular}}
\end{center}
\tablenotetext{a}{\citet{fre97}}
\tablenotetext{b}{This paper}
\tablenotetext{c}{Effectively a point source, $T_{\mathrm{b}}$ estimate ill-defined}
\tablenotetext{d}{Very poor long baseline $(u,v)$ coverage, fit parameters may not be robust.}
\tablenotetext{e}{See the text description of the absolute flux density uncertainty of our results. The second entry for this observation is from the image in \citet{fre02}
with preferential weighting to the long space--Earth baselines.}
\tablenotetext{f}{\citet{fre02}}
\tablenotetext{g}{\citet{osu11}}
\tablenotetext{h}{\citet{pet21}, Astrogeo VLBI FITS image database}
\tablenotetext{i}{$(u,v)$ short baseline coverage is degraded relative to 2014 February 18, much of the North Lobe flux density is missing. The true uncertainty cannot be
estimated.}
\end{table}

\begin{figure*}[htp!]
\begin{center}
\includegraphics[width= 0.85\textwidth,angle =0]{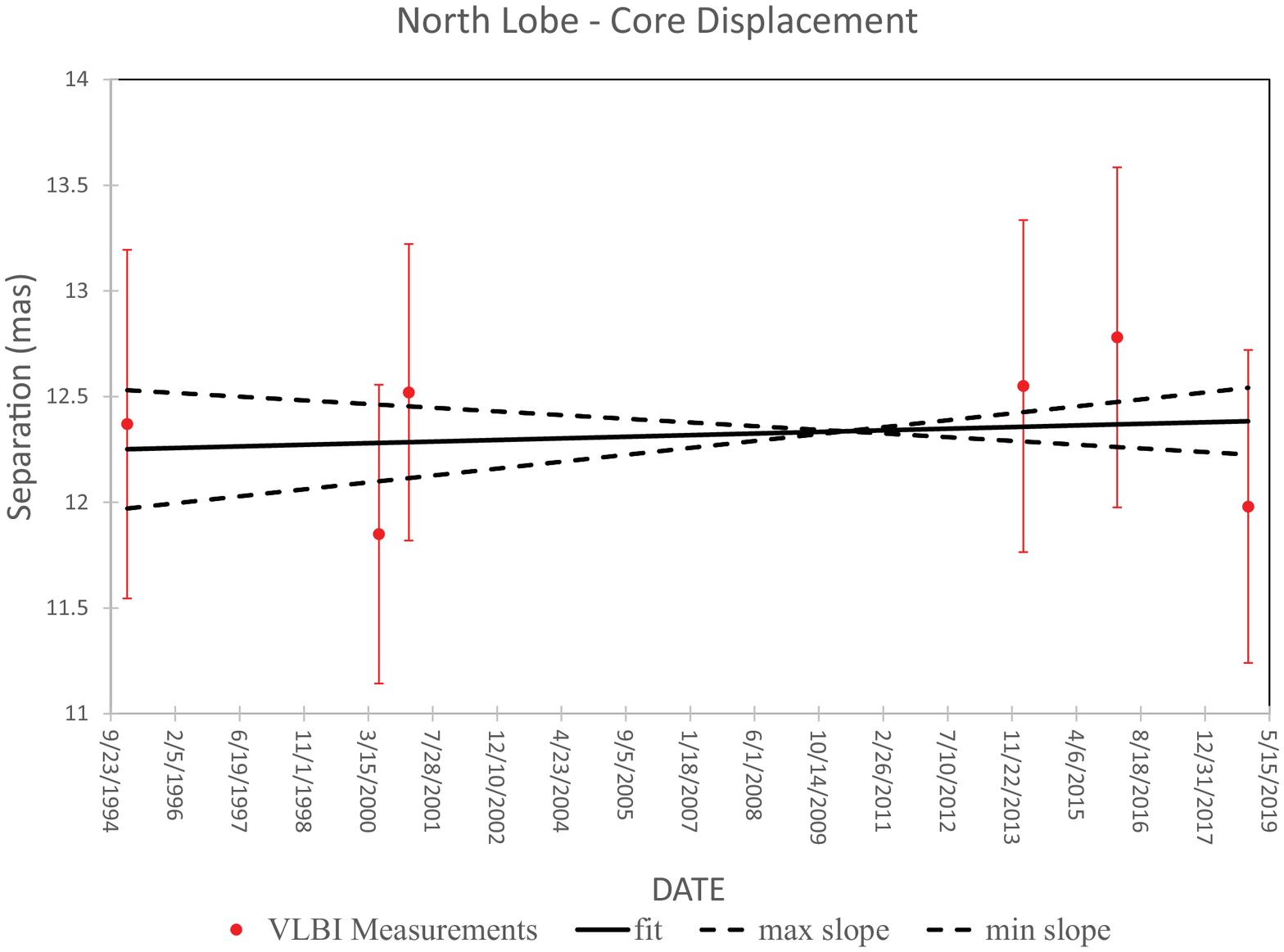}
\caption{There is formally no motion detected in the Northern Lobe in our 6 measurements of its position over 23 years. Due to the high redshift, the instrumental
uncertainties correspond to relatively large physical uncertainties in the true position thereby masking any slow motion. We try to bound the motion with a fit to the data
with uncertainty in the vertical variable \citep{ree89}. The fit is the solid line and the standard error of the fit is indicated by the dashed lines. The maximum apparent
velocity that is compatible with the standard error is $v_{\mathrm{apparent}}< 0.57\,c$.}
\label{fig:displacement}
\end{center}
\end{figure*}

The primary objective of the C-band Gaussian fits in Table~\ref{tab:vlbi-Cband-fit} is to quantify the apparent motion of the North Lobe. Hence, we have included some
Astrogeo data that likely have insufficient sensitivity to capture the diffuse lobe flux density accurately. However, they are adequate for finding the position of the North
Lobe. Based on Table~\ref{tab:vlbi-Cband-fit} and Figure~\ref{fig:Cband}, it seems to be stationary over a quarter century within uncertainties. Any apparent motion of the
Northern Lobe is masked by the relatively large uncertainties. These uncertainties control the constraints that we can put on an upper bound to the component motion.
Figure~\ref{fig:displacement} is a scatter plot of the separation versus time from the Gaussian fits in Table~\ref{tab:vlbi-Cband-fit}. The separation data in
Table~\ref{tab:vlbi-Cband-fit} were fit by least squares with uncertainty in the vertical variable in Figure~\ref{fig:displacement} \citep{ree89}. The standard error of the
fit is given by the dashed lines. Based on the fit, the lobe advances at $0.006\pm 0.018$\,mas~yr$^{-1}$, i.e. consistent with no motion. In our chosen cosmology, we have
7.31\,pc~mas$^{-1}$, so the apparent velocity of the North Lobe relative to the nucleus is $v_{\mathrm{apparent}}= (0.13\pm 0.44)\,c$. The uncertainties in the data do not
allow us to provide much more than an upper bound of $v_{\mathrm{apparent}}< 0.57\,c$.

\subsection{High Frequency VLBI}

High frequency VLBI observations have the resolution required to define the Southeast Component. The details of the four high frequency observations are described in
Table~\ref{tab:vlbi-high-fit}. The first observation is the X-band observation performed simultaneously with the first S-band entry in Table~\ref{tab:modelfit-Sband}
\citep{fey00}. The second observation to consider is the global VLBI X-band observation that was coincident with the S-band observation described in
Table~\ref{tab:modelfit-Sband} and Figure~\ref{fig:Sband}. The most curious feature is that the elongated core seen in the global VLBA observation of
2000 June 5 is resolved into a nuclear two component structure. There was an observation at 8.4 GHz with high background noise
that we do not include in the table from 2000 June 5 \citep{osu11}.

\begin{table}[htp!]
    \caption{Gaussian Fits to High Frequency VLBI Observations}
    \label{tab:vlbi-high-fit}
\begin{center}
   \tiny{ \begin{tabular}{ccccccccccc}
        \hline
        (1)  &(2) & (3) & (4)  & (5)  &  (6) & (7) & (8) & (9) & (10) & (11)\\
        Date  & Array/ & Restoring & Component  & Flux  & $r$ & Position & FWHM & Axial& $T_{\mathrm{b}}$ & Ref.\\
          & Frequency & Beam &  & Density  &  & Angle &  & & & \\
        &  & (mas) &         &      (mJy)    &        (mas)     & ($\degr$)       &  (mas)         &  Ratio/PA &  (K) &       \\
 \hline
    1997 Jan 11 & VLBA & $2.2 \times 1.0$ & Core  & $480\pm 48$ & 0 & ... & 0.50 & 0/$24\degr$ & ... & \tablenotemark{\tiny{a}} \\
               & 8.55 GHz & & North Lobe  & not  & detected & .... & ... & .. & ... &...  \\
                &  &  & Southeast Component\tablenotemark{\tiny{c}}  & $30\pm 9.3$  & $1.0\pm 0.3$ & $135$ & 0.9 & 0/$30\degr$ & ... & \\
        \hline
    2002 Jul 24 & Global VLBI & $1.75 \times 0.79$ & Core  & $466\pm 46$ & 0 & ... & 0.18 & 1& $1.07 \times 10^{12}$  & \tablenotemark{\tiny{b}} \\
               & 8.65 GHz & & North Lobe  & $11.8\pm 4.7$  & $11.5\pm 0.3$ & $-20.6$ & 2.1 & 1 & $2.62 \times 10^{9}$  &\tablenotemark{\tiny{e}}  \\
               &  &  & Nuclear Secondary\tablenotemark{\tiny{c}}  & $233 \pm 23$  & $0.65\pm 0.3$ & $-166.6$ & 0.33 & 1 & $1.64 \times 10^{11}$  &  \\
                &  &  & Southeast Component\tablenotemark{\tiny{c}}  & $37.9 \pm 8.5 $  & $1.2\pm 0.3$ & $141.0$ & \tablenotemark{\tiny{d}} & 1 & \tablenotemark{\tiny{d}} &
                \\
                 \hline
    1996 Jun 13 & VLBA & $1.16 \times 0.47$ & Core  & $529 \pm 53$ & 0 & ... & 0.75 & 0.19/$17.6\degr$  & $1.99 \times 10^{11}$  & \tablenotemark{\tiny{e}} \\
                & 15.4 GHz & & North Lobe  & not  & detected & .... & ... & .. & ... &  \\
         &  &  & Southeast Component  & $17.4 \pm 4.6 $  & $1.2\pm 0.2$ & $145.0$ & 0.0 & ... & ...  & \\
    \hline
   1996 Sep 7 & VLBA & $1.16 \times 0.47$ & Core  & $398 \pm 40$ & 0 & ... & 0.76 & 0.22/$18.7\degr$ & $3.61 \times 10^{10}$  & \tablenotemark{\tiny{e}} \\
                & 22.1 GHz & & North Lobe  & not  & detected & .... & ... & .. & ... & ... \\
         &  &  & Southeast Component  & $< 7.5$  &  & ... & ... & ... & ...& ...\\
    \hline
    \end{tabular}}
\end{center}
\tablenotetext{a}{\citet{fey00}. Flux density error estimate of Southeast Component uses the 26\% uncertainty computed in the 2002 observation added in quadrature with
roundoff error.}
\tablenotetext{b}{\citet{pus12}}
\tablenotetext{c}{Sparse $(u,v)$ coverage at baselines long enough to model the $\sim 1$ mas nucleus.}
\tablenotetext{d}{Effectively a point source, $T_{\mathrm{b}}$ estimate ill-defined}
\tablenotetext{e}{This paper}
\end{table}
\begin{figure*}[htp!]
\begin{center}
\textbf{VLBA 15.4 GHz \hspace{5.3cm} VLBA 22.1 GHz}\par\medskip
\includegraphics[width= 0.47\textwidth,angle =-90]{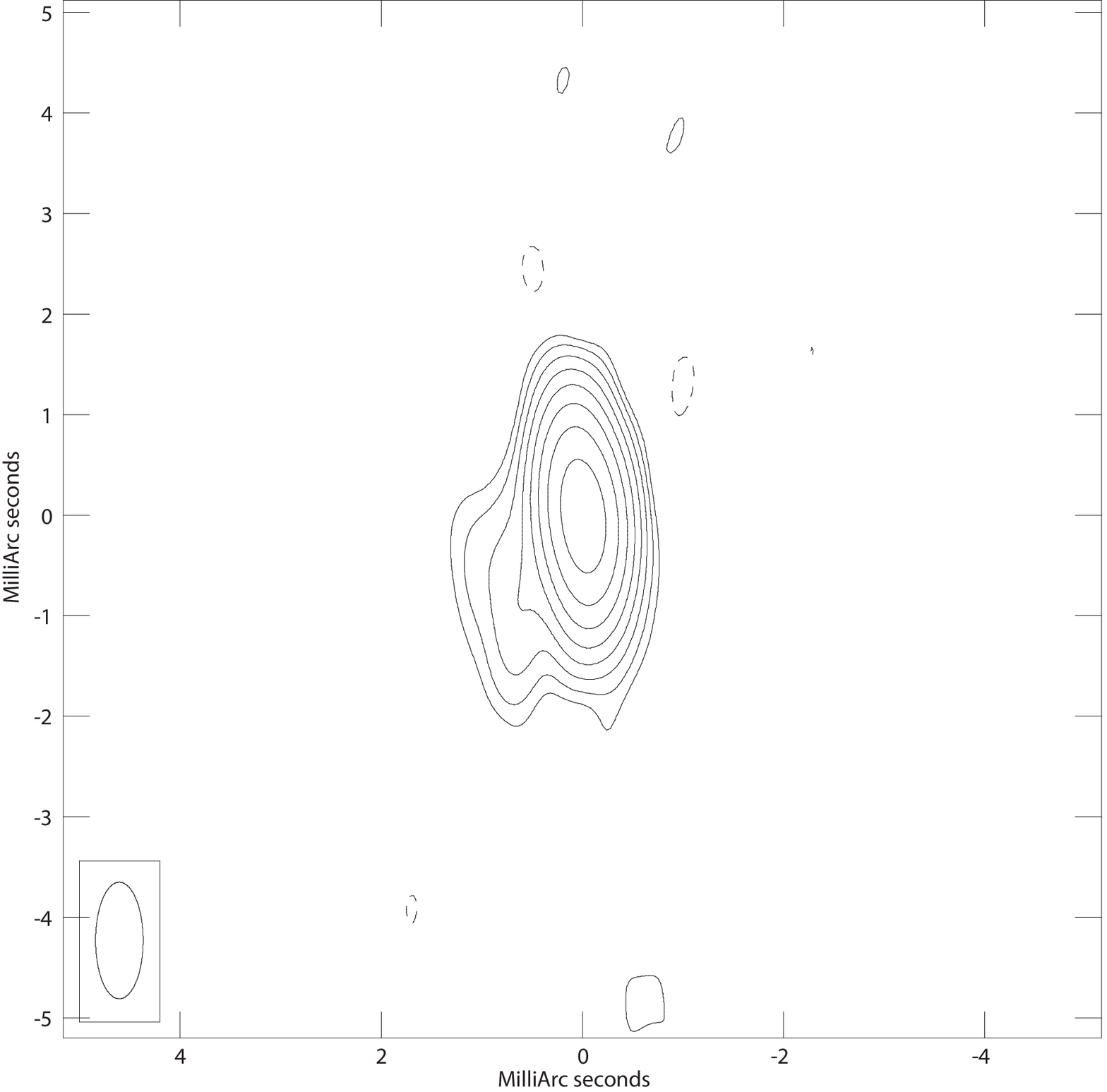}
\includegraphics[width= 0.47\textwidth,angle =-90]{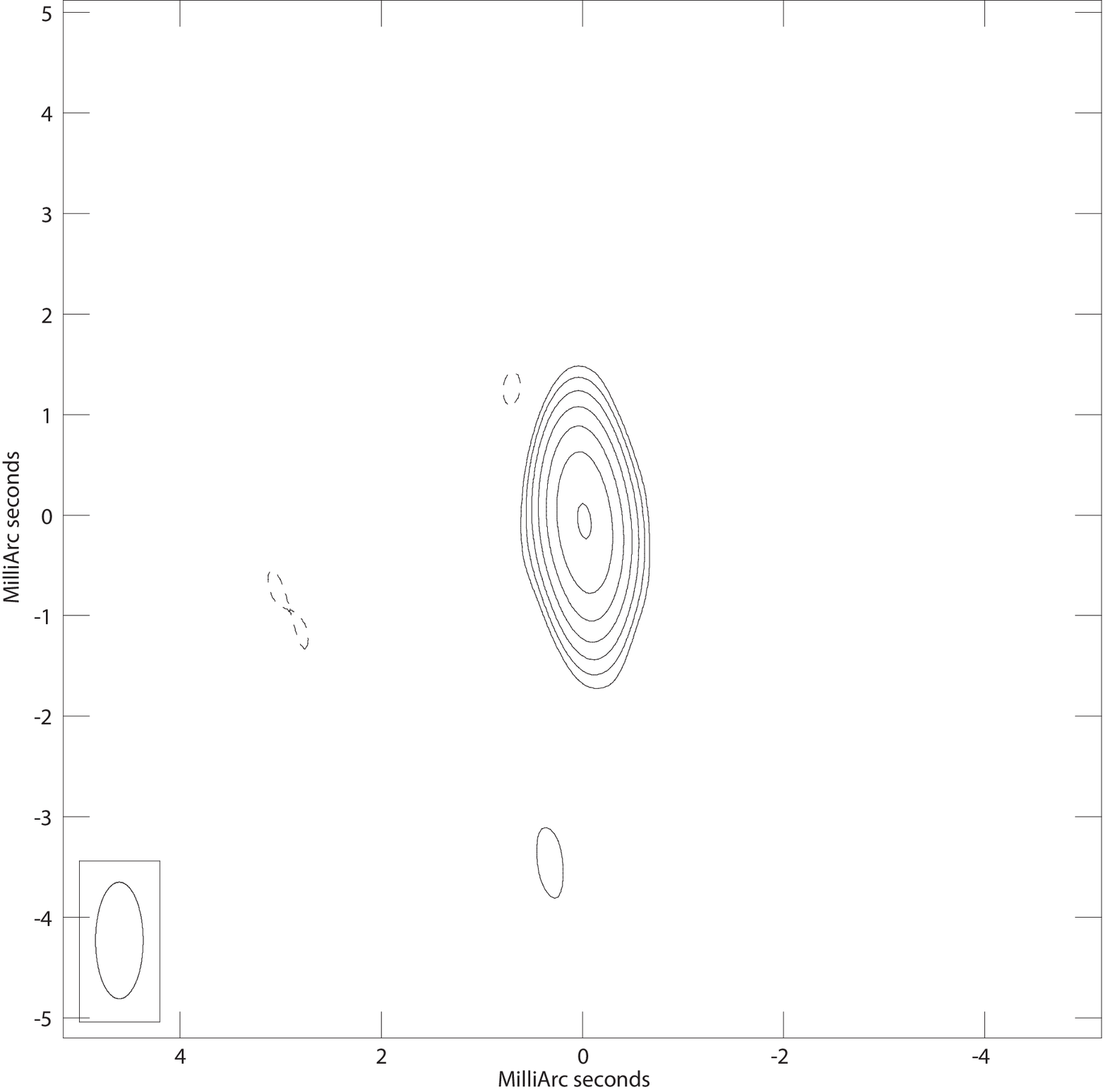}
\caption{On the left hand side is the 15.4 GHz VLBA image from project BK0042 on 1996 June 13. On the right hand side is the 22.1 GHz VLBA image from 1996 September 7. To aid
comparison, both images are created with the same restoring beam, the naturally weighted beam from the 15.4 GHz observation, $1.16\,\mathrm{mas} \times 0.47\,\mathrm{mas}$ at
$\mathrm{PA} = 1.9\degr$. The bottom contour is $\pm 3$ times the rms noise 1.95\,mJy~beam$^{-1}$ (4.47\,mJy~beam$^{-1}$) and the peak intensity is
405\,mJy~beam$^{-1}$ (298\,mJy~beam$^{-1}$) at 15.4 (22.1) GHz. The positive contours increase by factors of two. Even though the two observations occurred only 18.5 days
apart in the quasar rest frame, the prominent Southeast Component seen in the left panel is absent in the right panel.}
\label{fig:high-freq}
\end{center}
\end{figure*}

The two never before published observations from the VLBA archives are presented in Figure~\ref{fig:high-freq}. First of all, note that the three observations and the 2000
June 5 global VLBI observations at 5 GHz define a consistent axis for the nuclear region at $\mathrm{PA} \approx -165\degr$ and
based on the X-band resolution, it is directed toward the south. This explains the jet geometry. The inner jet is directed at $\mathrm{PA} \approx -165\degr$ for about
0.8\,mas. At this point it veers abruptly toward the Southeast Component at $\mathrm{PA} \approx 130\degr$ (an $\approx 65\degr$ rotation). We postulate that after which it
bends toward the north through the Knot in North Jet ($3-5$\,mas out), finally terminating at the North Lobe (12.5\,mas out). Such behavior up to and through the Southeast
Component is indicative of Doppler abberation and a nearly polar LOS \citep{lin85}.

\begin{figure*}[htp!]
\begin{center}
\includegraphics[width= 0.85\textwidth,angle =0]{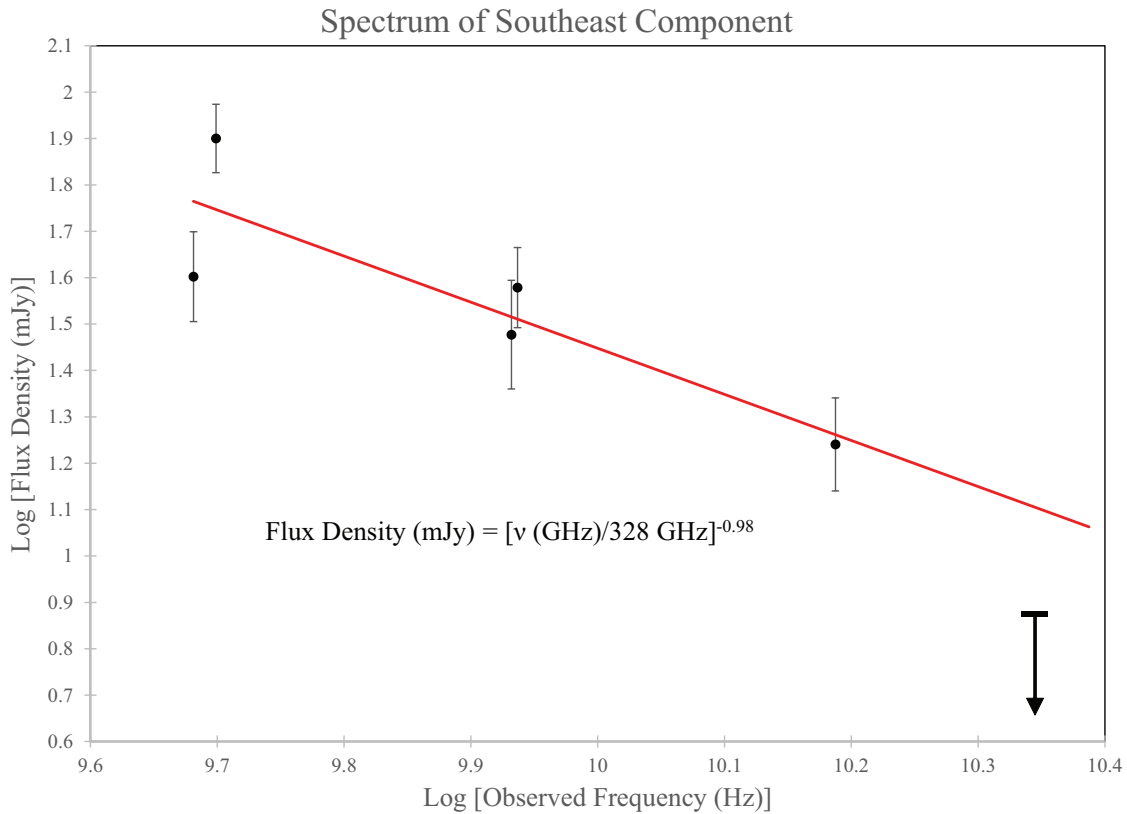}
\caption{\footnotesize{Based on all the available observations, this is our chosen characterization of the spectrum of the Southeast Component. It is a component only 1\,mas
from the core and only $\sim 3-5\%$ as bright. It is apparently variable as well, but this magnitude is difficult to determine as a result of the uncertainty of fitting a
relatively faint component so close to a bright core. Variations on the order $\pm 30\%$ seem consistent with the scatter to the fit of the data from 4.8 to 15.4 GHz. The
spectral fit is therefore, an average fit and it is clearly steep. We use the spectral index of $\alpha = 0.98$ in our theoretical models of the source. Our estimates of jet
power do not rely strongly on this quantity. We also note that there might be a spectral turnover at $\approx 15$~GHz.}}
\label{fig:spectrum}
\end{center}
\end{figure*}

Next, we use the high frequency observations to gain clues into the nature of the Southeast Component. The lack of a detection at 22.4 GHz is unexpected based on the apparent
sensitivity of the observations. The 9-h observation for VLBA project BN0003 during 1996 September 7--8 produced a total of 64 minutes of data on source PKS\,1351$-$018 with
32~MHz bandwidth in a single polarization. The phased-VLA (in D-array) was added to the array improving the (still sparse) \textit{u,v} coverage and improving the
sensitivity. We found an rms noise of $\approx 1.5$\,mJy~beam$^{-1}$. Based on the size found at 15 GHz in Table~\ref{tab:vlbi-high-fit}, we would have expected to have
detected a flux density as low as 7.5\,mJy at the 5$\sigma$ level. Figure~\ref{fig:spectrum} shows extrapolated power law fit from the lower frequency data. We expected
$10-15$\,mJy. There are three possible explanations:
\begin{enumerate}
\item There is a flaw in the radio observation introduced by a subtle calibration issue. Although there are no obvious signs of artefacts in the image, the surface
    brightness of the noise is 2.25 times higher at 22.1~GHz than at 15.4~GHz. We investigated this possibility by imaging 4C\,39.25, a very bright source observed during
    the same observing run. We would expect that such an
    error would be easier to spot in those data. Nothing was apparent. Further, we imaged the PKS\,1351$-$018 data using various subsets of the antennas, none of which
    indicated significant emission at the position of the Southeast Component, or significant changes in the observed morphology.
\item There could be strong synchrotron cooling that makes the spectrum curve downward sharply around 15 GHz.
\item The source might be highly variable. However, consider the following from Table~\ref{tab:vlbi-high-fit}. 18.5 days earlier in the quasar rest frame, the Southeast
    Component was detected with 22\,mJy at 15.4~GHz. 20 days later in the quasar rest frame, it was detected with $\approx 30$\,mJy at 8.55~GHz (similar to the estimated
    flux density in 2002).
\end{enumerate}
We simply do not have enough information to reach a definitive conclusion on the contribution of each of these possibilities. We assume that all of these are factors, the
K-band image is relatively noisy, and there is a spectral steepening (perhaps not severe). Furthermore, there is some significant (but not extreme) variability. In
Figure~\ref{fig:spectrum}, we look at a scatter plot of the flux density at the four frequency bands in order to try and organize the trend in the difficult to measure
quantity. A good measurement requires high resolution, high sensitivity, and high dynamic range. That being said, 15~GHz VLBA is our most reliable data. The only thing
stopping a very tight trend in Figure~\ref{fig:spectrum} are the two disparate C-band flux densities and the 22 GHz non-detection. The 22 GHz data are still a mystery
considering the detections of the Southeast Component within 3 weeks before and after (in the quasar rest frame). In the following, we will assume that the power law fit in
Figure~\ref{fig:spectrum} (spectral index $\alpha \approx 0.98$) applies to the data at frequencies below 15 GHz with some modest ($\sim 20-30\%$) variability.

\section{The Synchrotron Spectral Energy Distribution}
\label{sed}

This section compiles the radio to millimeter wave data that are used to construct possible physical models of the radio source later in this paper. The results are in
Table~\ref{tab:radio-data}. Since the data have relatively large measurement uncertainty compared to the magnitude of flux density variability of PKS\,1351$-$018, we have
averaged the historical data to reduce the uncertainty. Figure~\ref{fig:radiolightcurve} is the best example, but we also average at other frequencies when possible. For now,
we simply collect the data and create a synchrotron SED that extends from $\nu \sim 6\times 10^{8}$ Hz to $\nu \approx 10^{12}$ Hz in the quasar rest frame.

\begin{table}[htp!]
\begin{center}
    \caption{Radio Data for PKS\,1351$-$018}
    \label{tab:radio-data}
{\tiny\begin{tabular}{cccccc} \tableline\rule{0mm}{3mm}
$\nu_{\mathrm{o}}$ & $\log{\nu}$ &Flux & Telescope & Reference & Comments\\
Observed  &  Quasar Rest   & Density &  &  &  \\
Frequency   & Frame  &  &  &  &  \\
(MHz)   & (Hz)  & (mJy) & &  &  \\
\tableline \rule{0mm}{3mm}
$123^{+20}_{-16}$ & $8.76\pm 0.06 $ & $214.5 \pm 27.8 $ \tablenotemark{\tiny{a}} & MWA\tablenotemark{\tiny{b}} & \citet{way15}\tablenotemark{\tiny{c}}   & 5 bin average \\
150 &    8.76      & $185.0\pm 27.8$ \tablenotemark{\tiny{a}} & GMRT & \citet{int17,hur17}  & TGSSADR \\
$165\pm 15$ &   $8.89\pm 0.04$ & $246.4\pm 37.0$\tablenotemark{\tiny{a}} & MWA\tablenotemark{\tiny{b}} & \citet{way15}\tablenotemark{\tiny{c}} & 5 bin average \\
$208\pm 19$ &   $8.99\pm 0.04$     & $276.8\pm 41.5$\tablenotemark{\tiny{a}}  & MWA\tablenotemark{\tiny{b}} & \citet{way15}\tablenotemark{\tiny{c}} & 5 bin average   \\
330  &   9.19     & $329.7\pm 33$  & VLA & This paper  & A-array  \\
340 & 9.20 & $362\pm 50$  & VLA/VLITE & This paper, \citet{cla16,pol16}\tablenotemark{\tiny{d}} & 15 epoch average \\
365 & 9.24 & $371\pm 37$  & Texas Interferometer & \citet{dou96} &   \\
960 & 9.66 & $510\pm 51$  & RATAN-600 & \citet{kov99} &   \\
1400 & 9.82 & $733\pm 37$  & VLA D-array & \citet{con98} &  NVSS \\
1400 & 9.82 & $709\pm 35$  & VLA B-array & \citet{bec95} &  FIRST \\
1484 & 9.84 & $743\pm 37$  & VLA A-array & \citet{nef90} &   \\
2100 & 10.00 & $850\pm 85$  & ATCA & Calibrator Database\tablenotemark{\tiny{e}} &   \\
2700 & 10.10 & $897\pm 53$  & ATCA & Calibrator Database\tablenotemark{\tiny{e}} & 8 epoch average   \\
4900 & 10.36 & $905\pm 42$  & VLA and ATCA & Average of Figure 1 &   \\
8470 & 10.60 & $812\pm 70$  & VLA  & NVAS, this paper & 27 epoch average  \\
14900 & 10.85 & $669\pm 39$  & ATCA  & Calibrator Database\tablenotemark{\tiny{e}}& 8 epoch average  \\
22400 & 11.02 & $542\pm 35$  & ATCA  & Calibrator Database\tablenotemark{\tiny{e}} & 8 epoch average  \\
33000 & 11.19 & $373\pm 56$  & ATCA  & Calibrator Database\tablenotemark{\tiny{e}} &   \\
43000 & 11.31 & $400\pm 80$  & VLA  & Calibrator List\tablenotemark{\tiny{f}} &   \\
43000 & 11.31 & $303\pm 61$  & ATCA  & Calibrator Database\tablenotemark{\tiny{e}} &   \\
43000 & 11.31 & $373\pm 56$  & ATCA  & Calibrator Database\tablenotemark{\tiny{e}} &   \\
90000 & 11.63 & $177\pm 18$  &  IRAM 30-meter  & \citet{ste95} &   \\
93000 & 11.64 & $170\pm 50$  & ATCA  & Calibrator Database\tablenotemark{\tiny{e}} &   \\
230000 & 12.03 & $65\pm 13$  & IRAM 30-meter  & \citet{ste95} & \\ \hline
\end{tabular}}
\end{center}
\tablenotetext{a}{Uncertainty from \citet{hur17}}
\tablenotetext{b}{Murchison Widefield Array}
\tablenotetext{c}{GLEAM: \url{https://vizier.u-strasbg.fr/viz-bin/VizieR-3?-source=VIII/100/gleamegc}}
\tablenotetext{d}{VLA Low Band Ionospheric and Transient Experiment (VLITE). Data provided by Wendy Peters.}
\tablenotetext{e}{\url{https://www.narrabri.atnf.csiro.au/calibrators}}
\tablenotetext{f}{\url{https://science.nrao.edu/facilities/vla/observing/callist}}
\end{table}

\begin{figure*}[htp!]
\begin{center}
\includegraphics[width= 0.85\textwidth,angle =0]{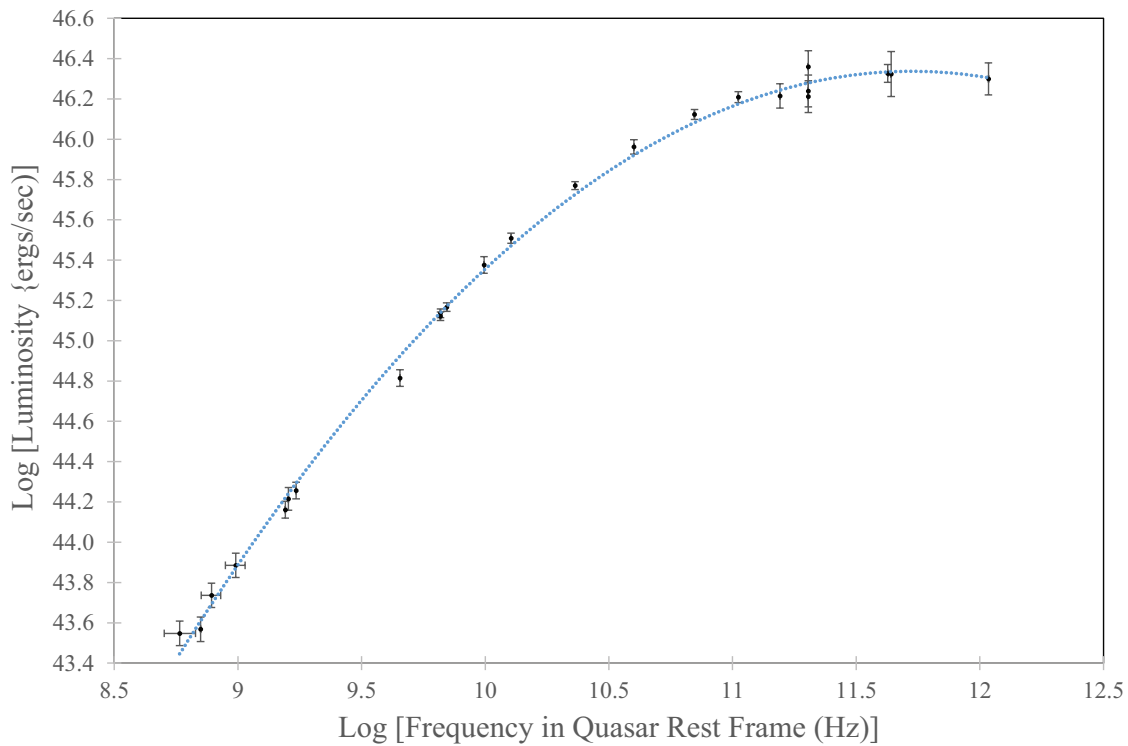}
\caption{The synchrotron SED created from the data in Table~\ref{tab:radio-data}. The luminosity of the low frequency portion of the SED (the plotted portion) has a
luminosity of $\gtrsim 7 \times 10^{46}$\,erg~s$^{-1}$. The full synchrotron luminosity most certainly exceeds $10^{47}$\,erg~s$^{-1}$. The peak of the SED is
$\nu_{\rm{peak}} \approx 5\times 10^{11}$\,Hz in the quasar rest frame.}
\label{fig:sed}
\end{center}
\end{figure*}

The first column of Table~\ref{tab:radio-data} is the observed frequency. This is converted into the logarithm of the frequency in the quasar rest frame in the next column.
We then give the flux density with its uncertainty which is considerable at very low and very high frequencies for individual measurements. We then list the telescope used,
the reference to the data and the comments in the final three columns. We note how many of the frequencies were able to be averaged over at least a modest set of
historical observations with the same telescope in the comments column. In our modelling, the low frequency data are very important. The $\nu_{\mathrm{o}}=107$\,MHz to
$\nu_{\mathrm{o}}=227$\,MHz Galactic and Extra-galactic All-sky MWA survey (GLEAM) data are very useful, but have much scatter due to the low flux densities
($\nu$ designates frequencies in the cosmological rest frame of
the quasar and $\nu_{\mathrm{o}}$ the observed frequencies; $\nu = (1+z) \nu_{\mathrm{o}}$). There were 15 channels and we averaged 5 at a time in order to get a more
robust flux density. Even so, the data do not agree well with the $\nu_{\mathrm{o}}=150$\,MHz Tata Institute for Fundamental Research (TIFR) Giant Metrewave
Radio Telescope (GMRT) Sky Survey Alternate Data Release (TGSSADR) flux density. The modest low frequency flux density of PKS\,1351$-$018
provides a challenge to survey observations, and TGSSADR has only one data point, so it is more difficult to check its consistency. We also were fortunate to obtain numerous P-band observations from JVLA and the one from the VLA. At $\nu_{\mathrm{o}}=340$\,GHz, we pick the mean and the standard deviation of 64 observations binned into 15 distinct epochs (each epoch has $< 10$\,d spread in the observations in the quasar rest frame) to be the values of the flux density and uncertainty, respectively. These observations span 5.5\,yr from 2015 to 2020. The synchrotron SED is plotted in Figure~\ref{fig:sed}. We added a log-parabolic fit. Unfortunately, the high frequency observations that define the peak are difficult and the uncertainty and scatter are larger than one would like. A carefully calibrated 43 GHz JVLA observation would be very useful. The SED peak is at $\nu_{\mathrm{peak}} \approx 5\times 10^{11}$\,Hz. Based on the ``blazar sequence'' this is at the low end of the peak frequency expected for a flat spectrum \textit{Fermi} detected quasar with a peak spectral luminosity of $\sim 2\times 10^{46}$\,erg~s$^{-1}$ \citep{ghi17}. Typically, they find $\nu_{\mathrm{peak}} \approx 2.0\times 10^{12}$\,Hz for a flat spectrum quasar that is \textit{Fermi} detected. The integrated luminosity of the SED up to $10^{12}$\,Hz in the quasar rest frame (the limit of our data) is $7\times 10^{46}$\,erg~s$^{-1}$, or $L_{\mathrm{synch}}> 7\times 10^{46}$\,erg~s$^{-1}$ since we do not include the high frequency side.

\section{Electromagnetic Signature of the Accretion Flow}
\label{energetics}

\begin{figure*}[htp!]
\begin{center}
\includegraphics[width= 0.75\textwidth,angle =0]{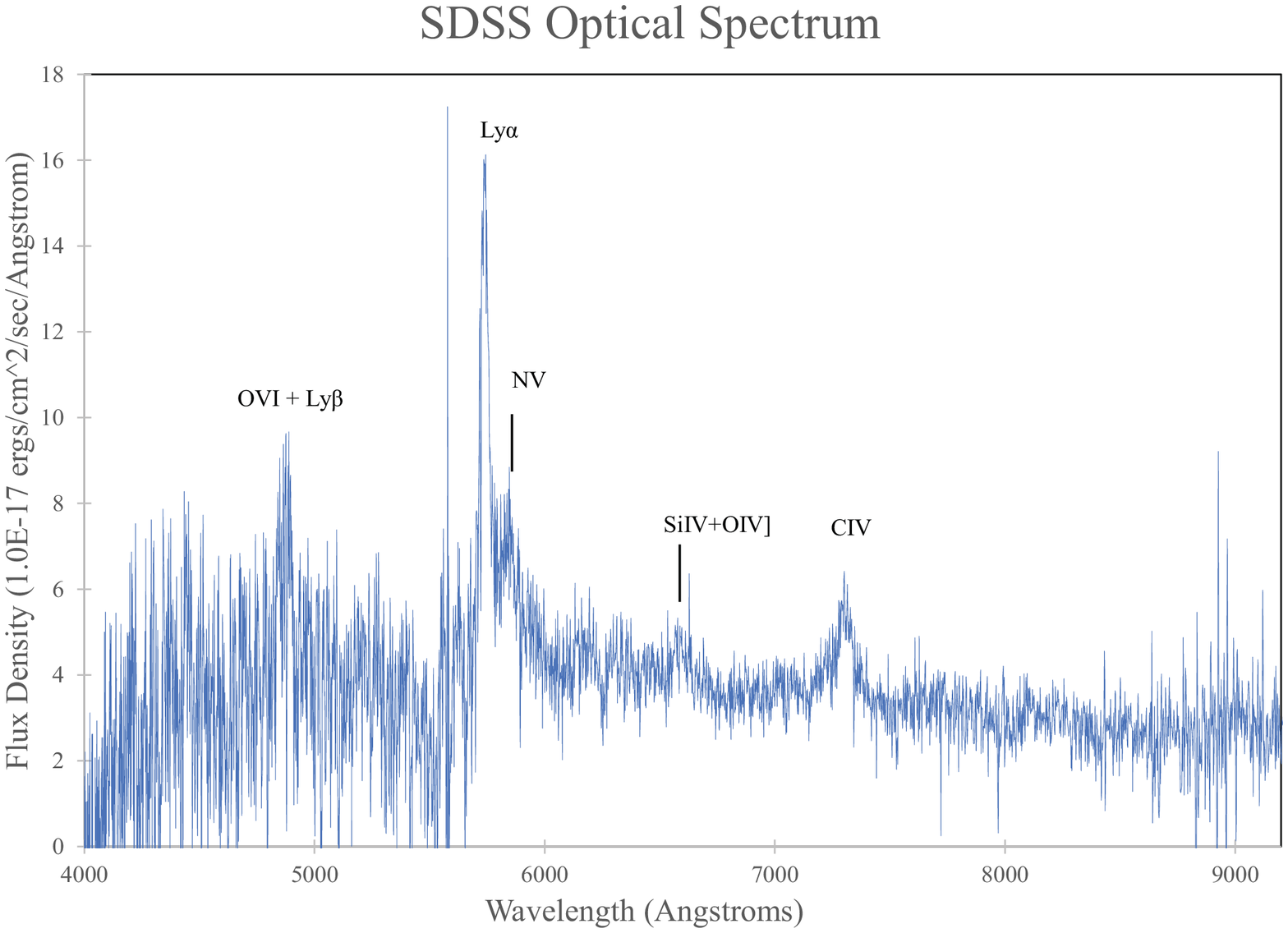}
\includegraphics[width= 0.75\textwidth,angle =0]{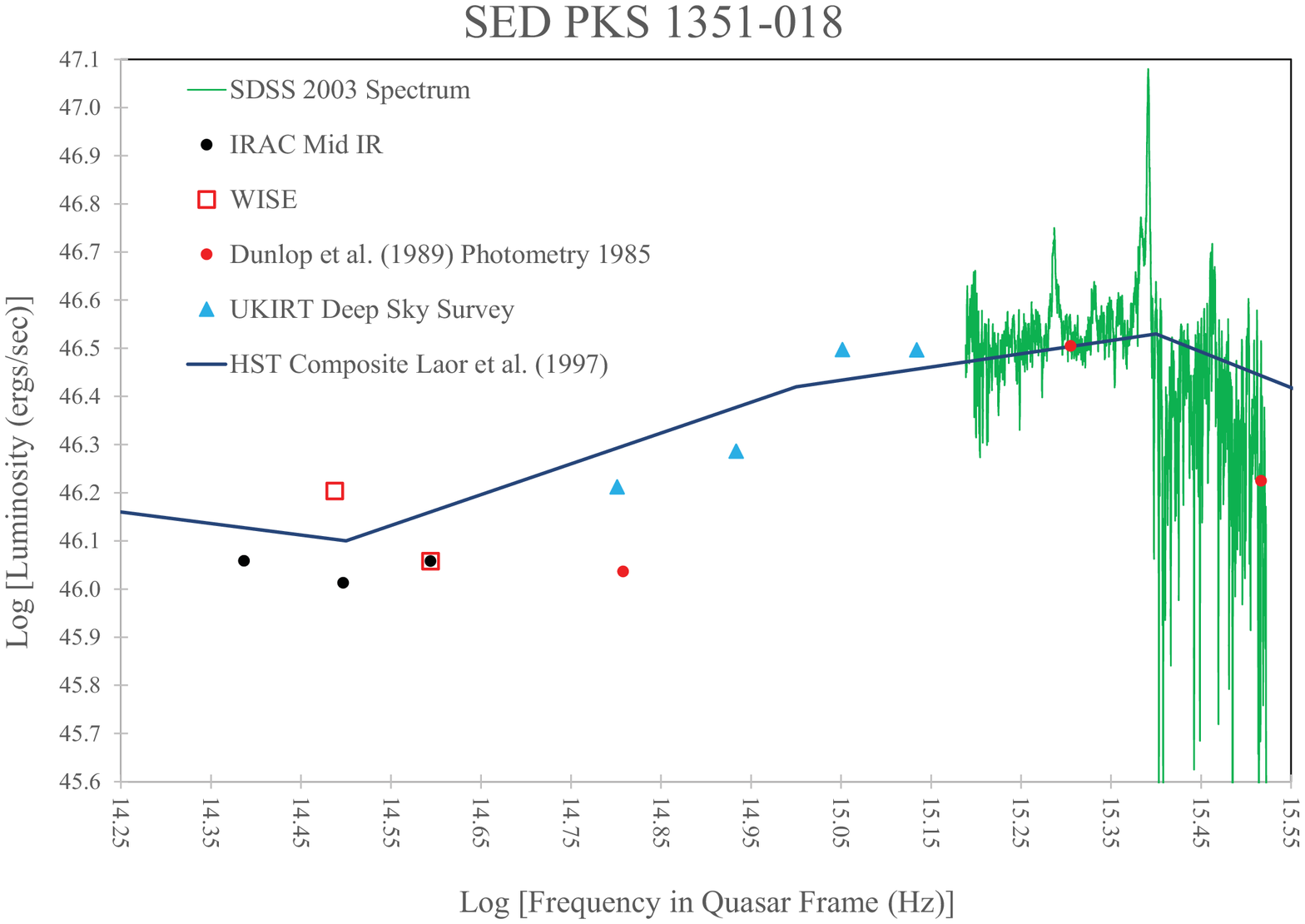}
\caption{The top panel is the SDSS optical spectrum. There is strong Ly$\alpha$ absorption short-ward of the Ly$\alpha$ BEL. The prominent BELs are indicated. The bottom
panel is the SED that includes some photometry points at different epochs. This is compared to the composite HST based spectrum \citep{lao97,tel02}. The rest frame near IR to
far UV SED is typical of a radio quiet spectrum. Optical and IR photometry at other epochs indicate very low variability compared to a blazar. This is consistent with the
apparent turnover of the synchrotron SED at $\nu_{\rm{peak}} \approx 5\times 10^{11}$ Hz indicated in Figure~\ref{fig:sed}.}
\label{fig:optical-spectrum}
\end{center}
\end{figure*}

In this section, we explore the optical spectrum, near IR and mid IR photometry, in order to define the thermal emission of the accretion flow from the rest frame UV to the
near IR (the signature of the quasar). There are two optical spectra of this quasar. The first observation was in 1985 with the Anglo-Australian Telescope
\citep[AAT,][]{dun89}. There is a 2003 March 23 Sloan Digital Sky Survey (SDSS) spectrum that we show in the top frame of Figure~\ref{fig:optical-spectrum}. It has
been corrected for Galactic extinction using the extinction values in the NASA Extragalactic Database (NED) applied to the models of \citet{car89}. There is very deep
Ly$\alpha$ absorption from intervening gas short-ward of the Ly$\alpha$ emission from PKS\,1351$-$018. The blue side of the Ly$\alpha$ broad emission line is completely
truncated. The continuum spectral index defined in terms of the flux density as $F_{\nu} \propto \nu^{-\alpha_{\nu}}$ is $\alpha_{\nu}\approx 0.78$ long-ward of Ly$\alpha$.
This is typical of a radio quiet quasar. The \textit{Hubble Space Telescope} (HST) composite spectral index was found to be $\alpha_{\nu}\approx 0.86$ \citep{zhe97}.
Thus, this has a very strong accretion signature with very little influence of the jet synchrotron emission long-ward of Ly$\alpha$ (see IR discussion below). This means that
there is no significant synchrotron dilution from the jet. Looking at the peak and turnover at $\approx 10^{12}$ Hz in the synchrotron SED in
Figure~\ref{fig:optical-spectrum} and the SDSS SED in the bottom panel of Figure~\ref{fig:optical-spectrum} this seems clear. The bottom panel of
Figure~\ref{fig:optical-spectrum} compares the SDSS data to the HST composite shape from \citet{lao97} and \citet{zhe97} to the spectrum in top panel after re-scaling to the
continuum level of PKS\,1351$-$018. We added photometry points from archival Mid-IR data in NED and IR observation found in the Data Release 11 of the United
Kingdom Infrared Telescope (UKIRT) Deep Sky Survey\footnote{\url{http://wsa.roe.ac.uk/index.html}} \citep{law07}. This shows the characteristic 1 micron dip in the quasar
spectrum that appears in the composite. Thus, there is no evidence of synchrotron dilution even on the up-slope from the dip. Note the three photometry points from
\citet{dun89}. The B-band and R-band photometry lie right on top of the SDSS spectrum indicating very little variability between 1985 and 2003. The K-band data point in
\citet{dun89} was taken with UKIRT, but disagrees with the value from the UKIRT Deep Sky Survey.

We can use the spectrum in Figure~\ref{fig:optical-spectrum} to estimate the bolometric thermal luminosity from the accretion flow, $L_{\rm{bol}}$. The desired estimate does
not include reprocessed radiation in the infrared from molecular clouds that are far from the active nucleus (not shown in the bottom panel of
Figure~\ref{fig:optical-spectrum}). This would be double counting the thermal accretion emission that is reprocessed at mid-latitudes \citep{dav11}. The most direct method is
to use the UV continuum as a surrogate for $L_{\rm{bol}}$. From the spectrum in Figure~\ref{fig:optical-spectrum} and the formula expressed in terms of quasar cosmological
rest frame wavelength, $\lambda_{e}$, and spectral luminosity, $L_{\lambda_{e}}$, from \citet{pun16},
\begin{equation}
L_{\mathrm{bol}} \approx (4.0 \pm 0.7)\lambda_{e}L_{\lambda_{e}}(\lambda_{e} = 1350 \AA)\approx (1.45 \pm 0.25) \times 10^{47} \mathrm{erg}~\mathrm{s}^{-1}\;.
\label{equ:Lbol}
\end{equation}
The bolometric correction was estimated from a comparison to HST composite spectra of quasars with $L_{\mathrm{bol}} \approx 10^{46}$\,erg~s$^{-1}$ \citep{zhe97,tel02,lao97}.

\begin{figure*}[htp!]
\begin{center}
\includegraphics[width= 0.65\textwidth,angle =0]{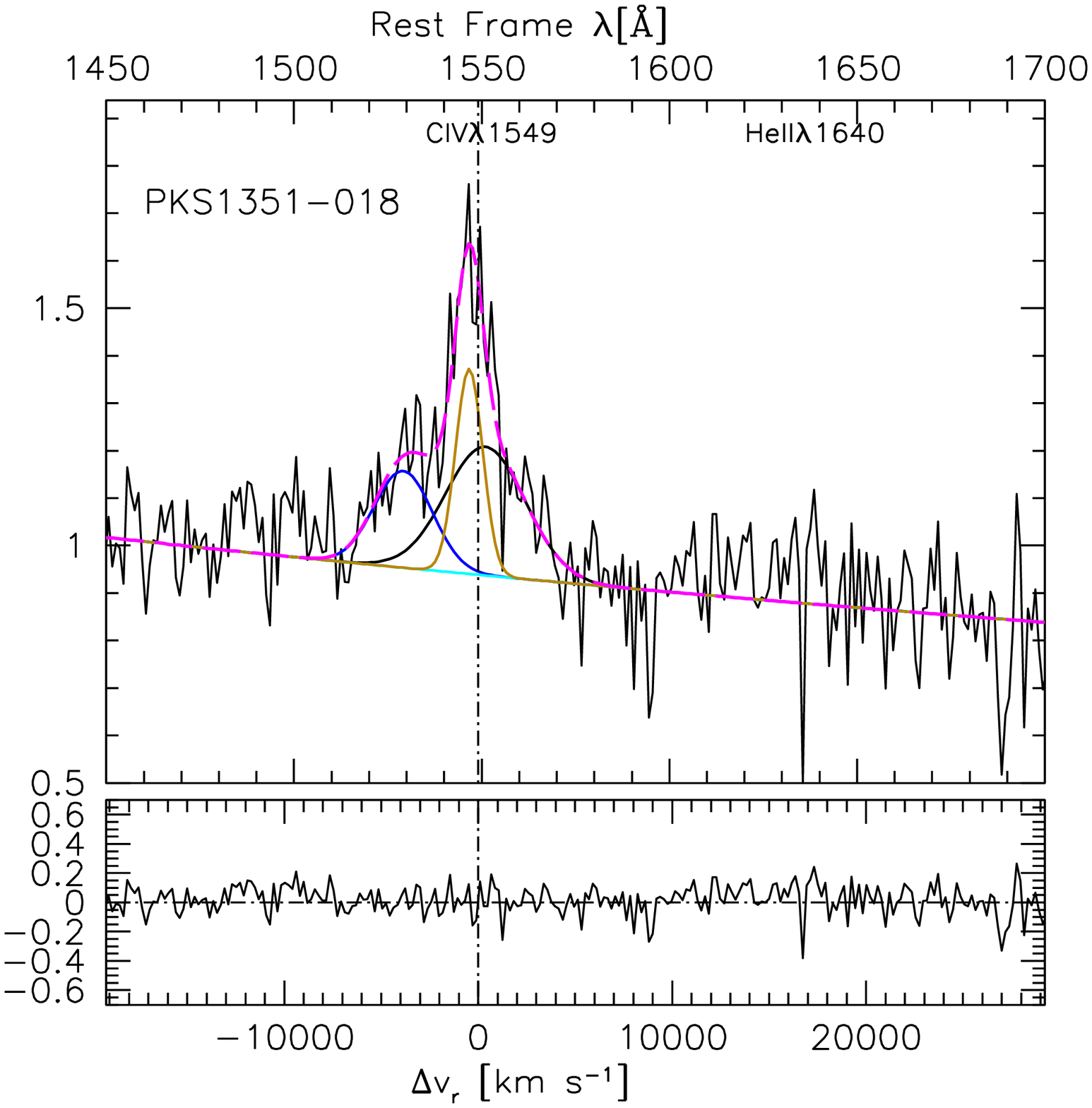}
\caption{Even though the SDSS spectrum is noisy, it is of sufficient signal to noise to reveal the nature of the \ion{C}{4} emission line. The vertical axis is the flux
density in the rest frame of the quasar in units of $3.7 \times 10^{-15}$\,erg~s$^{-1}$~cm$^{-2}$~\AA$^{-1}$, the continuum level at 1450\,\AA. The yellow Gaussian component
is the narrow line emission. There are two broad components that were fit. The broader, more luminous of the two is slightly redshifted and is plotted in black. There is a
highly blue-shifted Gaussian component plotted in blue. This is evidence of a strong, outwardly driven high ionization wind.}
\label{fig:CIV-line}
\end{center}
\end{figure*}

Since Ly$\alpha$ is truncated by the Ly$\alpha$ forest, the only strong broad emission line (BEL) in the SDSS spectrum is \ion{C}{4}. In Figure~\ref{fig:CIV-line}, we fit the
\ion{C}{4} emission line into a common decomposition format, two broad lines and one narrow line \citep{bro96,mar10,sul00}. The blue (red) broad Gaussian component has a line
center shifted $\approx 3900$\,km~s$^{-1}$ ($\approx -390$\,km~s$^{-1}$) with a FWHM of $\approx 3560$\,km~s$^{-1}$ ($\approx 4860$\,km~s$^{-1}$) and a luminosity of $\approx
9.3 \times 10^{43}$\,erg~s$^{-1}$ ($\approx 1.7 \times 10^{44}$\,erg~s$^{-1}$). The narrow line profile has a FWHM $\approx 1660$\,km~s$^{-1}$ and a luminosity of $\approx
9.3 \times 10^{43}$\,erg~s$^{-1}$. There are two odd things about this line. First, it is a relatively weak broad line. The rest frame equivalent width (EW) is $\approx
13$\,\AA\ and if we include the narrow line this only increases to $\textrm{EW} \approx 17$\,\AA. While the typical value from the HST sample of \citet{tel02} is $\sim
60$\,\AA. As discussed earlier this is not a consequence of synchrotron dilution. We note that such small EWs are not unheard of as documented in \citet{bal89,dia09}. The other odd
feature is the strong blue excess which is typical of high luminosity radio quiet quasars, while quasars with powerful radio jets and lobes tend to have a red excess
\citep{ric02,pun10}. The blue shifted Gaussian component is often considered evidence of a wind driven by the radiation pressure from the accretion flow
\citep{bro94,bro96,mur95,net10,sul17}. This behavior seems to be explained by the large $L_{\rm{bol}}$ found in Equation~(\ref{equ:Lbol}). The quasar might have a high
Eddington luminosity, but we have no reliable virial estimate in the absence of a broad low ionization line to
measure.

\section{$\gamma$-Ray Behavior}
\label{gammaray}

\begin{figure*}[htp!]
\begin{center}
\includegraphics[width= 0.8\textwidth,angle =0]{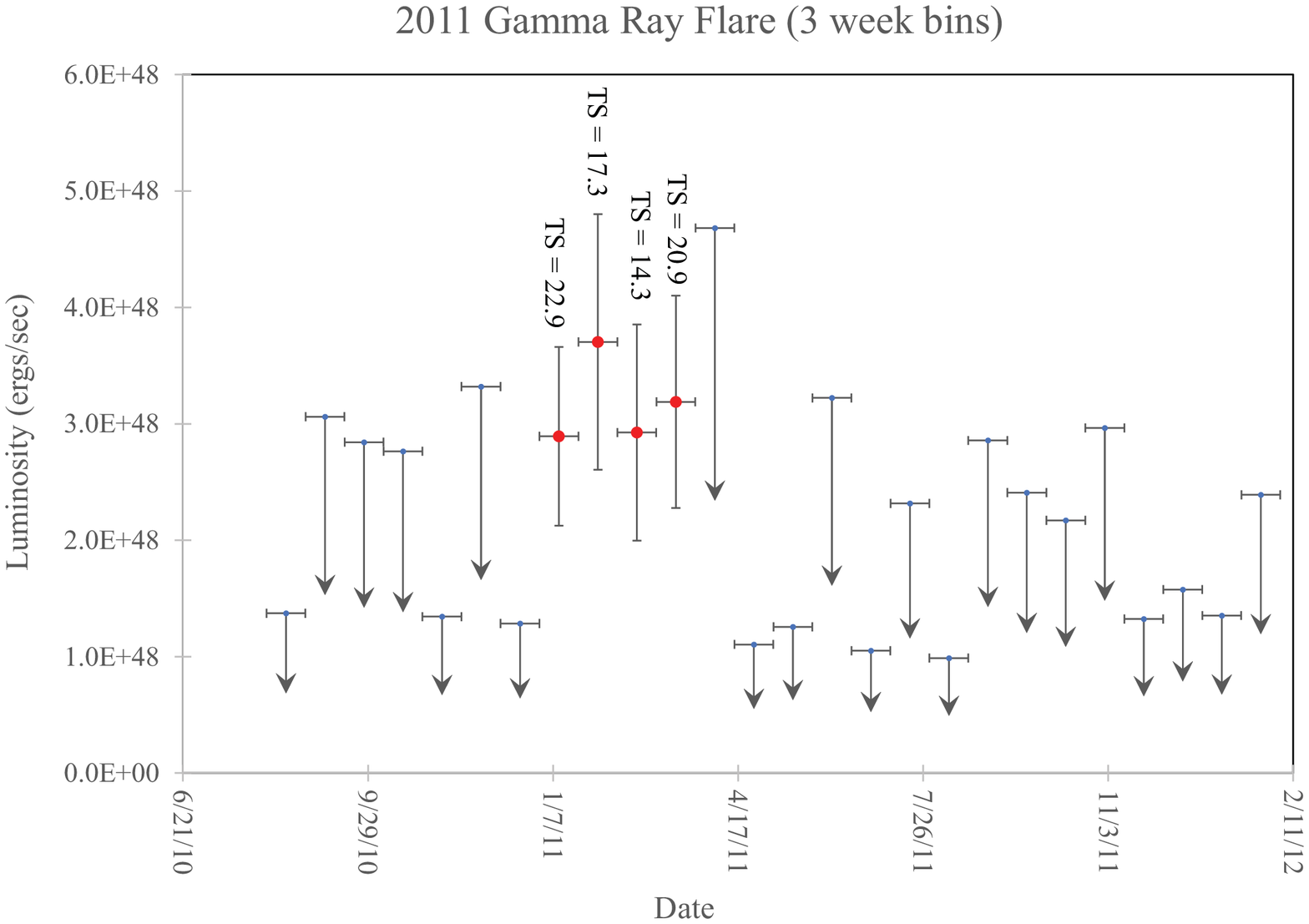}
\includegraphics[width= 0.8\textwidth,angle =0]{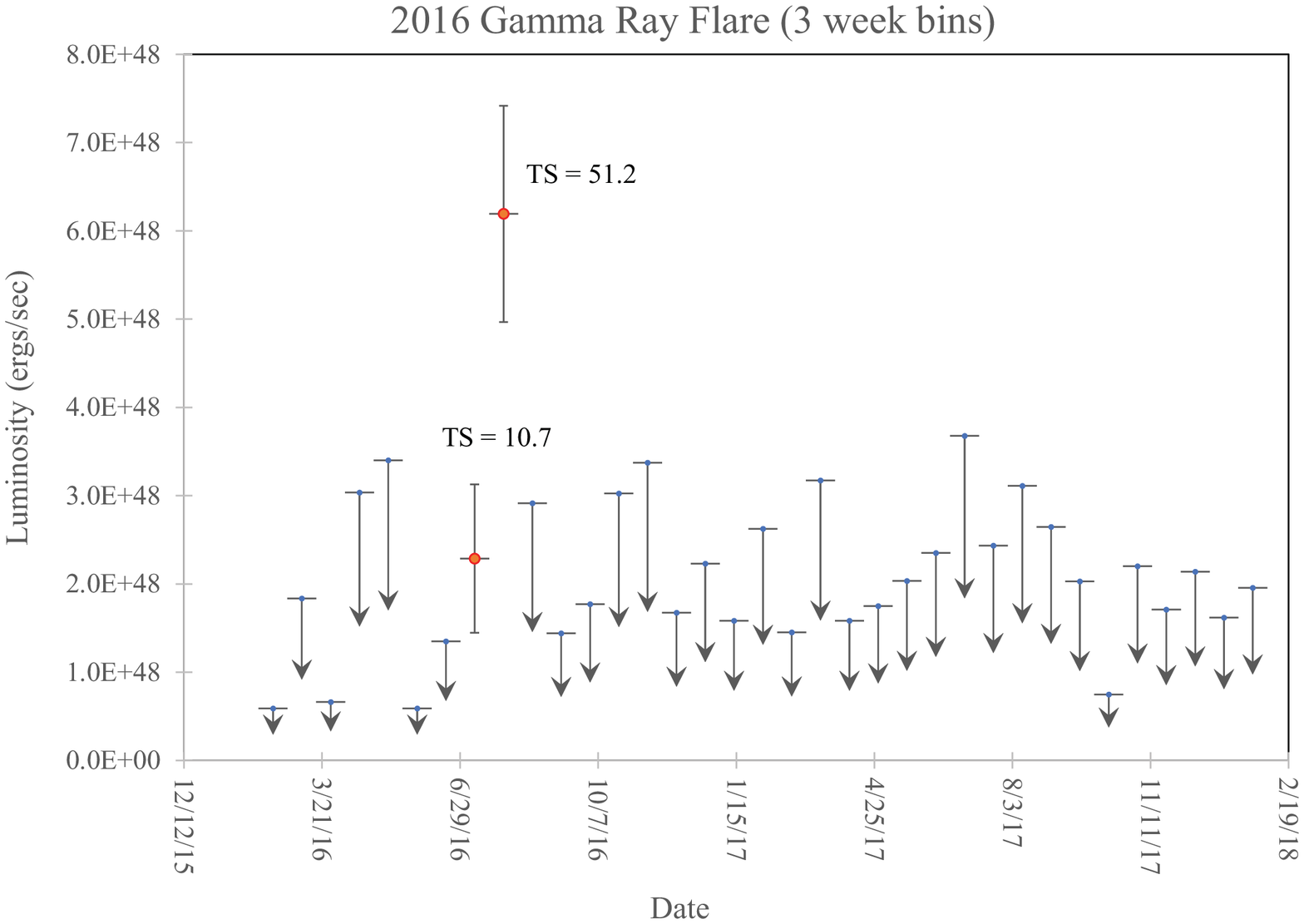}
\caption{The top panel is the light curve of the 2011 flare sampled in 3 week bins. It is detected in a few consecutive bins but with, $\mathrm{TS}<25$. This suggests a
prolonged faint flare. The vertical axis is the $0.1-100$\,GeV luminosity. The 2016 flare is plotted similarly in the bottom panel.}
\label{fig:gamma}
\end{center}
\end{figure*}

The ten year (from 2008 to 2018) average $\gamma$-ray luminosity detected by \textit{Fermi}-LAT from $0.1-500$\,GeV (observed energy) is $L_{\gamma} = 5.78 \times
10^{47}$\,erg~s$^{-1}$ \citep{sah20}. The temporal behavior of the $\gamma$-ray emission has been studied with low time resolution due to the low number statistics
\citep{li18}. In this section, we explore higher time resolution in order to ascertain the peak $\gamma$-ray luminosity. The data reduction method is defined in \citep{li18}. In summary, the publicly Fermi-LAT Pass 8 data (P8R3\_SOURCE\_V2) and the Fermitools were used to perform the data analysis. The data from 2008 August 4 to 2018 August 4 with the energy range from 100 MeV to 100 GeV was selected. We removed the $\gamma$-ray events with zenith angle greater than $90^{\circ}$ and the quality-filter cuts (DATA\_QUAL==1 \&\& LAT\_CONFIG==1) are applied. We selected photons set within a $10^{\circ}$ region of interest (ROI) and performed a unbinned likelihood analysis. The script {\tt make4FGLxml.py}\footnote{\url{https://fermi.gsfc.nasa.gov/ssc/data/analysis/user/make4FGLxml.py}} was used to generate the background model, which include all 4FGL-DR2\footnote{\url{https://fermi.gsfc.nasa.gov/ssc/data/access/lat/10yr_catalog/}} sources within 15$^{\circ}$ around the target as well as {\tt gll\_iem\_v07.fits} and {\tt iso\_P8R3\_SOURCE\_V2\_v1.txt}. The spectra of the point sources within $\rm 10^{\circ}$ around the center and the normalizations of the two diffuse emission backgrounds were set free. We determine the significance of the flare with the test statistic, TS \citep{mat96}. The test statistic is defined as TS = $2\ln{L/L_{0}}$, where $L$ and $L_{0}$ are the
maximum likelihood values for the model with and without target source, respectively. The quantity TS was identified with a statistical significance of $\sqrt{\mathrm{TS}}
=n\sigma$ in Equation (22) and Figure~3 of \citet{mat96}. We consider two likely flares in 2011 and 2016 that were previously identified with $\sim 5$ month time sampling
\citep{li18}. Since we already know that the flare is present, Figure~\ref{fig:gamma} considers higher time resolution of the \textit{Fermi}-LAT light curves at these epochs.
We consider 3 week bins or 4.5 days in the quasar rest frame.

The vertical axis in Figure~\ref{fig:gamma} is the $0.1-100$\,GeV luminosity. The first flare seems more prolonged, but the second flare was brighter. The peak flare
luminosity was $\approx 6 \times 10^{48}$\,erg~s$^{-1}$. With finer sampling (3 days), this increases to $\approx 10^{49}$\,erg~s$^{-1}$, but TS falls below 25, $\mathrm{TS}
\gtrsim 20$. The SED in the $\gamma$-ray region is very steeply decreasing power law ($\alpha \approx 2$) as expected for strong external inverse Compton cooling from the
existence of the luminous quasar environment that was described in the last section \citep{sah20,mar20}. This combined with the high redshift indicates that most of the
$\gamma$-ray luminosity is most likely at observed energies $<0.1$\,GeV. Thus, it is quite possible that the 2016 $\gamma$-ray flare is of comparable luminosity to some of
the strongest known $\gamma$-ray flares \citep{abd11}.

\section{Synchrotron Self Absorbed Power Law Fit to the Radio Data}
\label{radiofit}
A synchrotron self absorbed power law (SSA power law) for the observed flux density, $ S_{\nu_{\o}}$, is the solution to the radiative transfer in a homogeneous medium such as a uniform spherical volume \citep{gin65,van66}:
\begin{eqnarray}
&& S_{\nu_{\o}} = \frac{S_{\mathrm{o}}\nu_{\mathrm{o}}^{-\alpha}}{\tau(\nu_{\mathrm{o}})} \times \left(1 -e^{-\tau(\nu_{\mathrm{o}})}\right)\;, \; \; \tau(\nu_{\mathrm{o}})=\overline{\tau}\nu_{\mathrm{o}}^{(-2.5 +\alpha)}\;,
\label{equ:A5}
\end{eqnarray}
where $\tau(\nu)$ is the SSA opacity, $S_{\mathrm{o}}$ is a normalization factor and $\overline{\tau}$ is a constant. The wide spectral peak requires three SSA power law components. Adding a fourth SSA power law does not improve the fit. In Figure~\ref{fig:ssa}, we show the three components of the SSA power law fit that are naturally associated with the un-resolved nucleus, the Southeast Component and the North Lobe. The power laws for the core, Southeast Component and North Lobe are approximated by the data from $\nu_{\mathrm{o}}=5$ GHz to $\nu_{\mathrm{o}}= 22$ GHz in Table~\ref{tab:radio-data}, Figure~\ref{fig:spectrum}, and the VLBI data
in Tables~\ref{tab:modelfit-Sband}, \ref{tab:vlbi-Cband-fit} and \ref{tab:vlbi-high-fit}, respectively. Fine adjustment of the SSA power law parameters proceeds until the residuals (see Equation~(\ref{equ:variance})) of the fit to the total flux density from $\nu_{\mathrm{o}}=120$\,MHz to $\nu_{\mathrm{o}}=22$\,GHz are
minimized. The fit is based on two important assumptions that are motivated by the observations in Tables 1, 3 and 4 that are described below.
\subsection{Assumption 1: VLBI Does Not Capture All of the North Lobe Flux at High Frequency}
\begin{figure*}[htp!]
\begin{center}
\includegraphics[width= 1.0\textwidth,angle =0]{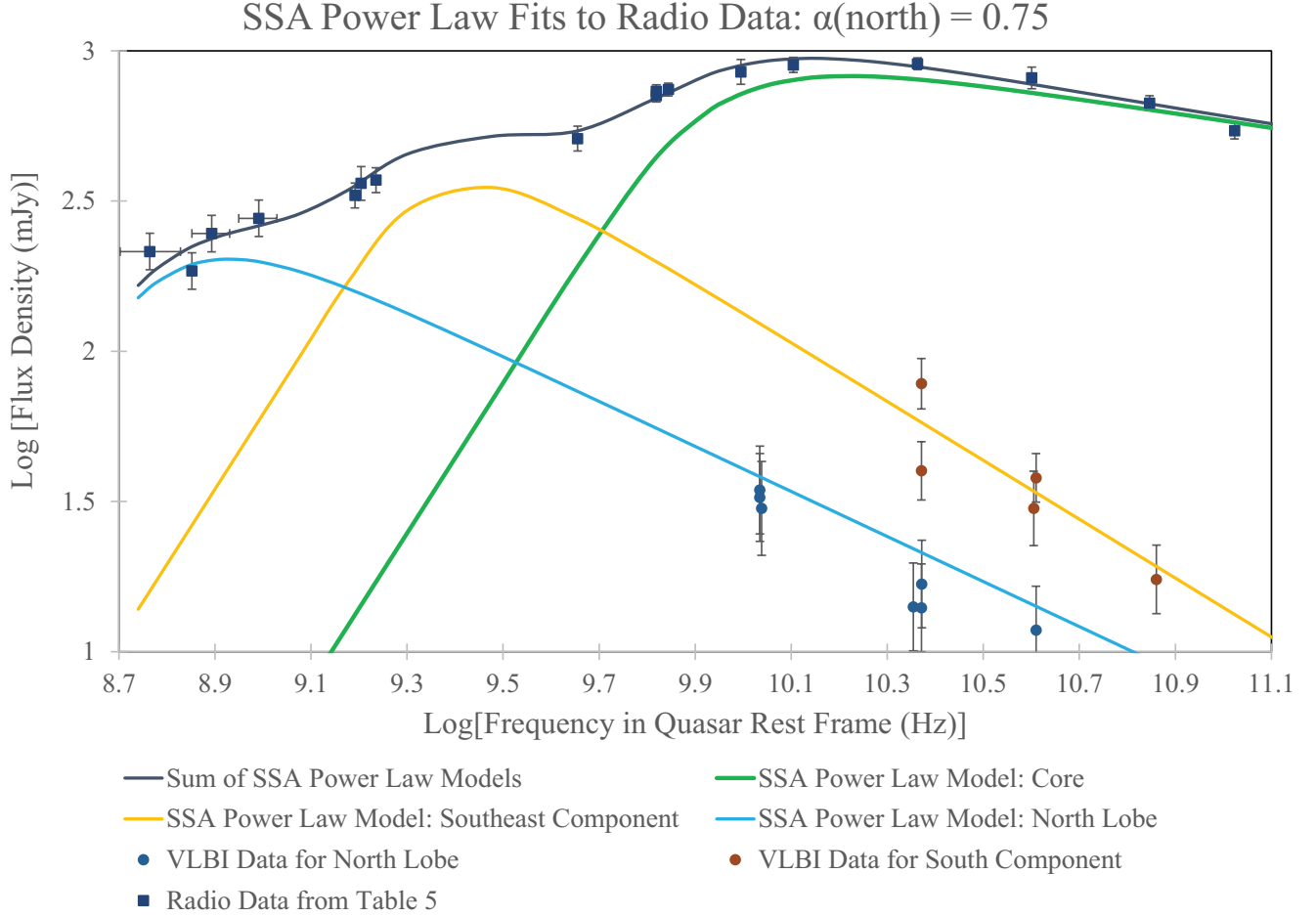}
\caption{\footnotesize{The figure shows the best fit of the three component SSA power law model to the total flux density of PKS~1351$-$018. A realistic fit to the North Lobe should be biased toward the top of the high frequency error bars for this diffuse structure due to the limited dynamic range of these VLBI data. } }
\label{fig:ssa}
\end{center}
\end{figure*}

Since the North Lobe is a diffuse, steep spectrum, component, there will be a tendency for VLBI to resolve out some of the diffuse emission due to limited dynamic range associated with imperfect (u, v) coverage. This effect is most pronounced at high frequency due to the lower flux density. This is clearly evident in the deep VLBA image at 15.4 GHz in Table 4 that does not detect the North Lobe, even though we expected at least 6.5 mJy to be present. Other evidence of this is the 2000 Jun 05 C-band observation in Table 3. This is the longest VLBI (which includes the VLBA baselines) observation (best (u, v) coverage) and it detects the largest flux density of the North Lobe of any C-band observation. To compensate for this effect, the fit to the North Lobe flux is biased towards the top of the error bars at high frequency. This is a valid compensatory device if the observations are sufficiently sensitive (i.e., large fractions of the flux are not resolved out). Thus, the three long duration C-band VLBI observations before 2014 in Table 3 are used in the fit to Figure 13 and the observations with short scans and poor (u,v) coverage from 2014 onward are ignored. A direct fit to the data without this biasing towards the top of the error bars yields a power law with $\alpha = 0.85$. We consider the bias towards the tops of the high frequency error bars in Figure~\ref{fig:ssa}, which yields $\alpha = 0.75$, to be a more plausible reconstruction of the physical source of lobe emission that is consistent with the observed data.
\subsection{Assumption 2: The Knot in the North Jet is Negligible to the Fit}
The Knot in the North Jet that appears in Table 3 is not considered as an important contributor to the total flux density at any frequency for the following  reasons.
\begin{enumerate}
\item The knot in the north jet is only detected at C-band. Thus, there is no spectral data and therefore no basis to extrapolate this to other frequencies.
\item According to Table 3, the feature is not detected at C-band in 1995 Jan 28 and only has 4 mJy in the full track VLBA observation in 2001 Jan 23.
\item The feature seems to move from 1.6 mas to $\sim 5.5$ mas when it is detected in Table 3 and might not be the same feature.
\item At C-band, it is much weaker than the nucleus and does not affect the fit in this region. The fit to the total flux density is determined by the core spectrum at all frequencies above C-band.
\item The Knot in the North Jet is at least one order of magnitude smaller than the North Lobe in Table 3. Being so compact, it is likely that the low frequency spectral turnover is at much higher frequency than the North Lobe \citep{van66,mof75,eze10}. Thus, the knot likely contributes insignificantly at frequencies below 350 MHz compared to the North Lobe (where the North Lobe is prominent).
\item As noted above, a fourth SSA power law does not improve the fit to the total flux density.
\end{enumerate}

\section{A Physical Model of the North Lobe SSA Power Law}
\label{northlobe}

Once an SSA power law is chosen for the North Lobe, we are interested in a physical model that is responsible. Complicated dynamics are most likely occurring in the radio
lobe \citep{blu00}. In general, there are fine-scale features such as shock fronts and filaments embedded within the diffuse lobe plasma. However, our image in
Figure~\ref{fig:Sband} is a structure-less plume that we fit as a circular Gaussian. There is no observational evidence to justify a model more complicated than a
homogeneous, spherical, single zone of plasma. Single zone spherical models are a standard technique even in blazar jet calculations out of practical necessity including
previous treatments of this source \citep{ghi10,sah20,mar20}. A simple homogeneous spherical volume model or plasmoid has historically provided an understanding of the
spectra and the time evolution of astrophysical radio sources \citep{van66}. We have used this formalism to study a panoply of phenomena, major flares in a Galactic black
hole, a $\gamma$-ray burst and flares in a radio quiet quasar \citep{pun12,pun19,rey09,rey20}. Most importantly, we used this method in \citet{pun20} to study the radio lobes in the super-luminous radio quasar, 3C 82 (which should be consulted for the details of the calculational method). The SSA turnover provides information on the size of the region that produces the preponderance of emission. This can be tied directly to the image in Figure~\ref{fig:Sband}. Furthermore, these models do not need to invoke equipartition in order to produce a solution.

We have established three physical constraints. Firstly, the FWHM of the North Lobe of the 2002 S-band observation (the best depiction
of the lobe) in Table~\ref{tab:modelfit-Sband} constrains the sphere radius, $R$,
\begin{equation}
R \approx 3.3\, \mathrm{mas}\;.
\label{equ:radius}
\end{equation}
Secondly, the apparent velocity, $v_{\rm{app}}$, is bounded by 23 years of C-band radio images as indicated in Figure~\ref{fig:displacement},
\begin{equation}
v_{\mathrm{app}} < 0.57\,c\;.
\label{equ:v-app}
\end{equation}
Thirdly, the light curve variability analysis of the flare in Figure~\ref{fig:radiolightcurve-1990} and Equation~(\ref{equ:theta-max}) provide an approximate constraint on
the line of sight to the jet axis, $\theta$,
\begin{equation}
\theta < \theta_{\rm{max}}\sim 10\degr\;.
\label{equ:theta}
\end{equation}
Note that this constraint is looser than Equation~(\ref{equ:theta-max}) because that derivation assumed no uncertainty arising from the fitting technique in
Figure~\ref{fig:radiolightcurve-1990}.
Equations~(\ref{equ:v-app}) and (\ref{equ:theta}) combine to give a constraint on $\delta$. From \citet{ree66} and \citet{gin69},
\begin{equation}
\beta_{\rm{app}} \equiv \frac{v_{\rm{app}}}{c} = \frac{\beta \sin{\theta}}{1-\beta\cos{\theta}}\;,
\label{equ:beta-app}
\end{equation}
where $\beta$ is the three-velocity of the moving plasmoid. Combining the definition of the Doppler factor, $\delta=\sqrt{1-\beta^{2}}/(1-\beta\cos{\theta})$, with
Equations~(\ref{equ:v-app}) and (\ref{equ:theta}) yields an equivalence to a constraint on the Doppler factor in our models. This can be emphasized by writing the Doppler
factor as $\delta(\beta_{\rm{app}}, \, \theta)$.

The value of minimum lepton energy, $E_{\mathrm{min}}$, is not constrained directly by observation. Values of $E_{\mathrm{min}}=m_{e}c^2$ and $E_{\mathrm{min}}=2.6m_{e}c^2$ are used in \citet{mar20} and \citet{sah20}, respectively, to fit the synchrotron peak and the inverse Compton spectrum from the nucleus. Here $m_{e}$ denotes the electron mass. Since this is the region with the most energetic electrons, we do not expect $E_{\mathrm{min}}$ to be larger in the less energetic North Lobe. We also note that \citet{cel08} have argued that $E_{\mathrm{min}}=m_{e}c^2$ based on fits to blazar jet spectra in the soft X-ray band. Thus, we initially consider $E_{\mathrm{min}}=m_{e}c^2$ and explore slightly higher values later. There is not a unique solution to the physical parameters of the North Lobe that recreate the fit in Figure 13. In this section we explore the solution space as calculated in \citep{pun20}.

\subsection{Kinematics of the Lobe Solution} We separate the energy content of the turbulent magnetized lobe into two pieces. The first is the kinetic energy of the protons,
$\mathcal{K}(\mathrm{protonic})$,
\begin{eqnarray}
 && \mathcal{K}(\mathrm{protonic}) = (\gamma - 1)Mc^{2}\;,
\label{equ:kin-energy}
\end{eqnarray}
where $M$ is the mass of the plasmoid and $\gamma$ is the Lorentz factor in the quasar rest frame. The other component is the lepto-magnetic energy, $E(\mathrm{lm})$, the
volume integral of the leptonic internal energy density, $U_{e}$, and the magnetic field energy density, $U_{B}$. In a spherical volume,
\begin{eqnarray}
 && E(\mathrm{lm}) = \int{(U_{B}+ U_{e})}\, dV = \frac{4}{3}\pi R^{3}\left[\frac{B^{2}}{8\pi}
+ \int_{\Gamma_{\mathrm{min}}}^{\Gamma_{\mathrm{max}}}(m_{e}c^{2})(N_{\Gamma}E^{-n + 1})\, dE \right]\;,
\end{eqnarray}
where in the proper frame, $B$ is the magnetic field, $N_{\Gamma}$ is the normalization of the number density power law and $\Gamma (m_{e}c^{2})$ is the lepton energy.
The corresponding energy density is $U(\mathrm{lm}) \equiv U_{e}+U_{B}$. The leptons also have a kinetic energy analogous to Equation~(\ref{equ:kin-energy}),
\begin{eqnarray}
 && \mathcal{K}(\mathrm{leptonic}) = (\gamma - 1)\mathcal{N}_{e}m_{e}c^{2}\;,
\label{equ:k-leptonic}
\end{eqnarray}
where $\mathcal{N}_{e}$ is the total number of leptons in the lobe.

There are protonic and leptonic energy fluxes due to bulk motion. The protonic energy flux is approximately the kinetic energy flux,
\begin{equation}
\mathcal{E}(\mathrm{proton}) = N(\gamma-1)\gamma v_{\mathrm{adv}}m_{p}c^{2}\;,
\end{equation}
where $m_{p}$ is the mass of the proton, $N$ is the proper number density, and $v_{\mathrm{adv}}$ is the advance speed of the lobe in the quasar rest frame. The magneto-leptonic energy flux is
\begin{equation}
\mathcal{K}(\mathrm{magneto-leptonic}) = N \gamma v_{\mathrm{adv}}\left[\gamma\mu c^{2}\right]\;,
\label{equ:magneto-leptonic}
\end{equation}
where $\mu$ is the specific enthalpy \citep{pun08}.
The specific enthalpy decomposes as
\begin{equation}
N\mu = U(\mathrm{lm}) + P \;,
\label{equ:enthalpy}
\end{equation}
where the relativistic pressure, $P \approx (1/3)U(\rm{lm})$ \citep{wil99}.

With the leptonic assumption, Equation~(\ref{equ:magneto-leptonic}) implies that the kinetic luminosity (jet power), $Q_{\mathrm{lm}}$,  is
\begin{equation}
Q_{\mathrm{lm}} = \int [\mathcal{K}(\mathrm{magneto-leptonic})] dA_{\perp} + L_{\mathrm{r}}\;,
\label{equ:q-lm}
\end{equation}
where $dA_{\perp}$ is the cross sectional area element normal to the jet axis and $L_{\mathrm{r}}$ is the energy flux lost to radiation. .

\begin{figure*}[htp!]
\begin{center}
\includegraphics[width= 0.47\textwidth,angle =0]{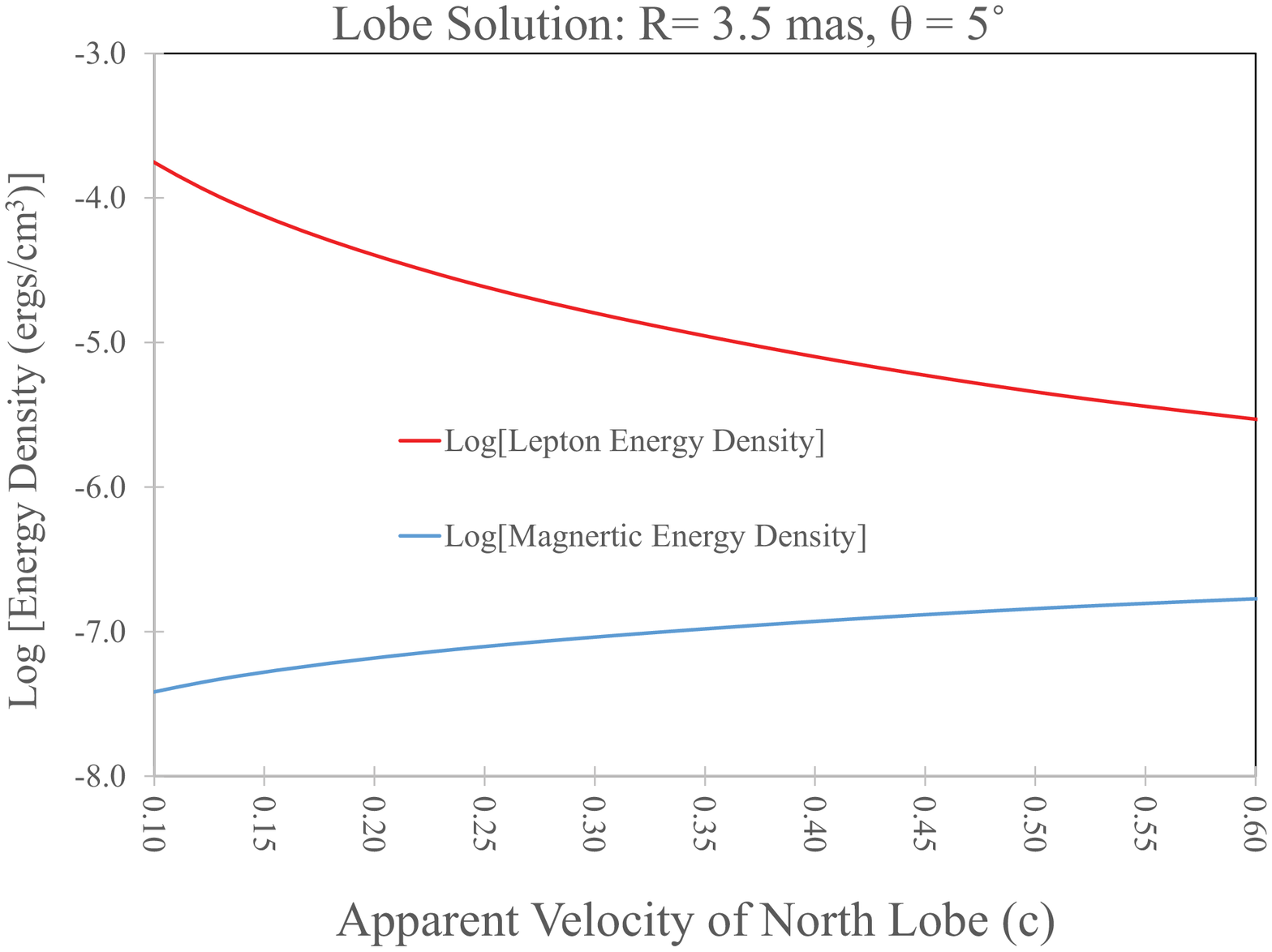}
\includegraphics[width= 0.47\textwidth,angle =0]{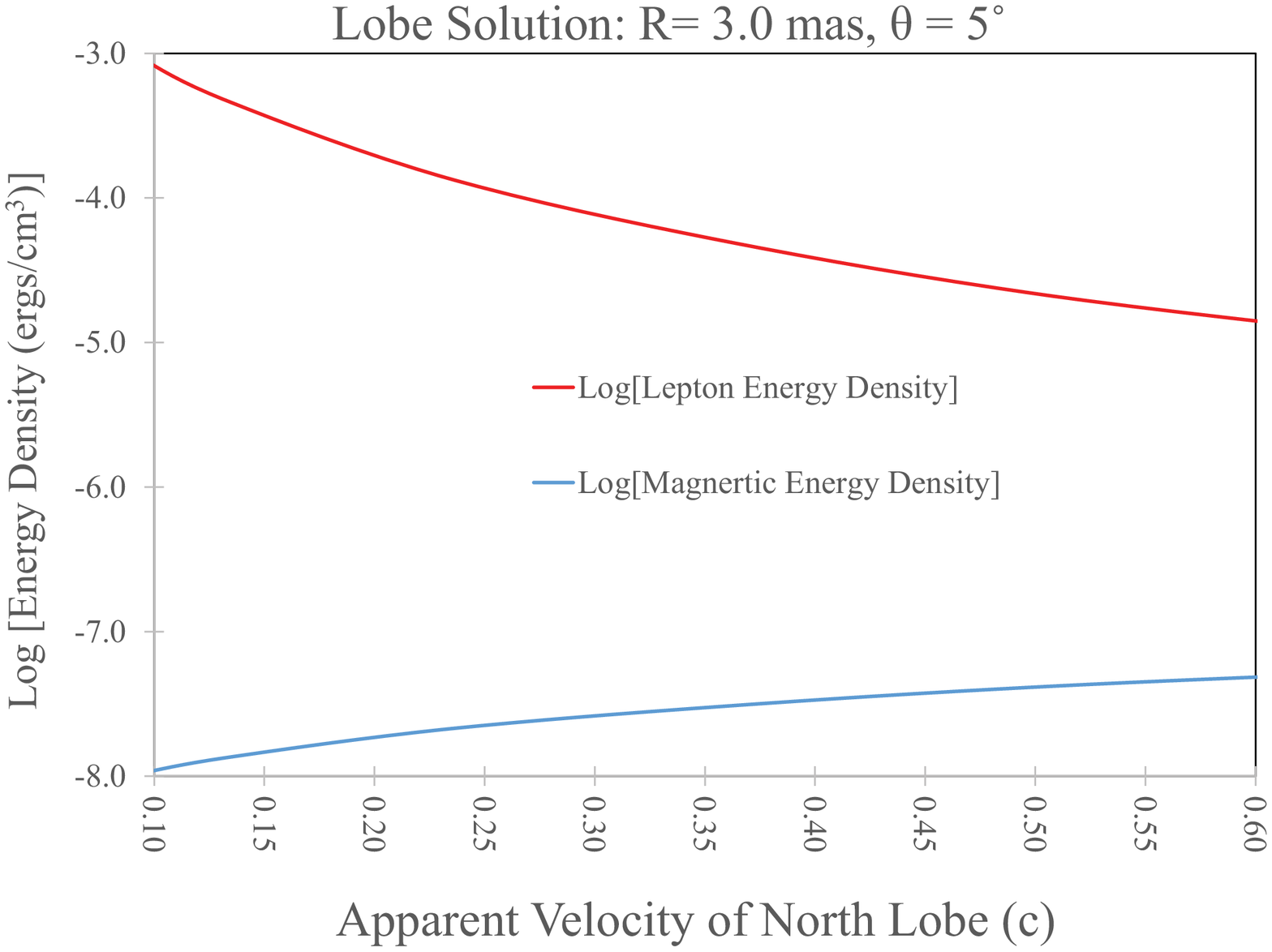}
\includegraphics[width= 0.47\textwidth,angle =0]{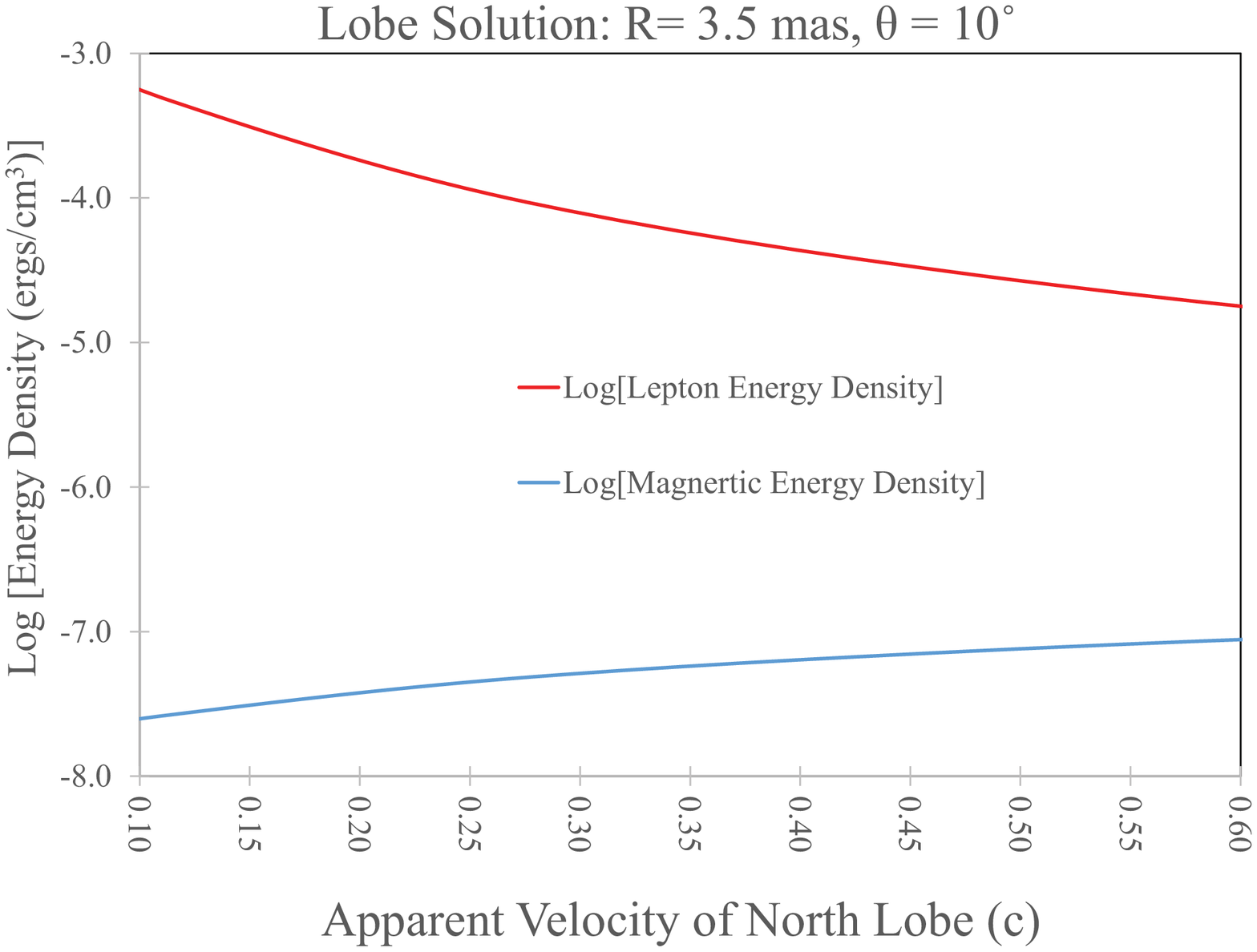}
\includegraphics[width= 0.47\textwidth,angle =0]{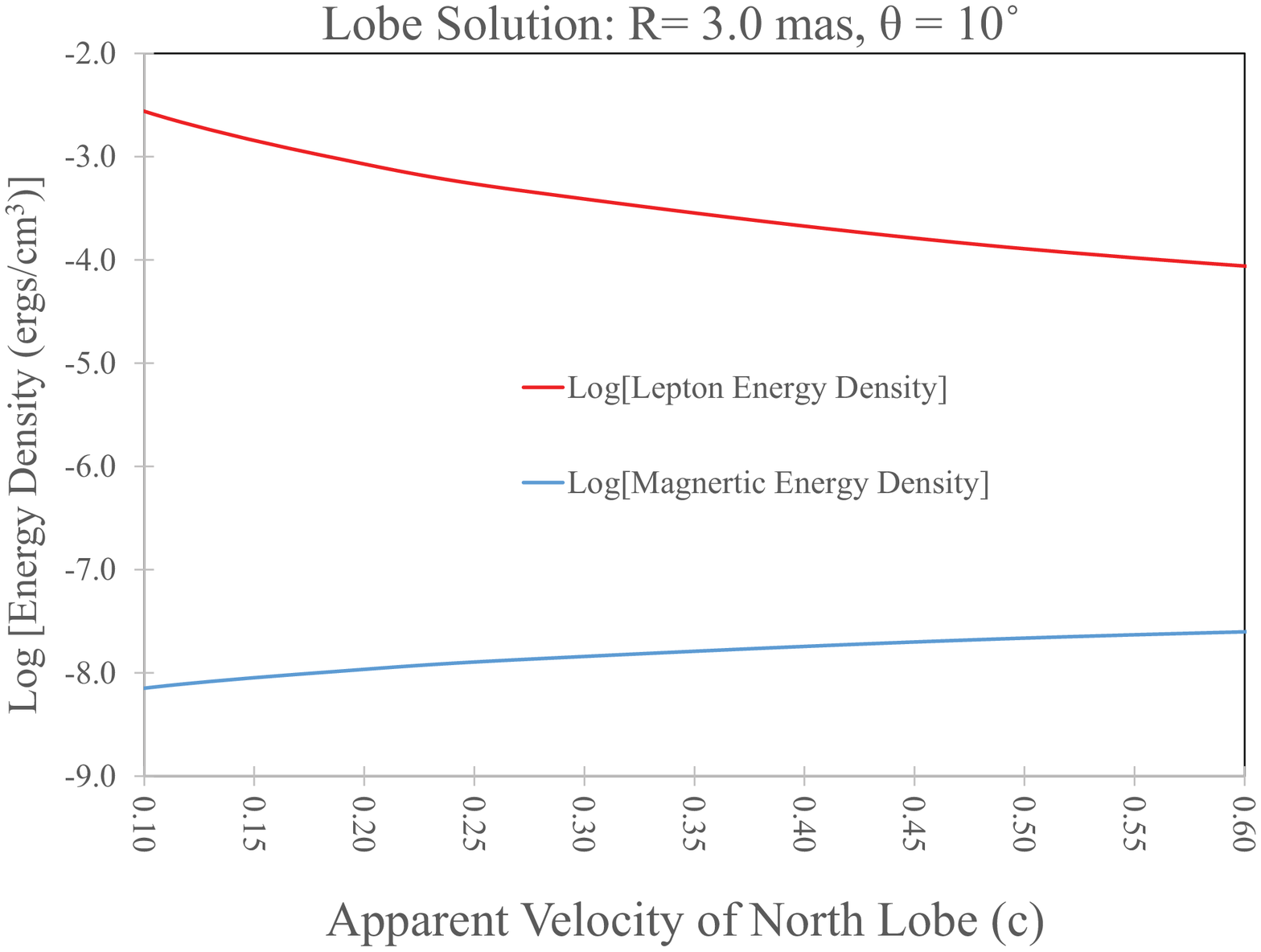}
\caption{This figure explores a large subset of the solution space for a spherical plasmoid that radiates the fit to the North Lobe Figure~\ref{fig:ssa}. Based on Equations~(\ref{equ:radius}) and (\ref{equ:theta}), we choose a rectangular subset $3.0 \, \mathrm{mas} < R < 3.5
\, \mathrm{mas}$ and $5\degr < \theta < 10\degr$ and bound its behavior by looking at the four corners of the rectangle. By plotting $U_{e}$ and $U_{B}$ as a function of
$\beta_{\mathrm{app}}$, we find no solutions that are near equipartition and obey the upper bound on $\beta_{\mathrm{app}}$ from Equation~(\ref{equ:v-app}). The closest
solution to equipartition is at $R = 3.5 \,\rm{mas}$ and $\theta = 5\degr$ in the upper left hand panel.}
\label{fig:lobe-solution}
\end{center}
\end{figure*}

The first thing that we noticed about the infinite set of solutions that conform to Equations~(\ref{equ:radius})--(\ref{equ:theta}) is that most of this domain yields
solutions that are extremely inertially dominated, $U_{e} \gg U_{B}$. The results are plotted in Figure~\ref{fig:lobe-solution} for the four corners of the rectangular domain of the two dimensional set of pre-assigned values of $3.0\, \mathrm{mas}< R < 3.5\, \mathrm{mas}$ and $5\degr < \theta < 10\degr$. There is no solution near equipartition with the constraint of Equation (11) imposed, $\beta_{\mathrm{app}}<0.57$.

These extremely inertially dominated solutions are disfavored on both theoretical and empirical grounds. Theoretically, the pair plasma is highly energetic with large random
velocities, and one would expect a relatively strong tangled magnetic field to form. Empirically, the radio lobes of powerful Fanaroff--Riley II (FR II) radio galaxies have
magnetic fields that tend to be near equipartition or slightly below this \citep{fr74,ine17,kat05}.  A large sample of FR II radio galaxy lobes was studied in X-rays and with
multi-frequency radio imaging \citep{ine17}. The X-ray observations were used to determine the inverse Compton emission (primarily of the Cosmic Microwave Background) and the
radio images were used to determine the synchrotron emission. From this they were able to estimate $U_{e}/U_{B}$. In Figure 2 of \citet{ine17}, they found,
$E_{\rm{equipatition}}<E(\mathrm{lm})< 7E_{\rm{equipatition}}$, in the pair plasma of the lobes, where $E_{\rm{equipatition}}$ is the equipartition lepto-magnetic energy. The
median value is $E(\mathrm{lm})\approx 2.4E_{\rm{equipatition}}$. We consider this range of possible values for the North Lobe of PKS\,1351$-$018 in the following analysis.
In order for this to be robust, we comment on our description of the northern component as a lobe. Recall that the MERLIN observation with a resolution $\sim 50-60$ mas could
not detect anything, except for a point source. The VLBI S-band image in Figure~\ref{fig:Sband} did not detect anything farther from the nucleus (within $\sim 30$ mas of the
nucleus) than the North Lobe. The results of Table~\ref{tab:vlbi-atca} indicate that the better S-band VLBI observations are consistent with minimal or no missing
flux density in the image. The North Lobe is therefore likely to be the furthest emission region from the nucleus. It appears to be at the end of a curving
continuous jet in Figure~\ref{fig:Cband}. Based on the Gaussian fit in Tables~\ref{tab:modelfit-Sband} and \ref{tab:vlbi-Cband-fit}, it is definitely wider than the jet,
indicating a difference in the physical composition. One could claim that it is a knot in a continuous jet that appears to be at its terminus. This would mean that it is
predominantly the hot spot in the lobe that is detected. There does not seem to be edge brightening in Figures~\ref{fig:Sband} and \ref{fig:Cband} that is characteristic of
the hot spot in FR II radio lobes \citep{fr74}. However, the lack of edge brightening could be an artifact of the imperfect $(u,v)$ coverage and insufficient resolution. We
do not think this to be the case since the North Lobe seems to be significantly inflated relative to the jet. In any event, our jet analysis does not depend on this
distinction. The hot spots in FR II radio lobes and the lobe plasma deviate similarly from equipartition. The hot spots have been found to have internal energies relative to
equipartition in a range very similar to the lobes \citep{kat05}.

\begin{figure*}[htp!]
\begin{center}
\includegraphics[width= 0.47\textwidth,angle =0]{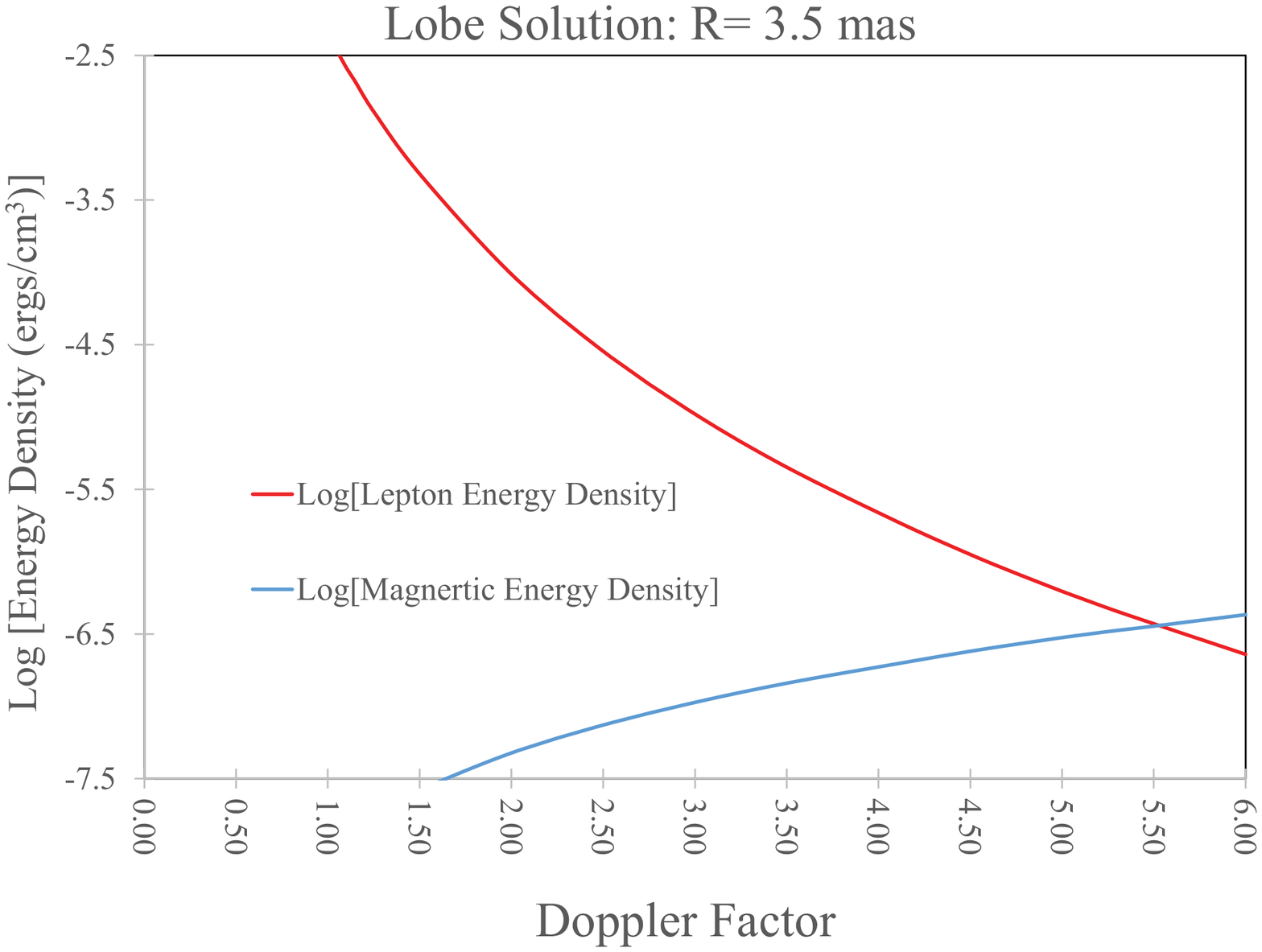}
\includegraphics[width= 0.47\textwidth,angle =0]{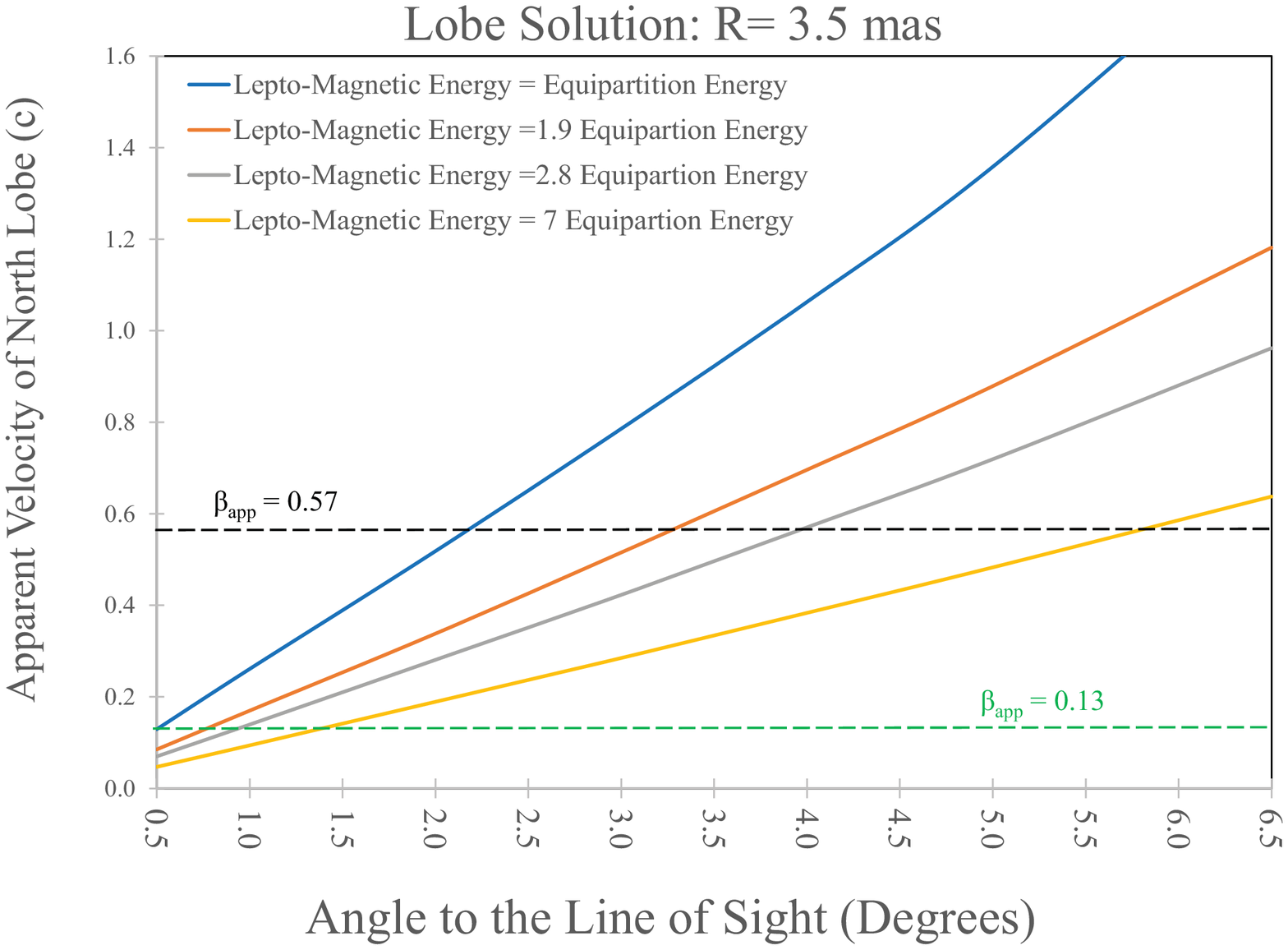}
\includegraphics[width= 0.47\textwidth,angle =0]{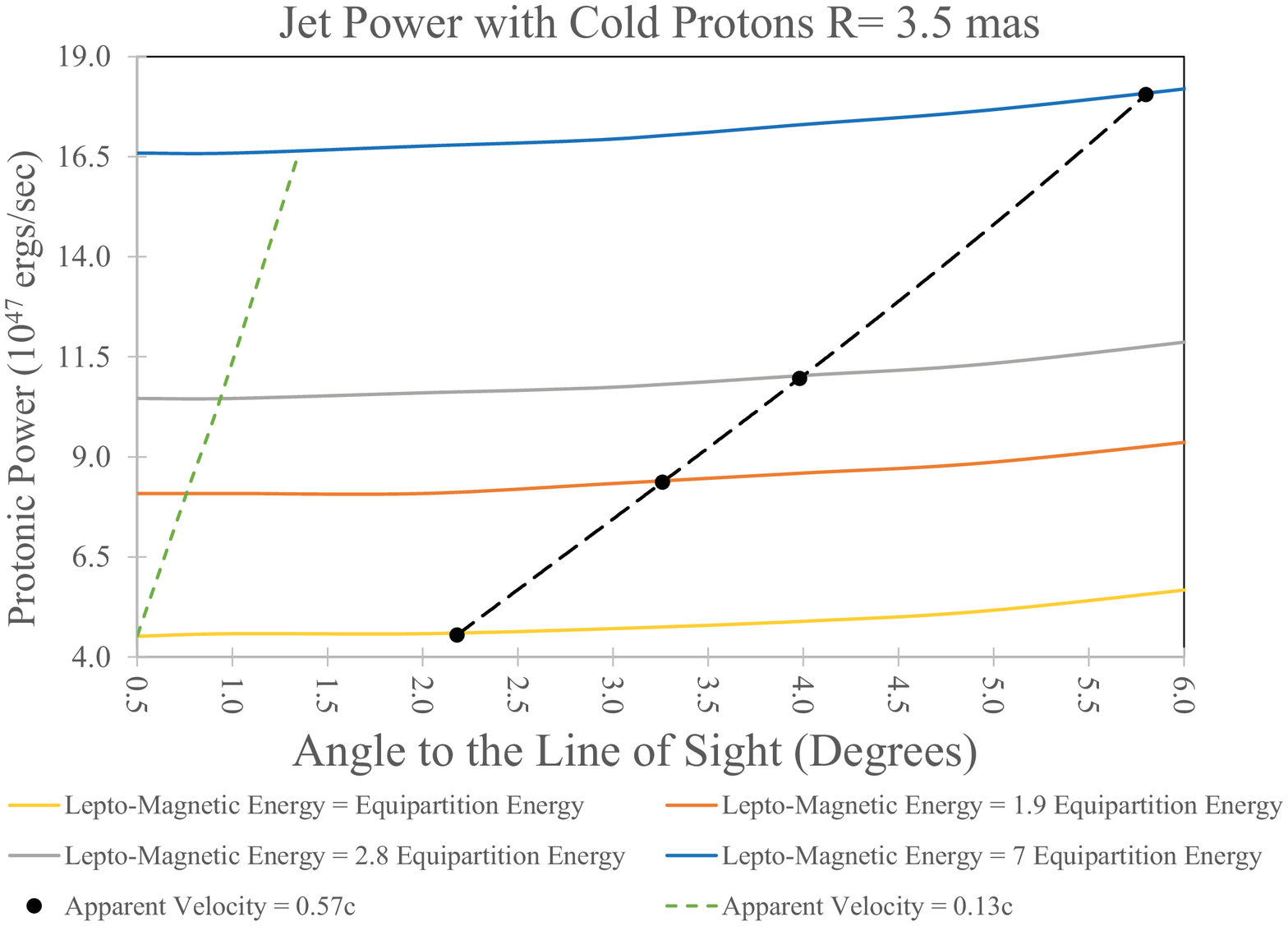}
\includegraphics[width= 0.47\textwidth,angle =0]{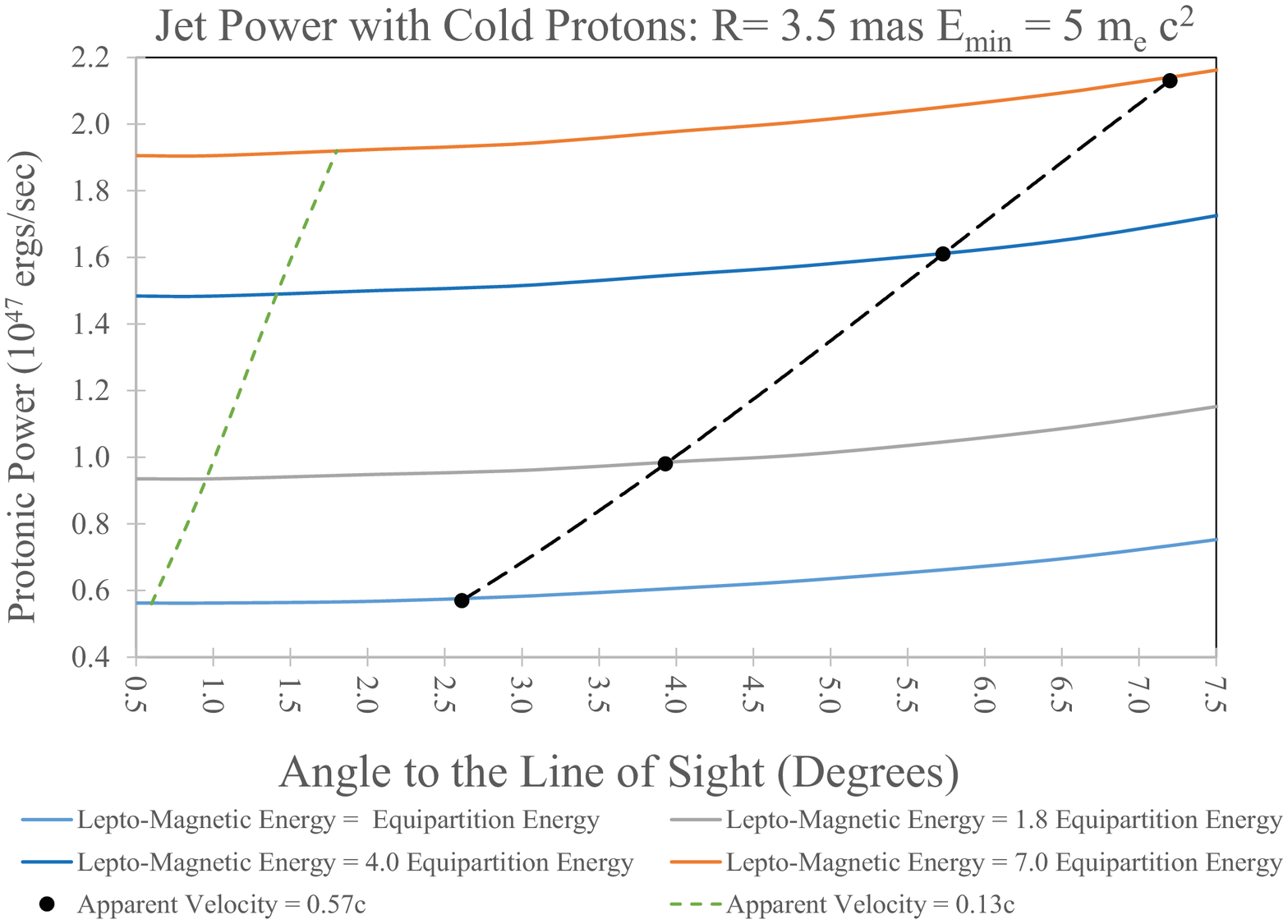}
\caption{The figure presents plots that describe the properties of the preferred solution for the lepton-positron lobe model. The top left hand panel shows the connection
between the energy density of the magnetic field and the energy density of the leptons as a function of $\delta$. Ostensibly, the plot appears to be independent of
$\beta_{\mathrm{app}}$ and $\theta$. However, as discussed earlier, that dependency is contained within the dependent variable, $\delta(\beta_{\mathrm{app}}, \, \theta)$. The
top right hand panel shows how $\beta_{\mathrm{app}}$ depends on $\theta$ for four families of solutions with energy densities consistent with the FR II energy densities
found in FR II radio lobes \citet{ine17}. The bottom left hand frame is the estimated power of the lepto-magnetic jet feeding the lobes as a function of the LOS for these
same
four families of solutions. Note the dashed black curve that represents the upper limit from Figure~\ref{fig:displacement} of $\beta_{\mathrm{app}}<0.57$. The only solutions
consistent with this observational constraint are to the left of the dashed curve. The green dashed curve are solutions with the best fit velocity from Figure 6. The bottom right hand panel is a similar plot with the condition $E_{\mathrm{min}}=
5m_{e}c^{2}$ instead of $E_{\mathrm{min}}= m_{e}c^{2}$ as was assumed in the other three panels. The change in the estimated jet power is $\approx 35\%$.}
\label{fig:preferred-solution}
\end{center}
\end{figure*}

Figure~\ref{fig:lobe-solution} indicates that the larger value of $R$ and the smaller value of $\theta$ moves the solutions closer to equipartition. Thus motivated, we look
at the $R= 3.5$\,mas case with $\theta <5\degr$ as a possible viable region of the solution space. The first thing that we explore is the dependence of $U_{e}$ and $U_{B}$ on $\delta$ in the top left
hand panel of Figure~\ref{fig:preferred-solution}. It is clear that for $E_{\mathrm{equipatition}}<E(\mathrm{lm})< 7E_{\mathrm{equipatition}}$, $\delta \lesssim 5.5$. This is
a large Doppler factor for the modest apparent velocity in Equation~(\ref{equ:v-app}). This is explained with Equation (\ref{equ:beta-app}) as a
consequence of a small LOS.

The top right hand panel of Figure~\ref{fig:preferred-solution} plots $\beta_{\mathrm{app}}$ as a function of the angle to the LOS under four
different constraints, in the range $E_{\mathrm{equipatition}}<E(\mathrm{lm})< 7E_{\mathrm{equipatition}}$. As $U_{B}/U_{e}$ is lowered, a larger LOS is consistent with
$\beta_{\mathrm{app}}<0.57$. Even so, the largest LOS angle in any of the plausible models is $\approx 5.8\degr$ for $E(\mathrm{lm})= 7E_{\mathrm{equipatition}}$. The
bottom left hand panel uses Equation~(\ref{equ:q-lm}) to plot the lepto-magnetic jet power, $Q_{\mathrm{lm}}$, as a function of the LOS angle for four cases in the
range, $E_{\mathrm{equipatition}}<E(\mathrm{lm})< 7E_{\mathrm{equipatition}}$. The black dashed curve represents the $\beta_{\mathrm{app}} = 0.57$ upper limit implied by
Figure~\ref{fig:displacement}. The only solutions consistent with observation are to the left of the black dashed curve. We also investigate the consequences of abandoning the
$E_{\mathrm{min}}= m_{e}c^{2}$ assumption. The bottom right hand panel is a plot of $Q_{\mathrm{lm}}$ as a function of the LOS angle, assuming that $E_{\mathrm{min}}=
5m_{e}c^{2}$. This value is motivated by the model of the $\gamma$-ray emission in \citet{sah20} which has $E_{\mathrm{min}}= 2.6m_{e}c^{2}$ in the nucleus. We intentionally
went above this value in order to bound a range of plausible assumptions. However it is not clear why $E_{\mathrm{min}}$ would exceed $m_{e}c^{2}$ and is included for the
sake of completeness. There is only modest variation over the entire plausible parameter range $ Q_{\mathrm{lm}} \approx (5.2 \pm 3.2) \times 10^{45}$\,erg~s$^{-1}$. Note
that the allowed LOS angle increases to $7.2\degr$ when $E_{\mathrm{min}}= 5m_{e}c^{2}$. The core synchrotron and inverse Compton emission ($\gamma$-ray emission) was
used in \citet{sah20} and \citet{mar20} to estimate $Q_{\mathrm{lm}} = 2.76 \times 10^{45}$\,erg~s$^{-1}$  and $Q_{\mathrm{lm}} = 4.24 \times 10^{45}$\,erg~s$^{-1}$,
respectively. Remarkably, the lepto-magnetic jet power estimates on sub-parsec scales agree with our estimate $\sim 2.5$ kpc (after de-projection) farther out in the North
Lobe. If there is a (Doppler de-boosted) counter-jet, the total power of the central engine would be $ Q_{\mathrm{lm}} \approx (1.04 \pm 0.64) \times 10^{46}$\,erg~s$^{-1}$.

We validate that the energy budget in the approaching jet derived from the North Lobe is sufficient to support the radiation losses. The intrinsic $\gamma$-ray luminosity,
$L_{\gamma}(\rm{intrinsic})=\delta^{-4}L_{\gamma}(\rm{apparent})$. From the estimated $\delta$ in the $\gamma$-ray emitting region from \citet{sah20} and the value of
$\delta$ from \citet{mar20}, we find $L_{\gamma}(\rm{intrinsic})=(20.47^{-4})5.78 \times 10^{47}$\,erg~s$^{-1}$ = $3.29 \times 10^{42}$\,erg~s$^{-1}$ and
$L_{\gamma}(\rm{intrinsic})=(15.4^{-4})5.78 \times 10^{47}$\,erg~s$^{-1}$ = $1.03 \times 10^{43}$\,erg~s$^{-1}$, respectively for the time averaged
$L_{\gamma}(\rm{apparent})$ from Section~\ref{gammaray}. Even if we look at the peak flare luminosity in 2016 in a 3 week bin in Figure~\ref{fig:gamma}, this only increases
by an order of magnitude. The
energy budget of the jet that is dissipated as $\gamma$-rays and the synchrotron peak is a negligible fraction of the total jet power.

We can also estimate the $P\Delta V$ work of inflating the lobe in the environment of the host galaxy. From Figure 2 of \citet{mat03}, we get an estimate of an external
pressure on the order of 1\,kpc from the nucleus (where most of the lobe propagation occurs), $P_{\mathrm{ext}} \sim 10^{-10}$\,dyn~cm$^{-2}$ if the host galaxy is a large
elliptical. From the expression for the pressure, below Equation~(\ref{equ:enthalpy}), the internal lobe pressures are $2\times 10^{-7}\,\mathrm{dyn~cm^{-2}} < P < 2 \times
10^{-6}\,\mathrm{dyn~cm^{-2}}$ for the solutions in Figure~\ref{fig:preferred-solution}. The North Lobe is highly over-pressurized relative to the environment. Much more jet energy is required to energize the plasma in the lobe volume than is required to push the enveloping gas away as this volume inflates. We conclude that the $P\Delta V$ work of inflating the lobe is insignificant in our jet power estimates.

\begin{figure*}[htp!]
\begin{center}
\includegraphics[width= 0.47\textwidth,angle =0]{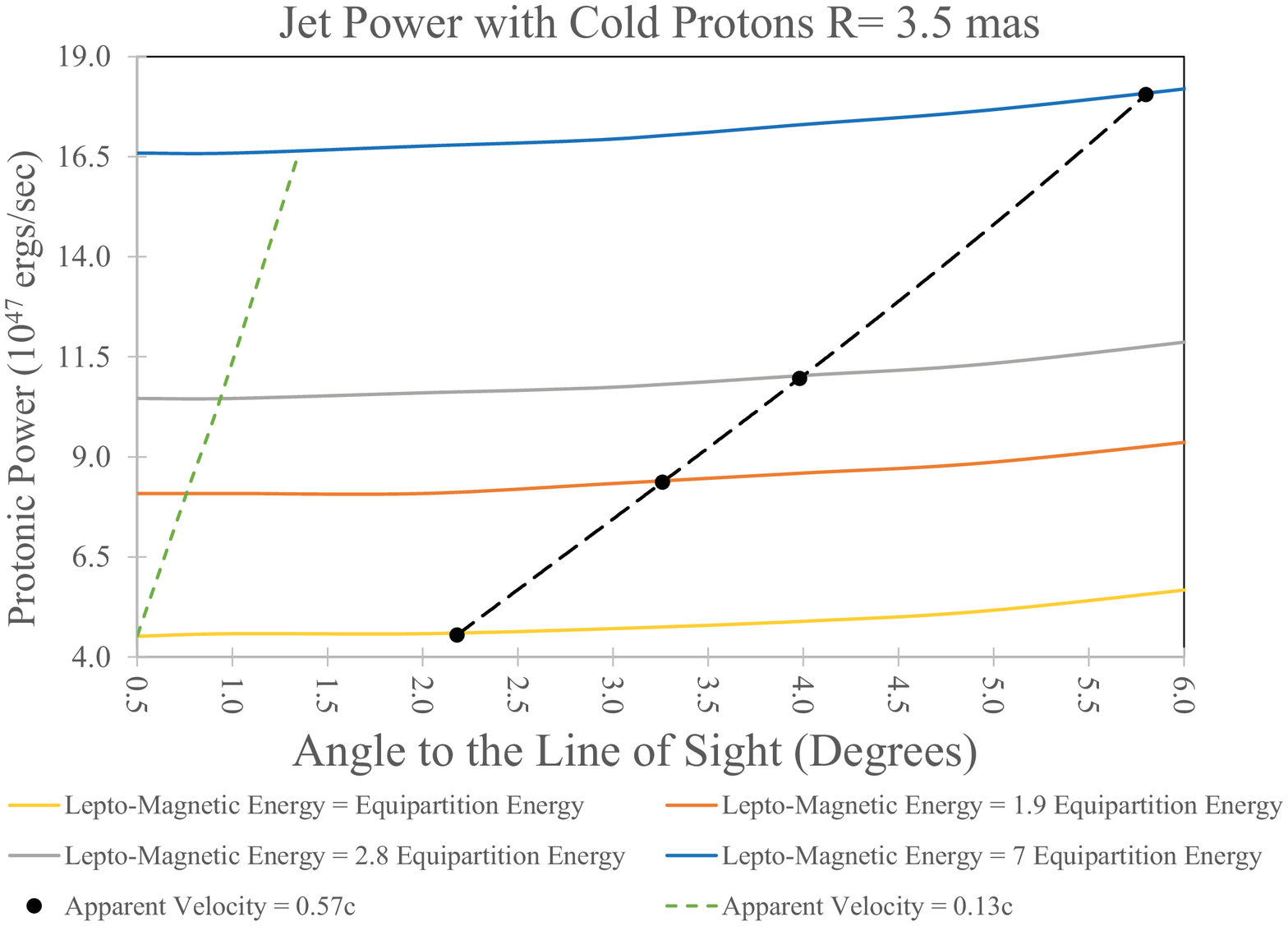}
\includegraphics[width= 0.47\textwidth,angle =0]{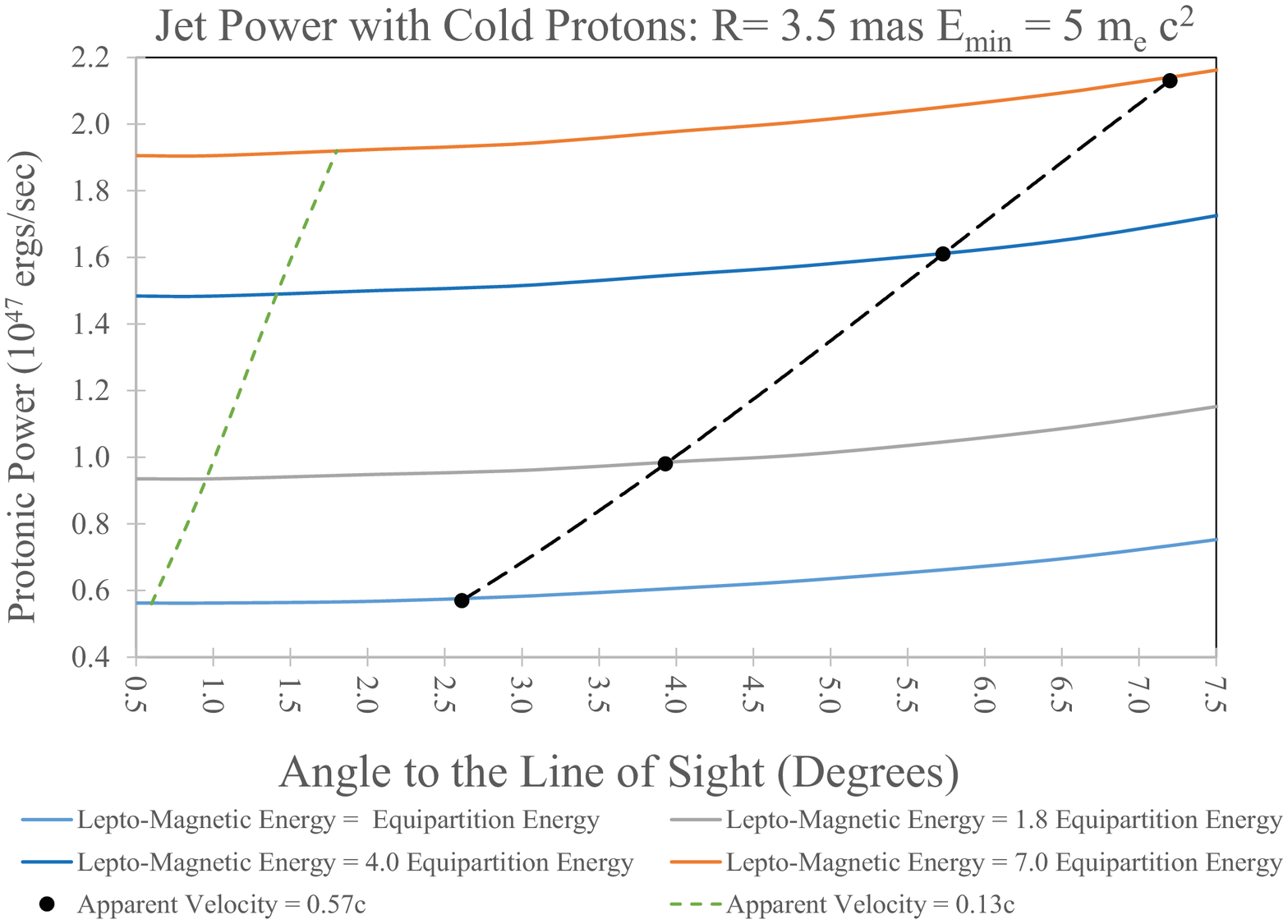}
\caption{The jet power computed under the assumption that the positive charges are cold protons. Otherwise, the plots are identical in format to the bottom two panels of
Figure~\ref{fig:preferred-solution}. The allowable physical solutions exist to the left of the black dashed curves. The green dashed curve represents solutions with the apparent velocity derived from the fit to the component separation in Figure 6.}
\label{fig:jet-power}
\end{center}
\end{figure*}

\subsection{A Protonic North Lobe}

In principle, the positive charges in the ionized lobes can be protonic matter instead of positronic matter. This was proposed by \citet{mar20} in their model of the
$\gamma$-ray emitting region. These are not thermal protons that add to the pressure, but cold protons. A significant thermal proton population in FR II radio lobes has been argued to be implausible \citep{cro18}. Based on Equations~(\ref{equ:k-leptonic}) and (\ref{equ:magneto-leptonic}), the kinetic energy of the protons would be much larger than $E(\mathrm{lm})$. The spectrum is created by electrons that radiate in the magnetic field. The spectrum is still depicted by the fit in
Figure~\ref{fig:ssa}. However, the kinetic luminosity of the moving lobe computed from Equation (17) is much larger in Figure~\ref{fig:jet-power} compared to the bottom panels of Figure~\ref{fig:preferred-solution}. We find no compelling reason to add these cold protons, but we cannot prove that this is not the case. One thing that is not encouraging is the kinetic luminosity of the protons in the lobe is an order of magnitude larger than \citet{mar20} estimated in the $\gamma$-ray region.

\section{Discussion and Concluding Remarks}
\label{discussion}

\par The quasar, PKS\,1351$-$018, is at the high end of the quasar synchrotron luminosity distribution, $>10^{47}$~ergs/sec (Section 4). The synchrotron spectrum displays very benign blazar characteristics. In particular, the radio flux density is mildly variable even at 100\,GHz in the quasar rest frame. But to the contrary, there were two strong $\gamma$-ray flares in 2011 and 2016 as shown in Section~\ref{gammaray}. We explore this dichotomous behavior through various analyses in this paper. In Section~\ref{lightcurve}, the $\nu_{\mathrm{o}} =5$\,GHz ($\nu =23.5$\,GHz) light curve was created. Even though the variability tends to be on the same order of magnitude as the measurement uncertainty, we were able to find a flaring event in 1990 that required a relativistic outflow with a LOS $<7.1^{\circ}$ to the jet. In Section~\ref{imaging}, we studied the VLBI images at various frequencies from 1995 to 2020. Combined with VLA and MERLIN images, we determine that the source is very compact, confined within $\sim 15$\,mas. There is a conspicuous lobe-like feature to the north at the end of a jet 12\,mas from the nucleus. It appears very steady in both position and flux density over 23 years. We use these properties to constrain a physical model of the lobe in Sections~\ref{radiofit} and \ref{northlobe}. The kinematics of the lobe are used to estimate the jet power, $Q \approx (5.2 \pm 3.2) \times 10^{45}$\,erg~s$^{-1}$, with quite possibly a similar energy flux directed in a counter jet. The physical model is corroborated by two independent results. The jet power agrees with the jet power estimated from studies of the $\gamma$-ray flares in the nucleus ($\sim 1-3$\,kpc away) based on completely different assumptions and physical conditions. Furthermore, a polar LOS $<5.8^{\circ}$ is required for any realistic physical model of the North Lobe, a derivation that is independent of the light curve analysis in Section 2.

We also note that there might be an interaction between the jet and the high ionization wind indicated from the spectral analysis in Section~\ref{energetics}. The jet is very highly curved based on
Tables~\ref{tab:modelfit-Sband}, \ref{tab:vlbi-Cband-fit} and \ref{tab:vlbi-high-fit}, even compared to most blazars \citep{bri08,kha10}. The curving jet is most clearly
illustrated in the right hand panel of Figure~\ref{fig:Cband-vsop}. Based on the elongation of the nuclear Gaussian in multiple epochs the jet begins in the southwest
direction in the first $0.5-0.7$\,mas, then it swings to the southeast at about 1.1\,mas from the nucleus. Apparently, after passing through the ``knot in the north jet''
about $2-3$\,mas from the nucleus in the northeast quadrant, it ends up slightly west of north in the North Lobe. The jet direction rotates $\sim 210\degr$. Most likely, the bending is enhanced by Doppler aberration, but this does not preclude some modest intrinsic bending. In fact there needs to be some intrinsic or seed bending that is
magnified by Doppler aberration. The shear layer between the jet and the denser, slower wind can decelerate the jet and possibly deflect the trajectory a few degrees. This could reduce the magnitude of the relativistic effects, thereby stabilizing the synchrotron luminosity and explain the large swing in the jet PA. This is speculative, but it does tie together three, otherwise coincidental, extreme behaviors of the source:
\begin{itemize}
\item A luminous high ionization wind moving at $\sim 4000$\,km~s$^{-1}$ is rare in radio loud quasars \citep{pun10,ric02}. Such a fast wind is rare even for luminous radio quiet quasars \citep{sul17}.
\item The parsec scale jet trajectory bends $210\degr$.
\item The preponderance of the enormous synchrotron flux is emitted by a region that has low variability.
\end{itemize}

\section{Acknowledgments}
We are grateful to the referee who had many useful comments that improved this work. Matt Lister provided many VLBI fits that benefitted the early stages of this work and motivated the path going forward. We are grateful to Lorant Sjouwerman, Jamie Stevens, and Natasha Hurley-Walker for help and guidance with the NVAS, ATCA and GLEAM data, respectively. Marco Berton and Matt Stevens generously provided useful JVLA data reductions. We thank Shane O'Sullivan for the 5 GHz VLBI images and fits. We were also fortunate to be helped by Narek Sahakyan, Lea Marcotulli and Vaidehi Paliya with the high energy data. Wendy Peters generously provided us with VLITE data. Basic research in radio astronomy at the U.S. Naval Research Laboratory is supported by 6.1 Base Funding. Construction and installation of VLITE was supported by the NRL Sustainment Restoration and Maintenance fund. The VLA is operated by the National Radio Astronomy Observatory (NRAO). We would like to thank Anita Richards of the MERLIN/VLBI National Facility for supplying the 5 GHz data. This work was supported by the National Radio Astronomy Observatory, a facility of the National Science Foundation operated under cooperative agreement by Associated Universities, Inc. This publication made use of the Astrogeo VLBI FITS image database (\url{http://astrogeo.org/vlbi$\_$images/}) maintained by Leonid Petrov. SF thanks the Hungarian National Research, Development and Innovation Office (OTKA K134213) for
support. ABP was supported by the Russian Science Foundation grant 21-12-00241.

\appendix
\section{The C-Band Data in Figures 1 and 2}
\begin{table}[htp!]
\begin{center}
\tiny{
\begin{tabular}{ccccc}
 Date  & Flux Density (mJy)  & Telescope &  Reference   \\
 \hline
 1968 Aug 03 & $940 \pm 47$  & Parkes 64m Telescope & \cite{wal73} \\
 1980 Nov 18 & $880 \pm 44$  & VLA  & \cite{per82}  \\
 1986 Oct 04 & $940 \pm 47$  & VLA  & \cite{dri97}  \\
 1989 Jan 14 & $863 \pm 86$  & VLA  & This Paper  \\
 1989 Mar 30 & $857 \pm 86$  & VLA  & This Paper  \\
 1989 May 28 & $878 \pm 88$  & VLA  & This Paper  \\
 1989 Dec 08 & $834 \pm 83$  & VLA  & This Paper  \\
 1990 Mar 23 & $854 \pm 43$  & VLA  & This Paper  \\
 1990 Apr 29 & $868 \pm 43$  & VLA  & This Paper  \\
 1990 Jul 08 & $987 \pm 99$  & VLA  & This Paper  \\
 1990 Aug 14 & $917 \pm 92$  & VLA  & This Paper  \\
 1990 Nov 04 & $1000 \pm 50$  & VLA  & This Paper  \\
 1991 Jun 15 & $905 \pm 91$  & VLA  & This Paper  \\
 1992 Nov 28 & $879 \pm 88$  & VLA  & This Paper  \\
 1993 Jun 12 & $920 \pm 92$  & VLA  & This Paper  \\
 1993 Sep 08 & $920 \pm 46$  & VLA  & This Paper  \\
 1994 Jan 08 & $936 \pm 47$  & VLA  & This Paper  \\
 1995 Sep 13 & $974 \pm 49$  & VLA  & This Paper  \\
 1995 Dec 15 & $954 \pm 95$  & MERLIN  & This Paper  \\
 1997 Jan 23 & $924 \pm 92$  & VLA  & This Paper  \\
 1997 Jun 25 & $948 \pm 95$  & VLA  & This Paper  \\
 1997 Jul 20 & $988 \pm 99$  & VLA  & This Paper  \\
 1998 Mar 24 & $914 \pm 91$  & VLA  & This Paper  \\
 1998 Apr 23 & $953 \pm 53$  & VLA  & This Paper  \\
 1998 Jun 02 & $911 \pm 46$  & VLA  & This Paper  \\
 1998 Jun 07 & $919 \pm 46$  & VLA  & This Paper  \\
 1998 Jun 11 & $904 \pm 45$  & VLA  & This Paper  \\
 1998 Aug 20 & $884 \pm 88$  & VLA  & This Paper  \\
 1998 Aug 31 & $883 \pm 88$  & VLA  & This Paper  \\
 1998 Dec 01 & $886 \pm 44$  & VLA  & This Paper  \\
 1998 Dec 05 & $894 \pm 89$  & VLA  & This Paper  \\
 2000 Jun 25 & $844 \pm 84$  & VLA  & This Paper  \\
 2000 Oct 16 & $862 \pm 43$  & VLA  & This Paper  \\
 2002 Mar 11 & $885 \pm 88$  & VLA  & This Paper  \\
 2004 Dec 22 & $902 \pm 90$  & VLA  & This Paper  \\
 2005 Jun 13 & $920 \pm 46$  & ATCA  & ATCA Calibrator Database  \\
 2007 Feb 05 & $952 \pm 49$  & ATCA  & ATCA Calibrator Database  \\
 2007 Feb 19 & $937 \pm 47$  & ATCA  & ATCA Calibrator Database  \\
 2008 May 28 & $952 \pm 48$  & ATCA  & ATCA Calibrator Database  \\
 2009 Feb 15 & $936 \pm 47$  & ATCA  & ATCA Calibrator Database  \\
 2010 Feb 13 & $935 \pm 47$  & ATCA  & ATCA Calibrator Database  \\
 2011 May 22 & $988 \pm 49$  & ATCA  & ATCA Calibrator Database  \\
 2012 Mar 18 & $959 \pm 48$  & ATCA  & ATCA Calibrator Database  \\
 2012 Apr 23 & $960 \pm 48$  & ATCA  & ATCA Calibrator Database  \\
 2012 Oct 26 & $945 \pm 47$  & ATCA  & ATCA Calibrator Database  \\
 2013 Feb 08 & $946 \pm 47$  & ATCA  & ATCA Calibrator Database  \\
 2013 May 30 & $960 \pm 48$  & ATCA  & ATCA Calibrator Database  \\
 2013 Sep 12 & $959 \pm 48$  & ATCA  & ATCA Calibrator Database  \\
 2014 Feb 01 & $920 \pm 46$  & ATCA  & This Paper  \\
 2014 Mar 25 & $946 \pm 47$  & ATCA  & ATCA Calibrator Database  \\
 2015 Jun 22 & $888 \pm 44$  & ATCA  & This Paper \\
 2015 Ju1 11 & $898 \pm 45$  & ATCA  & This Paper \\
 2015 Oct 15 & $929 \pm 46$  & ATCA  & This Paper \\
 2015 Dec 09 & $912 \pm 46$  & ATCA  & This Paper \\
 2016 Jan 13 & $946 \pm 47$  & ATCA  & This Paper \\
 2016 Jan 27 & $927 \pm 46$  & ATCA  &  ATCA Calibrator Database  \\
 2016 Feb 16 & $912 \pm 46$  & ATCA  & This Paper \\
 2016 Mar 05 & $894 \pm 45$  & ATCA  & This Paper \\
 2016 Apr 14 & $907 \pm 45$  & ATCA  & This Paper \\
 2016 May 08 & $898 \pm 46$  & ATCA  & This Paper \\
 2016 May 18 & $891 \pm 45$  & ATCA  & This Paper \\
 2016 Jun 08 & $877 \pm 44$  & ATCA  & This Paper \\
 2016 Aug 22 & $868 \pm 43$  & ATCA  & This Paper \\
 2016 Sep 29 & $880 \pm 44$  & ATCA  & This Paper \\
 2016 Nov 04 & $901 \pm 45$  & ATCA  & This Paper \\
 2016 Dec 03 & $886 \pm 44$  & ATCA  & This Paper \\
 2016 Dec 05 & $885 \pm 44$  & ATCA  & This Paper \\
 2017 Jan 21 & $865 \pm 43$  & ATCA  & This Paper \\
 2017 Feb 03 & $862 \pm 43$  & ATCA  & This Paper \\
 2017 Mar 02 & $866 \pm 43$  & ATCA  & This Paper \\
 2017 Apr 11 & $874 \pm 44$  & ATCA  & This Paper \\
 2017 Apr 12 & $873 \pm 44$  & ATCA  & ATCA Calibrator Database \\
 2017 May 14 & $856 \pm 43$  & ATCA  & This Paper \\
 2017 Jun 12 & $876 \pm 44$  & ATCA  & ATCA Calibrator Database\\
 2018 Oct 16 & $858 \pm 43$  & ATCA  & ATCA Calibrator Databaser \\
 2020 Feb 22 & $843 \pm 42$  & ATCA  & ATCA Calibrator Database \\
 2020 Apr 03 & $858 \pm 43$  & ATCA  & ATCA Calibrator Database \\
 2020 Sep 20 & $836 \pm 42$  & ATCA  & ATCA Calibrator Database

    \end{tabular}}
\end{center}
\end{table}
\pagebreak
\section{Error Analysis}
There are many VLBI observations analyzed Section 3. The observations are quite heterogenous, spanning 33 years. We are unable to access the visibility data for some of the observations reported in the literature. So we cannot estimate uncertainties in the Gaussian component flux density and position using the residuals of the fit to the visibility data, the post-fit rms noise, $\sigma_{\mathrm{rms}}$ \citep{fom99,lee08}. Our aim is assign uncertainties to the fitted components in a uniform manner. For example, if some references in literature assign liberal uncertainties to component positions and a conservative estimate is implemented on others this will affect the determination of component motion in a weighted least squares fit to the trajectory. So, we are motivated to use an approximate uncertainty in the distance between the nucleus and the components of $\sim 1/5$ of the projection of the elliptical Gaussian synthesized beam FWHM along this direction \citep{lis09,lis13}. This can be applied uniformly to all of the observations.

The uncertainty in the component flux densities for the VLBI measurements are of two varieties. Since the core has $>90\%$ of the total flux, we note that its uncertainty is approximated as the absolute flux density calibration uncertainty of $\approx 10\%$ associated with the VLBI observation \citep{hom02,pus12}. This approximation is applied uniformly to all the observations. There is a signal to noise ratio (SNR) driven uncertainty in the total flux density, $S$, of a component, $\sigma_{\mathrm{tot}}$, that is the dominant uncertainty for the other much weaker components \citep{fom99,lee08}. In particular, $\sigma_{\mathrm{tot}} \simeq S(1+\mathrm{SNR})^{0.5}/\mathrm{SNR}$, where $\mathrm{SNR} = \sigma_{\mathrm{rms}}/S_{\mathrm{peak}}$ and $S_{\mathrm{peak}}$ is the peak intensity of the component. But as stated above, $\sigma_{\mathrm{rms}}$ is not known or derivable in some cases. In order to derive a uniform estimate of the uncertainty of the flux density, we computed the SNR driven uncertainty for numerous cases for which we could determine $\sigma_{\mathrm{rms}}$. This uncertainty was added in quadrature with the 10\% absolute calibration uncertainty. There are three components (besides the core) that appear in our fits in Tables 1, 3 and 4. For various observations, we estimate a total uncertainty of $\sim 40\%$ in the North Lobe flux density, $S_{\rm{North\, Lobe}}$, $\sigma_{\mathrm{North\, Lobe}} \approx 0.4S_{\mathrm{North\, Lobe}}$. The other components are nearly point sources ($S$ measured in mJy $\approx S_{\mathrm{peak}}$ measured in mJy~beam$^{-1}$) and theoretically the uncertainty should scale approximately with flux density, $S$, as $(\sigma_{\mathrm{component}}/S) \approx 0.4 (S_{\rm{North\, Lobe}}/S)^{0.5}$. Our uncertainty estimates verify this to be true empirically as well. Using this prescription, we can uniformly apply these approximate uncertainties to all the Gaussian brightness distribution models in this paper. None of our results in Section~\ref{northlobe} depend strongly on the precise magnitudes of these uncertainties in flux density.

\end{document}